%% file: main.tex
\documentclass[a4paper,11pt]{article} 
\usepackage{jcappub} 
\pdfoutput=1 
\usepackage[utf8]{inputenc} 
\usepackage{subcaption} 
\usepackage{threeparttable} 
\usepackage[table]{xcolor} 
\usepackage[noabbrev]{cleveref} 
\Crefname{equation}{Eq.}{Eqs.} 
\usepackage{aasmacros} 

\title{\boldmath Bayesian inference methodology for Primordial Power Spectrum reconstructions from Large Scale Structure}

\author[1,2]{G. Martínez-Somonte}
\author[3]{A. Marcos-Caballero}
\author[1]{E. Martínez-González}
\author[4,5]{G. Cañas-Herrera}

\affiliation[1]{Instituto de Física de Cantabria, CSIC-Universidad de Cantabria,\\Avenida de los Castros s/n, E-39005 Santander, Spain}
\affiliation[2]{Departamento de Física Moderna, Universidad de Cantabria,\\Avenida de los Castros s/n, E-39005 Santander, Spain}
\affiliation[3]{Department of Theoretical Physics, University of the Basque Country UPV/EHU\\Barrio Sarriena s/n, Leioa, Vizcaya, Spain}
\affiliation[4]{European Space Agency/ESTEC
Keplerlaan 1, 2201 AZ Noordwijk, The Netherlands}
\affiliation[5]{Lorentz Institute for Theoretical Physics, Leiden University PO Box 9506, Leiden 2300 RA, The Netherlands}

\emailAdd{gmsomonte@ifca.unican.es}
\emailAdd{marcos@ifca.unican.es}
\emailAdd{martinez@ifca.unican.es}
\emailAdd{guadalupe.canasherrera@esa.int}

\abstract{
We use Bayesian inference to develop a non-parametric method to reconstruct the primordial power spectrum $P_{\mathcal{R}}(k)$ from Large Scale Structure (LSS) data. The performance of the method is assessed by testing it against simulations of the clustering of high-$z$ (QSOs) objects. Their clustering is derived from different templates of the primordial power spectrum motivated by models of inflation: the Standard Model power law characterized by the two parameters $A_s$ and $n_s$; a local feature template; and a global oscillatory template. The primordial power spectrum is reconstructed using $N$ knots in the log $\{k,P_{\mathcal{R}}(k)\}$ plane while sampling the cosmological parameters $\{H_0,\Omega_b, \Omega_c\}$. We use two statistical tests to examine the reconstructions for signs of primordial features: a global test comparing the evidences and a novel local test quantifying the power of the hypothesis test between the power law model and the marginalized probability over $N$ model. We also discuss results of an application to low-$z$ (ELGs) objects with two different photometric errors keeping the cosmology fixed. The method shows good performance in all scenarios considered. In particular, the tests show no feature detection for the standard power-law primordial power spectrum; yet, the method is able to detect power spectrum deviations at a percent level for all considered features, combining either the low-$z$ or the high-$z$ redshift bins. In addition, we include a test proof-of-concept application to real data from the Sloan Digital Sky Survey Luminous Red Galaxy Data Release 4 (SDSS LRG 04), finding no preference for deviations from the primordial power law. The method is flexible, model independent, and suitable for its application to existing and future LSS surveys.}


\begin{document}

\maketitle
\flushbottom 

\section{Introduction}
\label{sec:introduction}

Most cosmological observations support the hypothesis that the primordial fluctuations were adiabatic, Gaussian and quasi-scale invariant, and that the background universe was spatially isotropic and homogeneous \cite{Planck18Parameters}. These properties, together with several shortcomings of the standard Hot Big Bang scenario \citep{InflaGuth1981,InflaLinde1982}, provide strong motivation for cosmological inflation \cite{InflaGuth1981,InflaLinde1982,InflaBrout1978,InflaStarobinski1980,InflaAlbrechtSteinhardt1982,InflaLinde1983}, a hypothetical epoch of exponential expansion in the early universe. However, the nature and origin of the fields that drove inflation remain largely unknown and poorly constrained by current observations.

The primordial correlation functions encode very valuable information about the physical mechanism that generated the initial conditions for cosmic structure formation. Some well-motivated theoretical scenarios can produce distinctive features in those functions, such as the primordial scalar power spectrum of curvature perturbations\footnote{For now on we refer to it as `primordial power spectrum' only.} $P_\mathcal{R}(k)$. $P_\mathcal{R}(k)$ is a key quantity to probe the physics of the very early universe, allowing us to test and constrain different inflationary models. $P_\mathcal{R}(k)$ is usually parametrized by a simple power law with two parameters: the amplitude $A_s$ and the spectral index $n_s$ of the primordial comoving curvature perturbations.

The predictions of the simplest single field slow-roll models of inflation, of nearly-Gaussian and quasi-scale invariant power spectrum for scalar and tensor perturbations \cite{Starobinski1979,Linde1982}, are consistent with the latest results of the ESA Planck satellite that, in particular, provides a value for the scalar power spectrum parameters of $A_s = \left( 2.10_{-0.04}^{+0.03} \right) \times 10^{-9}$  and $n_s = 0.965 \pm 0.004$ \cite{Planck18Parameters}. Models as the Higgs inflation \cite{HiggsInflation} or the Starobinsky $R^2$ inflation \cite{StarobinskiInflation} are favoured by the latest Planck data \cite{JudgmentDayInflation} from a plethora of slow-roll inflationary models \cite{EnciclopediaInflation}.

Departures from the slow-roll scenario can have significant cosmological implications, such as the production of primordial black holes and the enhancement of the inflationary gravitational wave spectrum at small scales \cite{AnalyticAproachNonSlowRoll}. Therefore, it is important to look for observational signatures of slow-roll deviations \cite{MirandaHuModeloFeature,SearchingFeatures1,SearchingFeatures2} or modified scalar field dynamics during inflation \cite{InverseScalarDynamics}. There are various mechanisms acting in different inflationary models that can produce deviations from the power law form of $P_\mathcal{R}(k)$. For instance: logarithmic oscillations arising from non-Bunch-Davies initial conditions \cite{ModelMartin2001,ModelMartin2003,ModelBozza2003} or from axion monodromy \cite{ModelFlauger2017}; linear oscillations predicted by boundary effective field theory models \cite{ModelLinearOsci,ModelJackson2013}; localized oscillatory features induced by a step in the inflaton potential \cite{ModelAdams2001} or in the sound speed \cite{ModelAchucarro2010, ModelInflationSpeedSound}. Other models exhibit cutoffs at large scales \cite{ModelContaldi2003,ModeSinha2006} or more general modulations \cite{ModelDanielsson2002,ModelChen2012}.  
One way of exploring the presence of features is by looking to the Cosmic Microwave Background (CMB) anisotropies. Some anomalies in the CMB, that were originally detected in the NASA WMAP data \cite{WMAPAnomalies}, were confirmed later by Planck \cite{AnomaliesPlanck18}: power suppression at large scales \cite{AnomaliesSupression}, low variance \cite{AnomaliesLowVariance1, AnomaliesLowVariance2}, hemispherical power asymmetry \cite{AnomaliesHemispherical1, AnomaliesHemisphericalWMAP7}, preference for odd parity \cite{AnomaliesOddParity}, tension in the lensing parameter $A_l$ \cite{Planck15Parameters}, and the ``cold spot" \cite{AnomaliesColdSpot1, AnomaliesColdSpot2} are some of the more remarkable ones. The statistical significance of each of these anomalies is inconclusive, but it is worth studying them due to the relevance that their existence would have to uncover new physics beyond the $\Lambda$CDM model. We will refer to this model, including the primordial power law power spectrum, as the Standard Model (SM) of cosmology. 

Large Scale Structure (LSS) stage IV galaxy surveys \cite{JPASSpecifications,DESISpecifications,EuclidSpecifications} are expected to reveal more details than CMB experiments at intermediate scales, ranging from wavenumbers $k$ of about 0.01 $\text{h} \text{ Mpc}^{-1}$ to $1 \text{ h} \text{ Mpc}^{-1}$. The high signal to noise $S/N$ of their galaxy power spectra allows us to probe the potential existence of features in the primordial seeds of the LSS. Developing accurate methods to perform this analysis is crucial for extracting the maximum amount of information from these forthcoming surveys.

 The shape of the primordial power spectrum can be determined following two different approaches: parametrization and reconstruction. Parametrization relies on selecting a specific form or model of $P_\mathcal{R}(k)$ and constraining its parameters using CMB and/or LSS data. Various parametrizations of $P_\mathcal{R}(k)$ have been applied in the literature such as  \cite{ModelContaldi2003, ReconBridle,Parametric3Simon2005,Parametric4Bridges2006,Parametric2Sinha2006,Parametric6Covi2006,Parametric5Bridges2007,Parametric7Joy2009,Parametric8Paykari2010,ReconModulationWMAP,Parametric9Guo2011,Parametric10Goswami2013, Ballardini2, Ballardini3, EuclidSearchFeatures} and those used in the inflation analysis of the Planck data \cite{PlanckInflation18,PlanckInflation13,PlanckInflation15}. Also parametrizations of the two point correlation function can be found \cite{Ballardini1}. Reconstruction, on the other hand, does not assume any model or template for $P_\mathcal{R}(k)$, but rather infers its shape from the data. Several methods have been developed for model independent reconstruction of $P_\mathcal{R}(k)$ based on different inference methods, such as Bayesian inference, linear interpolation methods \cite{ReconBridle} \cite{ReconLinearInterpolation1}, combination of top hat functions \cite{ReconTopHat}, wavelet expansion of $P_\mathcal{R}(k)$ \cite{ReconWaveletReal, ReconWaveletSimulated}, smoothing splines \cite{ReconSplines,ReconSplinesWMAP1,ReconSplinesWMAP3}, fixed wavenumber knots joined with cubic splines \cite{PlanckInflation18,PlanckInflation15}, the critical filter method \cite{ReconFilter}, different spline techniques \cite{ReconOldReconstructions,WillPaper,WillPaper13} or placement of $N$ free knots in the $\{k$, $P_\mathcal{R}\}$ plane \cite{PlanckInflation18, PlanckInflation15,ReconOldReconstructions,WillPaper,WillPaper13}; penalized likelihood, using function space generalization of the Fisher matrix formalism \cite{ReconCMBlike} or a $P_{\mathcal{R}}(k)$ ansatz \cite{PlanckInflation18, PlanckInflation13, PlanckInflation15}; Principal Component Analysis, using expansion of a orthonormal set of basis functions \cite{ReconLSSOrthogonal}; and sparsity of the primordial power spectrum, using a sparsity-based linear inversion method \cite{ReconPRISM}. These works have been applied to CMB and/or LSS data to reconstruct $P_\mathcal{R}(k)$ and test for deviations from the standard power law form.

Most non-parametric methods of $P_{\mathcal{R}}(k)$ reconstruction do not find any statistically significant deviations from the primordial power law \cite{ReconBridle, ReconTopHat,ReconPRISM,ReconSplines,ReconSplinesWMAP1,ReconSplinesWMAP3,ReconLinearInterpolation1,WillPaper13}. The precision of these methods vary from subpercent to $30\%$, depending on the data and the technique used. The Planck collaboration applied three non-parametric methods to test the power law hypothesis and found no statistically significant evidence for deviations with a precision approaching $1\%$. The most notable deviation, although not statistically significant, was a deficit in power at $k \approx 0.001 \text{ h}\text{ Mpc}^{-1}$ ($\ell \approx 30$) \cite{PlanckInflation18,PlanckInflation13,PlanckInflation15}. Some methods have also focused on detecting features at scales between $k \approx 0.01 \text{ h}\text{ Mpc}^{-1}$ and $k \approx 0.2 \text{ h}\text{ Mpc}^{-1}$ \cite{ReconLSSOrthogonal,ReconWaveletReal,ReconWaveletSimulated,ReconFilter,ReconPRISM,WillPaper13}. In \cite{WillPaper}, models that can account for a lack of power at $k \approx 0.001 \text{ h}\text{ Mpc}^{-1}$ and in $k > 0.1 \text{ h}\text{ Mpc}^{-1}$ are slightly favoured against the power law parametrization in a Bayesian sense. However, all methods are consistent with a featureless tilted power law $P_\mathcal{R}(k)$ that agrees with Planck. Earlier studies with WMAP data have reported some features, such as a modulation around $k \approx 0.009 \text{ Mpc}^{-1}$ \cite{ReconModulationWMAP} and fine structure at $k \approx 0.002 \text{ Mpc}^{-1}$ and $k \approx 0.009 \text{ Mpc}^{-1}$ \cite{ReconFineStructure}. Moreover, the power law parametrization is claimed to be disfavoured against the Lasenby \& Doran model \cite{LDModel}, which produces a lack of power at scales $k \sim 10^{-4} \text{ Mpc}^{-1}$ \cite{ReconOldReconstructions}.

 Reconstruction techniques can identify coarse characteristics in $P_{\mathcal{R}}(k)$, but they are limited in detecting higher frequency features that are predicted by various physical mechanisms \cite{ChlubaFeatures}. The parametric approach can provide the necessary resolution to detect such features, but it depends heavily on the chosen model or template of $P_{\mathcal{R}}(k)$, which makes reconstructions more suitable for obtaining model independent information. In our non-parametric methodology, free placement of knots in the log $\{k,P_{\mathcal{R}}\}$ plane does not require any prior $k$-binning of $P_{\mathcal{R}}(k)$. As a result, it is sensitive to both global and local features in $P_{\mathcal{R}}(k)$. Model complexity is penalized by comparing evidences $Z$, which facilitates the comparison between different $N$ configurations when reconstructing $P_{\mathcal{R}}(k)$. This advantage is harder to obtain in other methods, such as those based on basis functions, top hats, or wavelet expansion. Moreover, our methodology is flexible and adaptable to additional data from diverse surveys, and it can increase the number of sampled knots if needed. However, one limitation of our methodology is the lack of smoothness introduced by the linear splines connecting the knots. Other approaches using smoothing splines or cubic splines offer greater sensitivity to curvature features in $P_{\mathcal{R}}(k)$. Additionally, our method may miss very sharp features, which is a common drawback of non-parametric approaches compared to parametric ones. Nevertheless, the discrete width at which the sampler evaluates the knots in the log $\{k,P_{\mathcal{R}}\}$ plane can be reduced, at the expense of a higher computational effort.      

Assuming a specific deviation from the standard inflationary model may be challenging due to the vast number of proposed models. Our objective is to detect features in $P_\mathcal{R}(k)$ from LSS galaxy clustering data, both simulated and real, with a non-parametric Bayesian method that does not impose any assumptions regarding any particular inflationary model. By reconstructing $P_\mathcal{R}(k)$ and quantifying statistical deviations from a power law, we aim to provide information about the very early universe.

The structure of the paper is as follows: \cref{sec:methodology}
 describes the methodology used to perform the $P_\mathcal{R}(k)$ reconstructions with nested sampling, along with the details of the tests used for the subsequent analysis. \Cref{sec:Simulations} describes the galaxy power spectrum simulations used to test our method, including the survey specifications and modelling. \Cref{sec:ApplicationSimulations} presents the results obtained from applying the methodology to simulated spectra, considering four different cases for the primordial power spectrum: the power law spectrum of the Standard Model (SM), a bump and an oscillatory feature from the same local template, and a global log-log oscillatory feature template. The smallest power deviations that can be detected are also given for each feature template. \Cref{sec:AppendixApplicationSDSSLRG04} focuses on the application of the methodology to real observational data from the SDSS LRG 04 catalogue, discussing the obtained results about possible deviations from the Standard Model. \Cref{sec:Conclusions} presents the conclusions derived from this work and some possible lines of future work. Finally, \cref{sec:AppendixSwitching,sec:AppendixPolyChord,sec:AppendixModifiedPPS,sec:AppendixPosteriors,sec:AppendixComparisonCosmology,sec:AppendixApplicationSDSSLRG04} include the appendices that cover the solution of the label switching problem, the PolyChord sample selection criterion, and a description of the feature model used for the bump/oscillatory local template.

\section{Methodology}
\label{sec:methodology}

We use a model independent and non-parametric approach to reconstruct the primordial power spectrum $P_{\mathcal{R}}$ by sampling $N$ knots freely in the log $\{k,P_{\mathcal{R}}\}$ plane using nested MCMC sampling. This approach follows \cite{WillPaper} and some of the inflation analyses of the Planck collaboration \cite{PlanckInflation18,PlanckInflation15}. In the first subsection we explain our procedure, priors and likelihood, and the method to represent the $P_\mathcal{R}(k)$ reconstructions. In the second subsection we present two tests for detecting primordial features in the reconstructed primordial power spectra.
\subsection{$P_{\mathcal{R}}(k)$ reconstructions}

\label{subsec:reconstructions}

We use a Bayesian framework in this work. Given a cosmological model $M$ characterized by a parameter set $\Theta$ and a set of cosmological data $D$, the Bayes Theorem allows us to update a prior $P (\Theta|M )$ to a posterior $P(\Theta | D, M)$ using the likelihood $\mathcal{L} = P(D | \Theta, M)$ and an evidence $Z$, which can be computed as:
	\begin{equation}\label{BayesianEvidence}
Z \equiv P(D | M)=\int P(D | \Theta, M) P(\Theta | M) \mathrm{d} \Theta.
	\end{equation}
To sample the posteriors and evidences for our reconstructions we use Cobaya \cite{CobayaMandatory1} \cite{CobayaMandatory2}: a framework for sampling and statistical modelling. In this framework we use the nested sampler PolyChord \cite{MandatoryPolyChord1,MandatoryPolyChord2}, as the Metropolis–Hastings sampler in Cobaya is insufficient due to the computational complexity of our posteriors and evidences. The PolyChord algorithm improves the sampling efficiency for our evidence computation in \cref{BayesianEvidence} and enables the acquisition of posterior samples, which present multi-modal behaviour. PolyChord has been used for many cosmological purposes, from which we highlight the reconstructions of the deceleration parameter $q(z)$ \cite{CC7DecelerationReconstructions} and the primordial power spectrum in a non-parametric Bayesian approach \cite{WillPaper}, a methodology followed in this work. The specific priors used are described in \cref{tab:Priors}. The evidences obtained from \cref{BayesianEvidence} will be normalized to have its maximum value equal to 1.

 	\begin{table}[h]
 		\centering
 		\input{PriorTable}
 		\caption{Priors used in this work. The sampled cosmological parameters $\{H_0,\Omega_b,\Omega_c\}$ follow Gaussian distributions from Planck DR 3 uncorrelated posteriors, while the knots are sampled from a uniform prior. The fiducial values for the cosmological parameters used for the analysis are also listed.}
			\label{tab:Priors}
	\end{table}

\begin{figure}[t]
	 \centering
	    \includegraphics[width=0.68\textwidth]{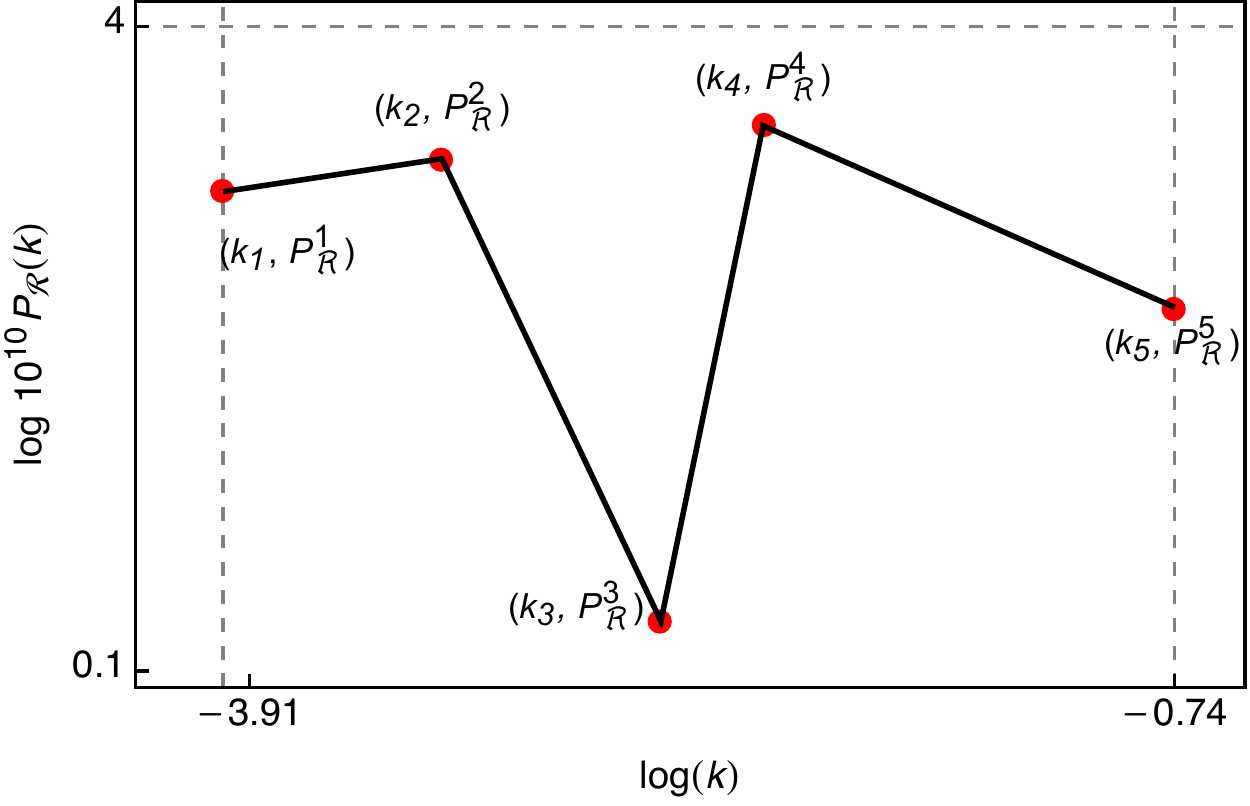}	 \caption{Illustration of the methodology for a 5 knots case ($N=5$). We fix both $k_1 = 0.02 \text{ h}\text{ Mpc}^{-1}$ and $k_{N} \equiv k_5 = 0.45 \text{ h}\text{ Mpc}^{-1}$, which correspond to the largest and smallest scales of the survey. The sampled parameters are the $\text{log}(x)$ (see next paragraph) for each \{$k_2$,$k_3$,$k_4$\} and the logarithms of \{$P_{\mathcal{R}}^1$,$P_{\mathcal{R}}^2$,$P_{\mathcal{R}}^3$,$P_{\mathcal{R}}^4$,$P_{\mathcal{R}}^5$\}, denoted as $\text{log}(y)$. We impose a flat prior with the upper and lower limits for all the $\{\log(x_i), \log(y_i)\}$ coordinates according to the dashed lines: scales within the survey limits and $P_{\mathcal{R}}$ from $1.0 \times 10^{-10}$ to $5.6 \times 10^{-9}$.}
	 \label{fig:MethodologyPlot}
	\end{figure}

In the present work the $P_\mathcal{R}(k)$ reconstructions are performed in a model independent approach. The underlying assumption is that the primordial power spectrum can be represented as a series of $N$ knots, connected by linear splines in logarithmic scale. The knots are pairs of coordinates $\{\text{log}(k_i), \text{log}(P_{\mathcal{R}}^{i})\}$ that are sampled with PolyChord. We consider different number of knots configurations with $N \in [2,9]$, that we can marginalize in order to achieve a non-parametric method. The number of knot parameters are $2N-2$, and together with the three cosmological parameters $\{H_0,\Omega_b,\Omega_c\}$ makes a total of $2N+1$ parameters to be sampled. The sampler PolyChord was used with $25$ live points for each sampled parameter, for a total of $50N+25$. We employed $\approx 5 \times 10^4$ CPU hours, using 128 CPU cores for each reconstruction. \Cref{fig:MethodologyPlot} illustrates the methodology for a case of $N = 5$. The knot position is determined by maximizing the likelihood of the knot-constructed spectrum given the galaxy power spectrum data from a specific LSS survey. To achieve this, a model to relate the knot-constructed primordial power spectrum into a galaxy power spectrum is required. The likelihood is given below in this section, and the model is described in the next section.

We assume that the $P_{\mathcal{R}}^{(0)}(k)$ is given in the scale range $k \in [0.02,0.45] \text{ h}\text{ Mpc}^{-1}$, divided in 125 elements with uniform logarithmic steps $\text{Log}(\Delta k) = 0.02$. For low-$z$ objects, that are also discussed at the results, this range is reduced to $k = 0.2 \text{ h}\text{ Mpc}^{-1}$, with the same step.\footnote{For high-$z$ the non-linearities are expected to occur at smaller scales according to a smaller growth factor $D(z)$. By comparing the typical amplitude of fluctuations in a sphere of radius $R$, $\sigma_R(z)$, at the two redshifts considered, we obtained that the same amplitude of fluctuations occur for high-$z$ at scales approximately 2.35 times smaller than for low-$z$, resulting in a scale $k$ approximately 2.35 times bigger.} We adopt flat priors for both $P_{\mathcal{R}}$ in a wide range $P_{\mathcal{R}} \in [0.10,5.60] \times 10^{-9}$ and $x \in [0,1]$. For the cosmological parameters $\{H_0,\Omega_b,\Omega_c\}$ we use Planck DR 3 uncorrelated Gaussian posteriors. This decision is motivated to reduce possible degeneracies between the knots and the cosmological parameters when exploring the parameter space and, therefore, decrease the computational cost. In the future, primordial power spectrum reconstructions will benefit, most likely, from a wide range of cosmological observations, including the full Planck likelihood.

We reconstruct $P_{\mathcal{R}}(k)$ by maximizing a likelihood function that quantifies the agreement between the theoretical model of the monopole galaxy power spectrum $P^{(0)}_{g}(\mbox{model})$ and the one derived from the data catalog $P^{(0)}_{g}(\mbox{data})$. We evaluate the spectra for different bins of redshift $z$ and for each value of the survey scale grid. The likelihood $\mathcal{L}$ used in this analysis follows a multi-variable Gaussian form:
		\begin{equation}\label{likelihood}
- 2 \mbox{Log}(\mathcal{L}) = \sum_{i=1}^{i=Z} \sum_{j=1}^{j=K}\left[ x_{ij}^2 \frac{1}{\sigma^2_{ij}} + \mbox{Log}\left( \sigma^2_{ij}\right)\right],
	\end{equation}
where $x_{ij} \equiv [P^{(0)}_{g,ij}(\mbox{model}) - P^{(0)}_{g,ij}(\mbox{data})]$; $i$ is the index for the different redshift bins, with $Z$ the total number of redshift bins for the survey; $j$ is the index accounting for the scales, with $K$ the total number of scales for that survey; and $\sigma_{ij}^2$ is the variance calculated from $P^{(0)}_{g,ij}(\mbox{model})$, as explained in the next section. In general a covariance matrix is needed. Since we consider uncorrelated scales and redshift bins, it is diagonal and we can reduce it to its variance elements. The reconstruction of $P_{\mathcal{R}}(k)$ are obtained from the values of $P^{(0)}_{g,ij}(\text{model})$ that maximize $\mathcal{L}$.

We summarize our sample selection criteria in \cref{sec:AppendixPolyChord}. After the sampling is done and a first set of chains is chosen, we get a large number of the sampled primordial power spectra for each number of knots configuration, each one denoted as $P_{\mathcal{R},j,N}(k)$. Each primordial spectrum has associated its own normalized importance weight $\omega_{j,N}$.

We can merge all the primordial spectra coming from any $N$ configuration by reconstructing a marginalized probability over $N$ of $P_{\mathcal{R}}(k)$. Since PolyChord calculates the individual evidences $Z_N$, the marginalization is done by combining all spectra from all $N$ with a new set of evidence-dependent weights $q_{j,N}$:
	\begin{equation}\label{newweights}
q_{j,N} = \omega_{j,N} \frac{Z_N}{\sum_{M}Z_M}.
	\end{equation}
Each weight $q_{j,N}$ can be interpreted as the joint probability of both the chain $j$ and the $N$ knots configuration. This joint probability is the product of two factors: the conditional probability of the chain $j$ given $N$, $\omega_{j,N}$, and the probability of the $N$ configuration given by the normalized evidence.
Their corresponding constructions of the primordial power spectrum $P_{\mathcal{R},{j,N}}(k)$ have an associated probability $q_{j,N}$, from which we determine the confidence intervals of $P_{\mathcal{R}}(k)$ according to the $q_{j,N}$ distribution. 

We test our method on simulated data with different $P_\mathcal{R}(k)$, based on the power spectrum templates given in \cref{ModifiedPPS} and \cref{ModifiedPPSLogLog} and described in \cite{PlanckInflation18,ModelMartin2001,ModelMartin2003,ModelBozza2003}. These templates cover a broad range of theoretical motivated models. We also apply our methodology to real observational data obtained from the Sloan Sky Digital Survey Luminous Red Galaxy data release 4 (SDSS LRG 04) in \cref{sec:AppendixApplicationSDSSLRG04}.

We focus on the small scale regime where the LSS surveys can achieve higher precision than the CMB experiments. For simplicity we consider uncorrelated bins of $k$. An estimate of these correlations was obtained by performing simulations based on the mock generation code \texttt{lognormal\_galaxies} \cite{LognormalGalaxies} for the redshift bins and volumes considered in this paper, finding an upper limit for the non-diagonal terms of $\approx 30$\% at the smallest scales.

Two statistical tests are used to analyze the reconstructed spectra and assess possible deviations from the power law assumption, as explained below.

\subsection{Feature detection from $P_{\mathcal{R}}(k)$ reconstructions}

To detect features in the reconstructed $P_{\mathcal{R}}(k)$, we perform two tests for each reconstruction. The first test is a global one based on the comparison of the evidence ratios and the second test is a novel local one based on a hypothesis test. Both tests are applied to the comparison of the $N=2$ and the marginalized probability over $N$ reconstructions.

\subsubsection{Global test}

Evidence comparison for model selection and comparison with the SM is widely used \cite{BayesianEvidencesPaper}. In this work we use the evidence ratio $Z_{\text{max}}/Z_2$, where $Z_2$ is the evidence of the primordial power law reconstruction and $Z_{\text{max}}$ is the highest value among the evidences $Z_N$ except $Z_2$. The $Z_{\text{max}}/Z_2$ ratio measures how a power law primordial spectrum compares with a different knot number configuration spectrum with the highest evidence. 
We have checked that the sampling errors in the estimates of the evidences are less than 1\% for values above 1\% of the maximum evidence. In this way, we are confident in the application of the Jeffreys criterion \cite{Jeffreys1998theory} for quantifying the feature detection status. 

This global test provides an effective way of measuring how different $N$ configurations compare when reconstructing $P_{\mathcal{R}}(k)$. The detection of a feature can be claimed, but with only this analysis we could not identify at which scales the deviations are more significant. To locate the deviations at the scales they occur, we use a local test, explained below.

\subsubsection{Local test}

We perform another complementary analysis of the $P_{\mathcal{R}}(k)$ reconstructions. This analysis is a novel approach for detecting features in $P_{\mathcal{R}}(k)$, providing insights about possible features present in $P_{\mathcal{R}}(k)$ in a localized way, since it enables us to identify the scales at which the deviations occur.

The local test is a hypothesis test applied to the distributions of the $N = 2$ and the marginalized probability over $N$ reconstructions for each value of the survey grid scale. The details of the hypothesis test can be found in \cite{Cowan}. The significance level considered for all the tests is $\alpha = 0.05$. The power of the test $1-\beta$ quantifies the separation between both distributions providing a measure of the significance of the deviation from the power law.

We establish the feature detection status using the following thresholds for the power of the test: $1-\beta > 0.95$ at any $k$ indicates that both distributions are separated enough to consider the feature detection at that scale; $0.95 > 1-\beta > 0.5$ shows a hint of a feature; and $0.5 > 1-\beta$ shows no evidence of the presence of a feature.

The global test is a statistically more robust test for feature detection, since it compares evidences computed for all the scales and it is less sensitive to fluctuations, but the local test enables us to locate features at certain scales $k$, providing an indicator for the existence and location of features in $P_{\mathcal{R}}$. We apply both test to all our reconstructions, exploiting the complementary information that both methods offer.

\section{Galaxy power spectrum simulations: survey specifications and modelling}
\label{sec:Simulations}

In order to test the methodology we apply it to simulations. In this work we simulate galaxy power spectra using galaxy clustering modeling. Recent galaxy clustering models account for various effects, such as Baryon Acoustic Oscillations (BAO) modelling \cite{SDSSLRG04, EuclidForecastModel}, non-linear corrections \cite{Ballardini2, BOSSDR12, AgnosticBOSS}, the Fingers of God \cite{FoGPaper} and Alcock-Paczynski \cite{APEffectOriginal, APEffect} effects, or complex shapes of the redshift-space power spectrum \cite{RSDKaiser, RSDScoccimarro, TNS1, TNS2}. A simple BAO modelling and non-linear corrections will be considered for the SDSS LRG real data application in \cref{sec:AppendixApplicationSDSSLRG04}.

All the simulations described in this section have been done assuming a linear regime using a Kaiser redshift-space power spectrum, encompassing both the Fingers of God and the Alcock-Paczynski effects. Although this model resembles those used in the forecasts of upcoming surveys \cite{EuclidForecastModel}, we focus on testing our methodology, without aiming for a state-of-the-art modelling of the simulated data. A comparison of the deviations w.r.t. non-lineal modelling \cite{EuclidForecastModel, BOSSDR12} at scales $k \sim 0.1 \text{ h}\text{ Mpc}^{-1}$ shows an underestimation of power of $\approx 30 \%$. However, this underestimation in the modelling at the smallest scales is not biasing our results since it is present coherently in both the model and the simulations. Therefore, we consider that for our analysis, scales up to $k = 0.4 \text{ h}\text{ Mpc}^{-1}$ are adequate enough for our purposes. When dealing with low-$z$ objects we reduce the smallest scales considered to $0.2 \text{ h}\text{ Mpc}^{-1}$, since non-linear effects happen at larger scales. We explain how our simulations are done in the rest of this section.

\subsection{Simulated surveys specifications}
\label{sec:Surveys}

We differentiate between low and high redshift objects:
Luminous Red Galaxies (LRGs) and Emission Line Galaxies (ELGs) are examples of the former, while Quasi-Stellar Objects (QSOs) of the later. Typical redshifts covered by current planned surveys for galaxies are $z\lesssim 1$ whereas for QSOs $z\sim 2$. In this work we show the results for high-$z$ objects, for which we choose a redshift binning with a step $\Delta z = 0.2$ and evaluate the power spectra at the central values of the bins.

We estimate mean values of the photometric redshift error $\delta_z$ from J-PAS representative data \cite{MiniJPAS, QueirozQSOMocks, QueirozPreparation} by weighting the $\delta_z$ associated to each redshift bin with the densities derived below (see \cref{tab:Densities}). For high redshift objects a single value $\delta_z = 0.0036$ is obtained considering all the QSOs, and for low-$z$ objects, a low photometric error of $\delta_z = 0.0038$ is obtained using the 30\% best determined redshifts, and a high photometric error of $\delta_z = 0.039$ is derived when the 50\% best ones are considered.

The observed sky fraction assumed for the simulations is $f_{\mbox{sky}} = 0.2575$, equivalent to an area of $10620 \text{ deg}^2$, an intermediate value between expected $f_{\mbox{sky}}$ of future LSS surveys, such as J-PAS with an area of $8500 \text{ deg}^2$ \cite{JPASSpecifications}, DESI with $14000 \text{ deg}^2$ \cite{DESISpecifications}, or Euclid with $15000 \text{ deg}^2$ \cite{EuclidSpecifications}.

\subsubsection{Bias model}

Galaxy biasing is a complex process that needs to be modeled to extract cosmological information from current and future LSS data. Different tracers of the LSS exhibit different biases, which can be derived from bottom-up or top-down approaches, with local or non-local bias, and considering non-linearities. In this work we focus on spectra with scales larger than $0.45 \text{ h}\text{ Mpc}^{-1}$. For these scales we assume scale independent bias models and neglect the effects of scale-dependent deviations in the bias. 

The QSO bias model is obtained by matching the integrated correlation function for QSOs with the expected one for the WMAP/2dF cosmology \cite{Croom2005}. The model is parametrized as \cite{BiasHighZ}:
	\begin{equation}\label{biasQSO}
b(z) = 0.53 + 0.289 (1+z)^2.
	\end{equation}
For the low-$z$ galaxies we adopt the relation in \cite{Fry1996}: $b(z) = \frac{b_0}{D(z)}$ with an ELG-like biasing, with $b_0 = 0.84$ following \cite{RescoMaroto} and consistent with \cite{JPASSpecifications}.
 
Galaxy clustering observables require non-linear and non-local bias models for a broad range of scales, as linear and scale independent bias models are inadequate \cite{LinearBiasInsufficient}. For instance, the power spectrum multipoles and the bispectrum of BOSS CMASS DR11 galaxies were analyzed using such models \cite{RoyMcDonald2009,Beutler2014,GilMarin2015a}, and they are also necessary for more recent galaxy clustering data. However, the main goal of this work is to test our reconstructions methodologically rather than to model galaxy clustering with the latest techniques.

\subsubsection{Number density of galaxies}

To estimate the uncertainties in the monopole galaxy power spectrum, we need the densities of the survey objects, which
depend on both the redshift and the photometric error, i.e. $\hat{n}(z,\sigma_z)$, where $\sigma_z = \delta_z (1+z)$.
We use realistic values of densities $\hat{n}$ for both low-$z$ and high-$z$ power spectrum simulations, based on the expectations of the J-PAS collaboration \cite{JPASSpecifications}. To estimate the densities we follow \cite{MiniJPAS} for low redshift (ELGs) and \cite{QueirozQSOMocks, QueirozPreparation} for high redshift (QSOs). The values of the densities that we use in our power spectrum model are obtained by fitting the density data with the following function for $\hat{n}(z)$:
	\begin{equation}\label{sourcegalaxyfunction}
\hat{n}(z) = C_0 z^\alpha \text{e}^{-\left(\frac{z}{z_0}\right)^{\beta}}.
	\end{equation}
By evaluating that function at the central values of the $z$-bins for each object type, we obtain the densities required. The parameters $\{C_0, z_0, \alpha\}$ are considered as free parameters, with $\beta = 3/2$ \cite{RescoMaroto}. \Cref{tab:Densities} shows the values of the densities used for each $z$-bin.

 	\begin{table}[t]
 		\centering
 		\input{DensitiesLow.tex}
 		\hfill
 		\input{DensitiesHigh.tex}
 		\caption{Densities $\hat{n}$ in units of $10^{-3} \text{ h}^3 \text{ Mpc}^{-3}$ of low (left table) and high (right table) redshift objects for each bin of redshift $z$. The values of the densities depend on the photometric error $\delta_z$, and are estimated for low-$z$ \cite{MiniJPAS} and high-$z$ \cite{QueirozQSOMocks, QueirozPreparation} objects using \cref{sourcegalaxyfunction}.}
			\label{tab:Densities}
	\end{table}

\subsection{Construction of the monopole galaxy power spectrum}

Our observable of interest is the monopole galaxy power spectrum $P_{g}^{(0)}(k,z)$, from which we reconstruct $P_{\mathcal{R}}(k)$ using the likelihood of \cref{likelihood}. We estimate the mean value $\bar{P_{g}}^{(0)}(k,z)$ following the model in \cite{RescoMaroto}, which is based on \cite{SeoEisenstein} and inspired by the Kaiser model \cite{RSDKaiser}. This model accounts for redshift space distortions at a basic level, with a Gaussian uncertainty in its radial position. $\bar{P_{g}}^{(0)}(k,z)$ is obtained through the matter power spectrum, which is computed with CAMB \cite{CAMB}. We describe the construction of the galaxy power spectrum below.

We start by choosing a primordial power spectrum $P_{\mathcal{R}}(k)$. We will consider either the SM power law $A_s \left(\frac{k}{k_0}\right)^{n_s-1}$ or various templates with deviations from the power law, all based on physically motivated models, such as the ones in \cref{ModifiedPPS}, that we explain in \cref{sec:AppendixModifiedPPS}, or in \cref{ModifiedPPSLogLog}.

The linear matter power spectrum $P_m(k)$ reads as:
	\begin{equation}\label{MatterPS}
P_m(k) = 2 \pi^2 h^3 T^2(k) k P_{\mathcal{R}}(k).
	\end{equation}
A simple way to estimate a galaxy power spectrum $\bar{P_g}(k)$ is to assume that it is proportional to the matter power spectrum. The proportionality is taking into account through the bias $b$: 
	\begin{equation}\label{GalaxyPS}
\bar{P_g}(k) =  b^2 P_{m}(k).
	\end{equation}
This simple method is not accurate enough to describe a galaxy power spectrum \cite{RoyMcDonald2009}, not even in the linear regime. We use a more complex description of the galaxy power spectrum, incorporating the Kaiser model for the redshift space distortions and both the Fingers of God and the Alcock-Paczynski effects:
	\begin{equation}\label{GalaxyPSRescoMaroto}
\bar{P_g}(k,\mu,z) = \frac{D_{A,\text{fid}}^2 E(z)} {D_A^2 E_{\text{fid}}(z)} F_{\text{FoG}}(k',\mu',z) \left[b(z)+f(z) \mu'^{2}\right]^{2} P_{m}(k',z) e^{-k^{2} \mu^{2} \sigma_{z}^{2}(z)},
	\end{equation}
 with $\mu$ being the cosine of the angle between the wavevector $\vec{k}$ and the line of sight, $b(z)$ the bias described in the previous subsection and $f(z)$ the growth function. The subscript `fid' indicates that a variable is evaluated at a fixed fiducial cosmology. The model has an exponential term $e^{-k^{2} \mu^{2} \sigma_{z}^{2}(z)}$ that accounts for a reduction of power due to uncertainties induced by the photometric error of the measurements through the parameter $\sigma_{z} = \frac{\delta_z (1+z)}{H(z)}$, with $H(z) = H_0 E(z)$ the Hubble function. The power suppression results from the convolution of the line of sight comoving position $r$ with a Gaussian $\text{e}^{\frac{-(\Delta_r)^2}{2 \sigma_{z}^2}}$. The function $E(z)$ is defined as:
\begin{equation}\label{EdeZ}
E(z) = \sqrt{\Omega_m (1+z)^3+(1-\Omega_m)},
	\end{equation} 
the angular distance $D_A(z)$ can be constructed from the comoving radial distance $\chi(z)$:
\begin{equation}\label{AngularDistance}
D_A(z) = \frac{\chi(z)}{1+z},
	\end{equation} 
and the growth function $f(z)$ is estimated as: 
\begin{equation}\label{GrowthFunction}
f(z) = \left(\Omega_m (1+z)^3 \frac{1}{E^2(z)}\right)^\gamma, 
	\end{equation}
being $\gamma = 0.545$ the growth index.

The Fingers of God effect is taken into account with the function $F_{\text{FoG}}(k,\mu,z)$, which is modelled as a Lorentzian following \cite{FingersOfGodPercival2004}:
\begin{equation}\label{FingersOfGod}
F_{\text{FoG}}(k,\mu,z) = \frac{1}{1+ \left[f(z) \mu k \sigma_{p,\text{fid}}(z)\right]^2}, 
	\end{equation}
where the dispersion parameter $\sigma_{p,\text{fid}}(z)$ is obtained integrating the linear matter power spectrum as:
\begin{equation}\label{DispersionParameter}
\sigma^2_{p,\text{fid}}(z) =  \frac{1}{6 \pi^2} \int P_{m,\text{fid}}(k,z) \text{ d}k.
	\end{equation}
In practice, we perform this integral with finite boundaries $k_\text{min} = 10^{-5} \text{ h } \text{Mpc}^{-1}$ and $k_\text{max} = 10^{3}  \text{ h } \text{Mpc}^{-1}$.

The Alcock-Paczynski effect \cite{APEffectHistorico} distorts the scales $k$ and $\mu$ as:
\begin{equation}\label{AlcockPaczynskiKDistortion}
k' = Q k,
	\end{equation}
 \begin{equation}\label{AlcockPaczynskimuDistortion}
\mu' = \frac{E}{E_{\text{fid}}Q} \mu,
	\end{equation}
with:
\begin{equation}\label{AlcockPaczynskiQ}
Q = \frac{\sqrt{E^2 \chi^2 \mu^2-E_{\text{fid}}^2 \chi_{\text{fid}}^2\left(\mu^2-1\right)}}{E_{\text{fid}} \chi}.
	\end{equation}
A factor $\frac{D_{A,\text{fid}}^2 E(z)} {D_A^2 E_{\text{fid}}(z)}$ must be included in the galaxy power spectrum for properly modelling the
Alcock-Paczynski effect.

The $\ell$-multipoles of $\bar{P_{g}}\left(k, \mu, z \right)$ can be computed as:
	\begin{equation}\label{GalaxyMultipoles}
\bar{P_{g}}^{(\ell)}(k,z)=\frac{(2 \ell+1)}{2} \int_{-1}^{1} \bar{P_{g}}\left(k, \mu, z \right) \mathcal{L}_{\ell}(\mu) d \mu,
	\end{equation}
with $\mathcal{L}_{\ell}(\mu)$ the Legendre polynomial of degree $\ell$. The monopole ($\ell=0$) is: 
\begin{equation}\label{GalaxyMonopole}
\bar{P_{g}}^{(0)}(k,z)=\frac{1}{2} \int_{-1}^{1} \bar{P_{g}}\left(k, \mu, z \right) d \mu.
\end{equation}
Measurements of higher order multipoles, as the quadrupole $\bar{P_{g}}^{(2)}(k)$ in \cite{BOSSDR12}, are usually significantly noisier than those of the monopole. While these measurements can potentially provide valuable cosmological information, we believe its impact on the determination of the $P_{\mathcal{R}}(k)$ shape will be small. Therefore our focus remains primarily on the monopole. 

\Cref{GalaxyPSRescoMaroto} does not capture the redshift space power spectrum accurately, as it neglects the non-linear effects \cite{NonLinear1, NonLinear2} and the smoothed out BAOs \cite{BoyleKomatsuSmoothing}. More advanced models take into account both the non-linearities, such as the ones in \cite{RSDKaiser, RSDScoccimarro, TNS1}, and the BAO smoothing, as in \cite{EuclidForecastModel}. We will not consider these two effects here, but would be required for state of the art surveys.

The expected value of the monopole galaxy power spectrum at wavenumber $k$ and redshift $z$ is given by \cref{GalaxyMonopole}. To estimate its uncertainties, its variance matrix has to be computed. We follow the method in \cite{CovarianceMatrix}, which requires an additional term: the Fourier number of modes assigned to the $k$-shell $N_k(z)$. For this, we define the comoving radial distance as:
	\begin{equation}\label{ComovingDistance}
\chi(z) = \frac{1}{H_0} \int_0^z\frac{1}{E(z')}d z'.
	\end{equation}
Then we estimate the volume in a redshift bin of central value $z$ and upper and lower limits $z_+$ and $z_-$:
	\begin{equation}\label{VolumeAlpha}
V_\alpha(z) = \frac{4}{3}\pi f_{\mbox{sky}} \left[\chi^3(z_+)-\chi^3(z_-)\right].
	\end{equation}
The number of Fourier modes $N_k$ inside a $k$-shell between $k_+$ and $k_-$ for the redshift bin $z$ are:
	\begin{equation}\label{NumberOfModes}
N_k(z) = \frac{V_\alpha(z)}{(2\pi)^3} \frac{4}{3}\pi(k_+^3-k_-^3),
	\end{equation}
where $k_+$ and $k_-$ represent the upper and lower limits of the bin associated to $k$.
The number of Fourier modes contributes to the sampling variance effect. The shot noise is included by the inverse of the density $\hat{n}$. Then, the variance accounting for both effects can be written as:
	\begin{equation}\label{CovMatDefinition}
\sigma^2(k,z) = 2\frac{[\bar{P_{g}}^{(0)}(k,z)+\frac{1}{\hat{n}}]^2}{N_k(z)}.
	\end{equation}

The standard deviations are considered as the uncertainties in our observable monopole galaxy power spectrum, i.e. $\delta \bar{P_{g}}^{(0)}(k,z) = \sigma(k,z)$.

The sample volume at the largest scales and the dominance of non-linear effects at the smallest ones determine the scales at which we apply our methodology. A scale range of $k \in [0.02-0.2] \text{ h}\text{ Mpc}^{-1}$ for low-$z$ objects and $k \in [0.02-0.45] \text{ h}\text{ Mpc}^{-1}$ for high-$z$ ones is used, as non-linear effects appear at smaller scales for high redshifts. The noise properties also vary with scale $k$: the sampling variance is higher at large scales and the shot noise tends to dominate at small ones. The variation of the signal to noise ratio with scale $k$ can be seen in \cref{fig:SN}, presented in the next subsection for different redshifts.

 We use this model of the monopole galaxy power spectrum to generate simulations and test our methodology for the reconstruction of the primordial one in \cref{sec:ApplicationSimulations}.

\subsection{Signal to noise analysis}

In the previous section we have described the observed power spectrum model, the physical properties of the objects and the characteristics of the survey. We now study the expected signal to noise ratio $S/N$ for each clustering scenario. The sensitivity of our methodology to power deviations will depend on the $S/N$ of $P_g^{(0)}(k_i,z_j)$ for the different objects, given by
$S/N(k,z) = \frac{\bar{P}_g^{(0)}(k,z)}{\sigma(k,z)}$.

\begin{figure}[t]
\centering 
\includegraphics[width=.49\textwidth]{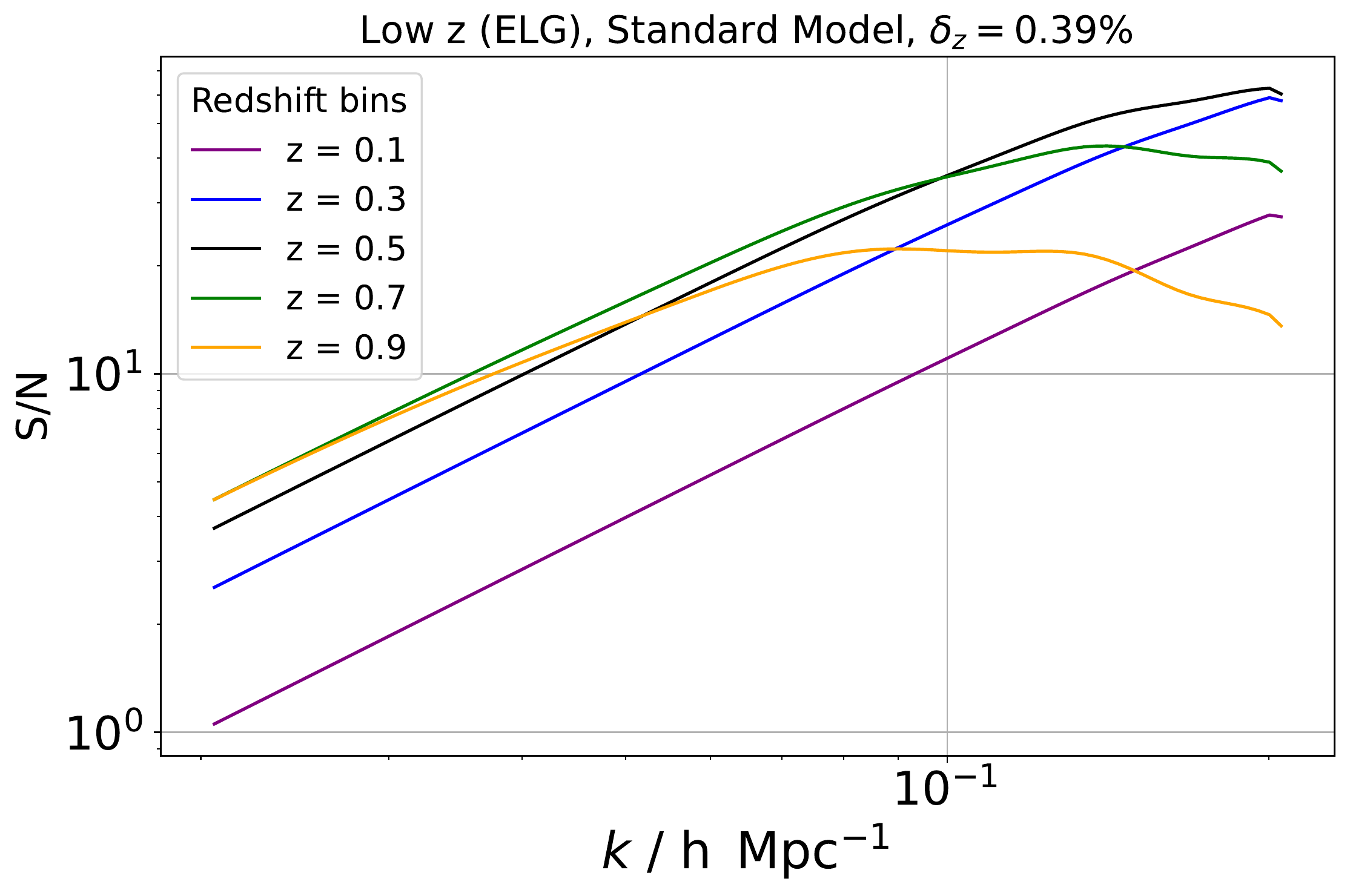}
\hfill
\includegraphics[width=.49\textwidth]{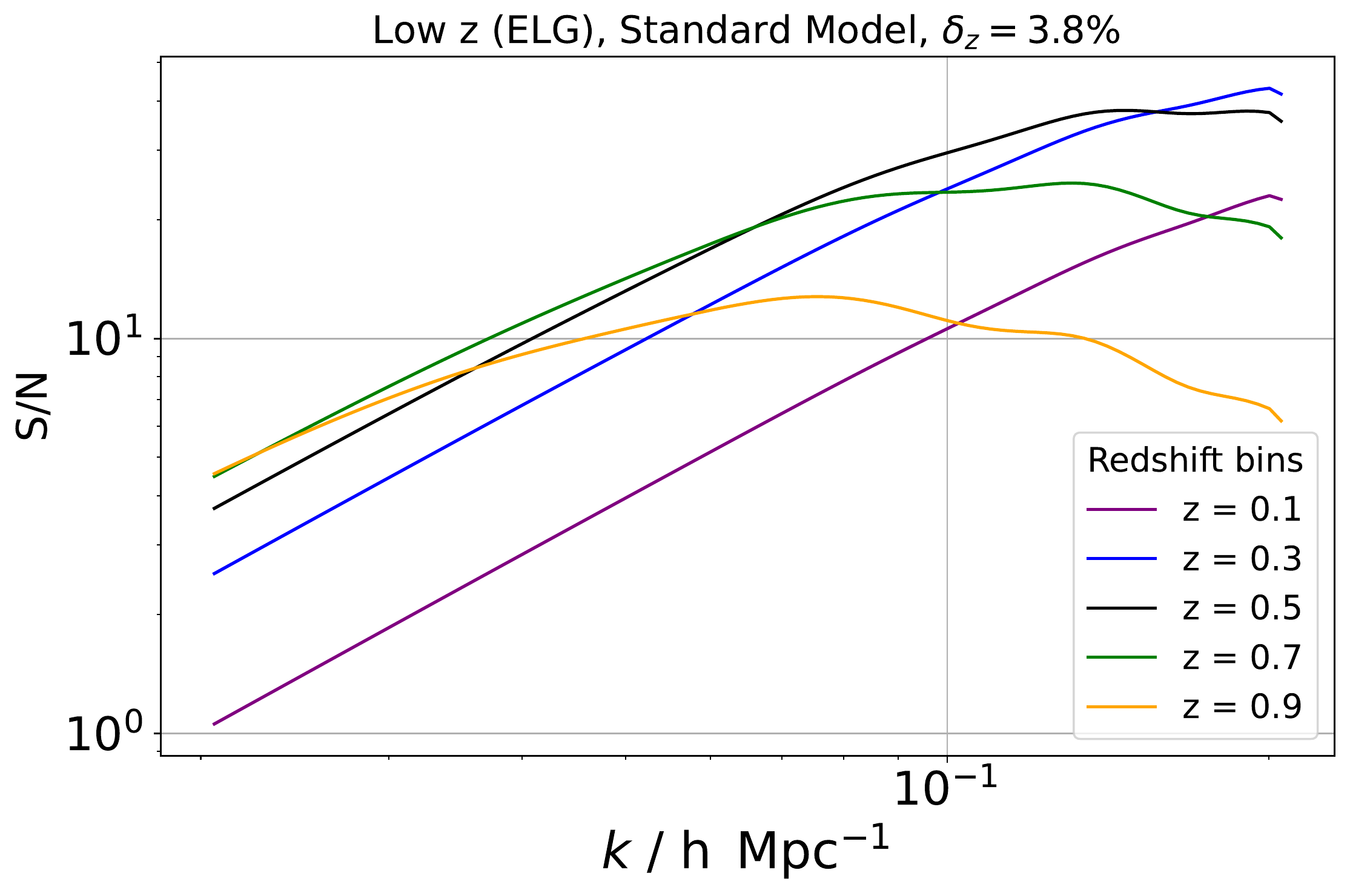}
\hfill
\includegraphics[width=.49\textwidth]{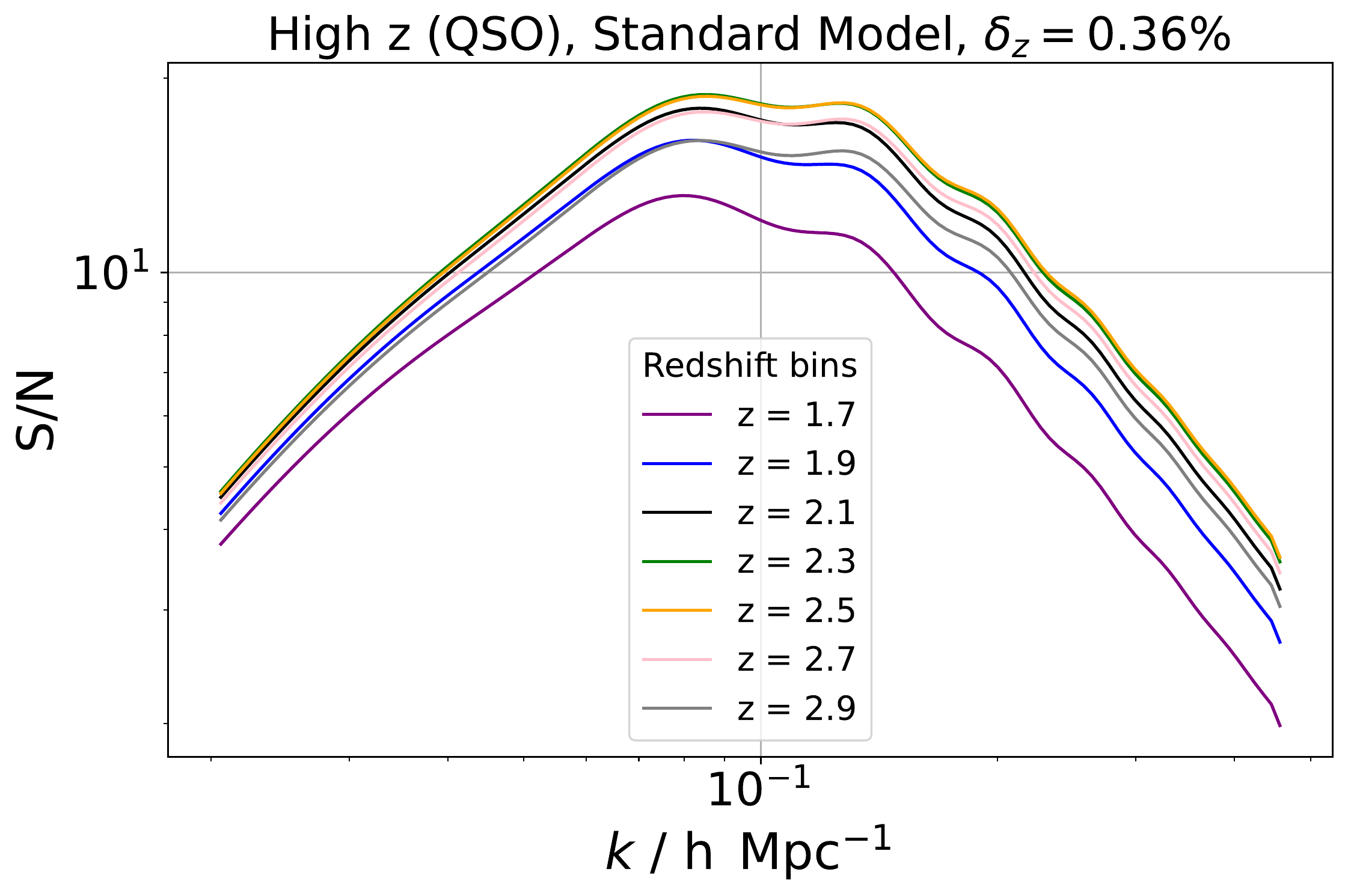}
\caption{Signal to noise ratios $S/N(k)$ for the SM at different redshift bins for different scenarios. Top panels: for low-$z$ objects with low (left) and high (right) $\delta_z$. Bottom panel: for high-$z$ objects.}
	 \label{fig:SN}
\end{figure}

In \cref{fig:SN} the $S/N$ as a function of the scale $k$ is represented for the SM. This figure reflects the dominance of the sampling variance for most of the redshift bins of the low-$z$ objects (top panels) and the shot noise dominance for the high-$z$ objects (bottom panel). For the high-$z$ objects (plotted in the bottom panel of \cref{fig:SN}) the shot noise dominates at all scales due to their lower densities compared to the low-$z$ ones. As a consequence, those bins with the highest densities will have the smallest noise and the highest S/N. The volume differences between redshift bins are much smaller than in the case of the low-$z$ objects, implying similar sampling variance among them. In the low-$z$ scenario the shot noise term $\frac{1}{\hat{n}}$ in \cref{CovMatDefinition} is subdominant for the majority of redshift bins, except for the highest ones that have the lowest densities (listed in \cref{tab:Densities}) at small scales. The number of modes $N_k(z)$ for a certain $k$ leads to a very different sampling variance between redshift bins, explaining the low $S/N$ at the smallest bins. In the case of low-$z$ with high-$\delta_z$ (top right panel), the shot noise has a slightly higher effect than for low $\delta_z$ due to the reduction of $\bar{P}_g^{(0)}$ (implying a reduction in both the signal and the sampling variance), caused by the higher photometric error (see \cref{GalaxyPSRescoMaroto}).

In \cref{sec:ApplicationSimulations} we apply the methodology to  high-$z$ simulations where we consider the favorable bin $z = 2.3$ having the highest $S/N$ of all the bins up to $k = 0.15 \text{ h}\text{ Mpc}^{-1}$, and for smaller scales is slightly lower than for the $z = 2.5$ bin where the $S/N$ is maximum. In this way we can obtain the minimum amplitude of features that the method is able to detect at low or high redshift.

\section{Application to simulated spectra}
\label{sec:ApplicationSimulations}

In this section we apply the methodology to simulations of different observational scenarios, characterized by the monopole galaxy power spectrum model $\bar{P_g}^{(0)}(k_i,z_j)$ at a given scale $k_i$ and redshift bin $z_j$, and the survey and object specifications described in \cref{sec:Simulations}. We simulate the ``observed" monopole power spectrum at a scale $k_i$ and redshift bin $z_j$, $P_g^{(0)}(k_i,z_j)$, by drawing random samples from a Gaussian distribution with mean $\bar{P_g}^{(0)}(k_i,z_j)$ and standard deviation $\delta \bar{P_g}^{(0)}(k_i,z_j) = \sigma(k_i,z_j)$ obtained from the variance matrix given by \cref{CovMatDefinition}. We consider a high-$z$ QSO survey with a relatively low $\delta_z$ (see \cref{sec:Surveys}). Also we discuss the obtained results for a low-$z$ galaxy survey with two photometric redshift errors, a low and a high $\delta_z$ with a fixed cosmology. 

First we test the methodology with SM simulations, i.e. assuming a power law for the primordial power spectrum $P_{\mathcal{R}}(k)$. Then we apply the methodology to simulations with primordial features. We consider two features templates, a local and a global one. From the local template given in \cref{ModifiedPPS}, we first generate a feature that consists of a 20\% bump in power at $k \approx 0.4 \text{ h} \text{ Mpc}^{-1}$ (see \cref{fig:LocalBumpFeature}). Using the same template, we also generate a second local feature having an oscillatory behaviour with excess and deficit of power of approximately 10\% at the maximum and minimum of the oscillation (see \cref{fig:LocalOscillatoryFeature}). 
We also apply the methodology to a different feature template, namely a global log-log oscillatory feature with oscillations at all scales (see \cref{ModifiedPPSLogLog}). We consider three amplitudes for the global template, corresponding to approx. 10\%, 3\% and 1.5\% power deviations from the featureless primordial power law. We have checked that different realizations of the same observational scenario provide very similar reconstructions leading to the same feature detection results derived from the global and local tests.

\subsection{Standard Model}

\begin{figure}[t]
\centering 
\includegraphics[width=.79\textwidth]{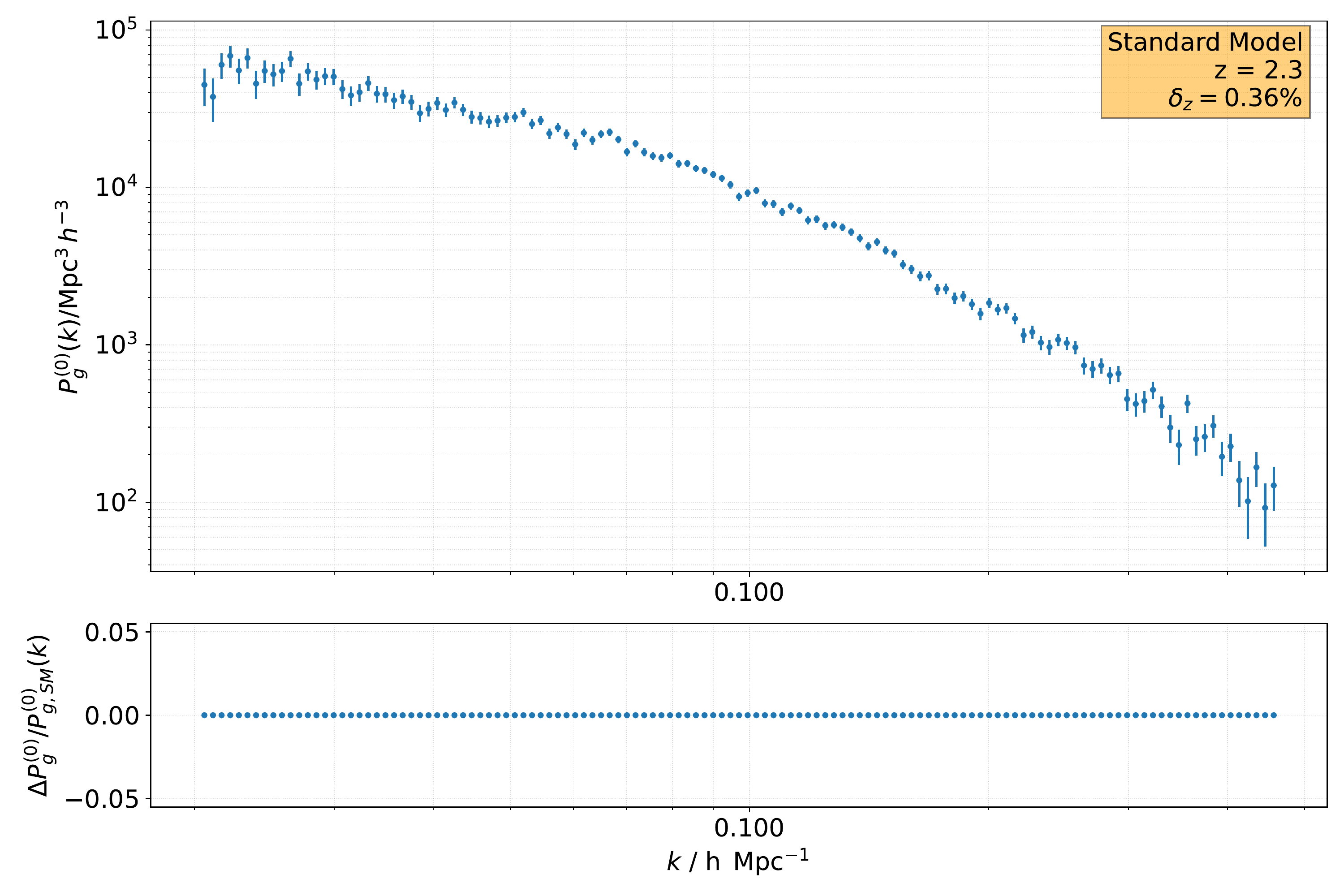}
\caption{Top panel: Simulated monopole galaxy power spectrum $P_g^{(0)}(k)$ of high-$z$ objects performed for the SM, i.e. with its $P_{\mathcal{R}}(k)$ being the power law. Bottom panel: relative difference with respect to the Standard Model monopole galaxy power spectrum $P_{g,\text{SM}}^{(0)}(k)$.}
	 \label{fig:SMRealization}
\end{figure}

The Standard Model simulations are performed considering $P_{\mathcal{R}}(k)$ as a power law $\mathcal{P}_{\mathcal{R}}^{\text{SM}}(k) = A_s \left(\frac{k}{k_0}\right)^{n_s-1}$. We consider simulations of high-$z$ objects $P_g^{(0)}(k)$ at the $z = 2.3$ bin, a favorable case since it has the minimum shot noise and the highest $S/N$ values for almost all the scale range (see \cref{fig:SN}). The $P_g^{(0)}(k)$ is represented in \cref{fig:SMRealization}.

\begin{figure}[h]
\centering 
\includegraphics[width=.55\textwidth]{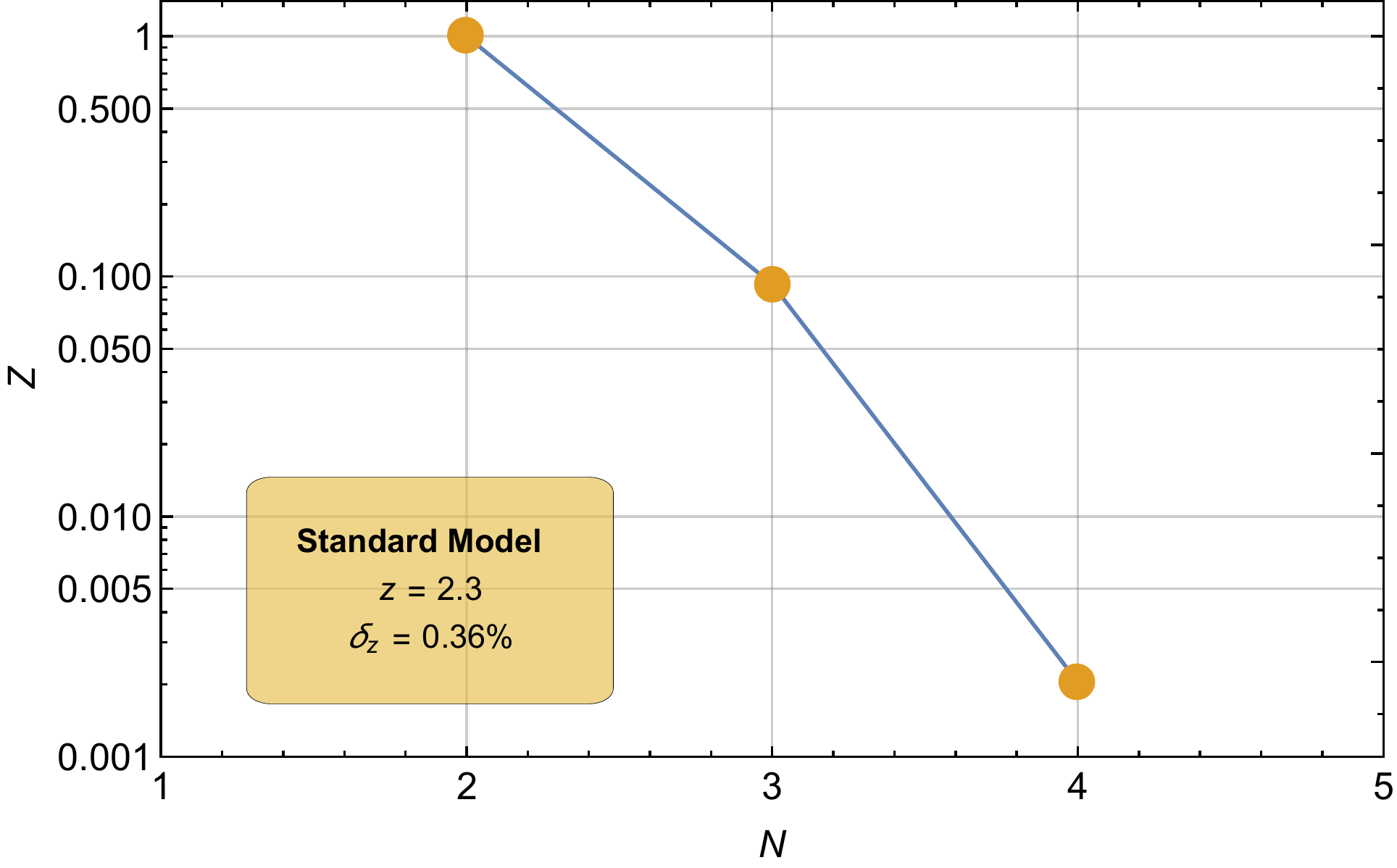}
\caption{Evidence $Z$ of the Standard Model reconstructions for each of the $N$ knots configuration considered for the high-$z$ case.}
	 \label{fig:SMEvidences}
\end{figure}

As expected, the evidences of the reconstructed spectrum peak at the 2 knots case, $Z_2$, and exhibit a quasi-exponential decay with increasing $N$, as shown in \cref{fig:SMEvidences}. The recovery of the power law is clear, with negative feature detection as summarized in the second column of \cref{tab:TableLocalFeature}.

\begin{figure}[h]
\centering 
\includegraphics[width=.79\textwidth]{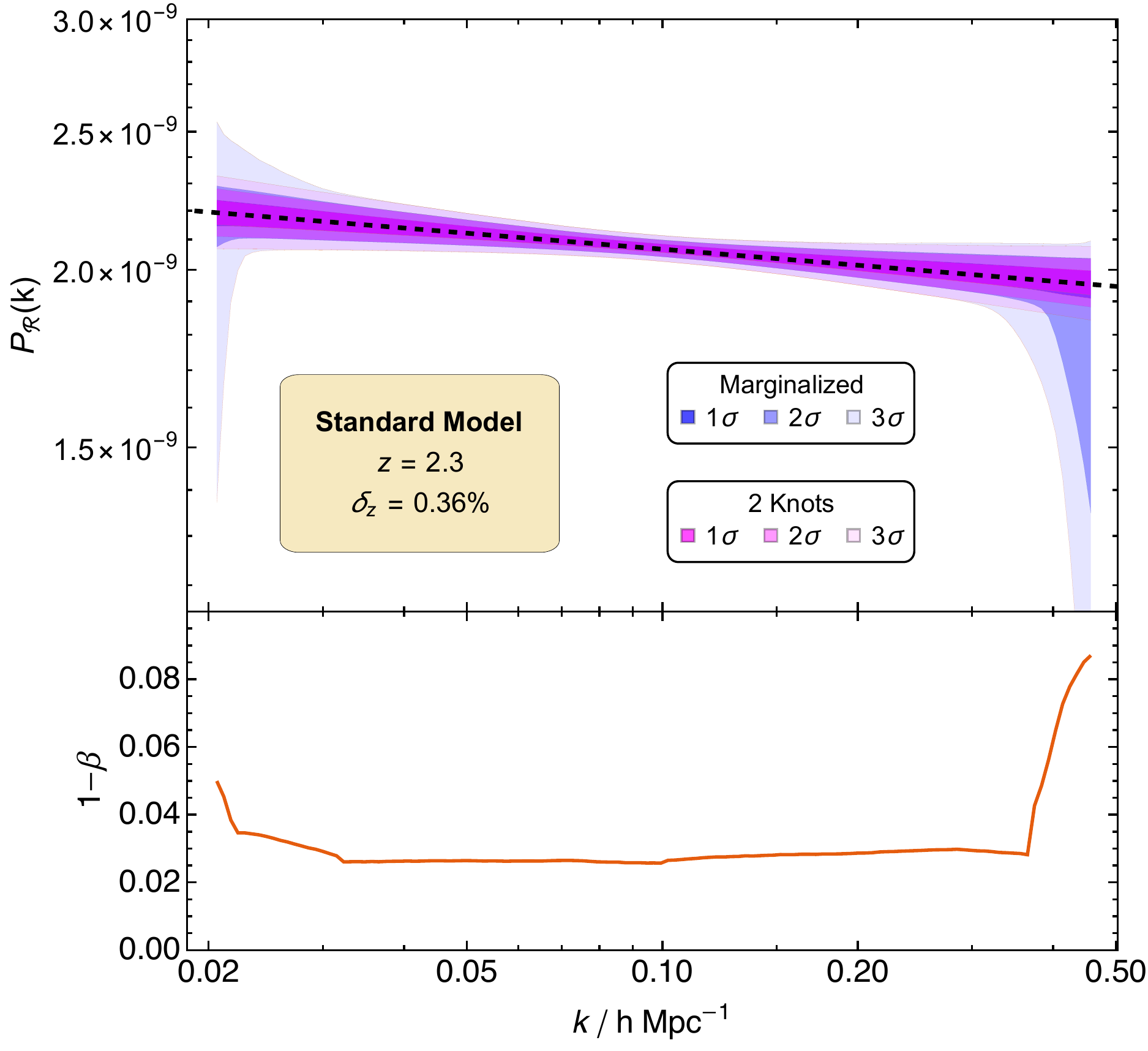}
\caption{Contours of the high-$z$ $P_{\mathcal{R}}(k)$ reconstructions for the SM simulations in the $N = 2$ case (magenta) and for the case of marginalized probability over $N$ knots (blue). We include in dashed black the power law that sourced the simulations. In the bottom panel the values of the power $1 - \beta$ obtained from the hypothesis test are plotted.}
	 \label{fig:SMContoursAndHypoTest}
\end{figure}

The value of $1 - \beta$, is very small ($<0.05$) at all scales, showing a very good consistency between the $N = 2$ and the marginalized reconstruction. The contours displayed in the top panel of \cref{fig:SMContoursAndHypoTest} are narrower at the largest scales and wider at the smallest ones $k \approx 0.5 \text{ h}\text{ Mpc}^{-1}$ due to the dominance of the shot noise. The hypothesis test do not show any local deviations between the marginalized and the $N = 2$ case, showed at the bottom panel of \cref{fig:SMContoursAndHypoTest}.

The results obtained for the case of the SM show the preference for the power law, with a good recovery of the input values for the parameters $\{A_s,n_s\}$ with the global test and no deviation detected with the local one, validating the performance of the proposed methodology.

\subsection{Local bump feature}
\label{subsec:local_bump_feature}

\begin{figure}[h]
	 \centering
	    \includegraphics[width=0.65\textwidth]{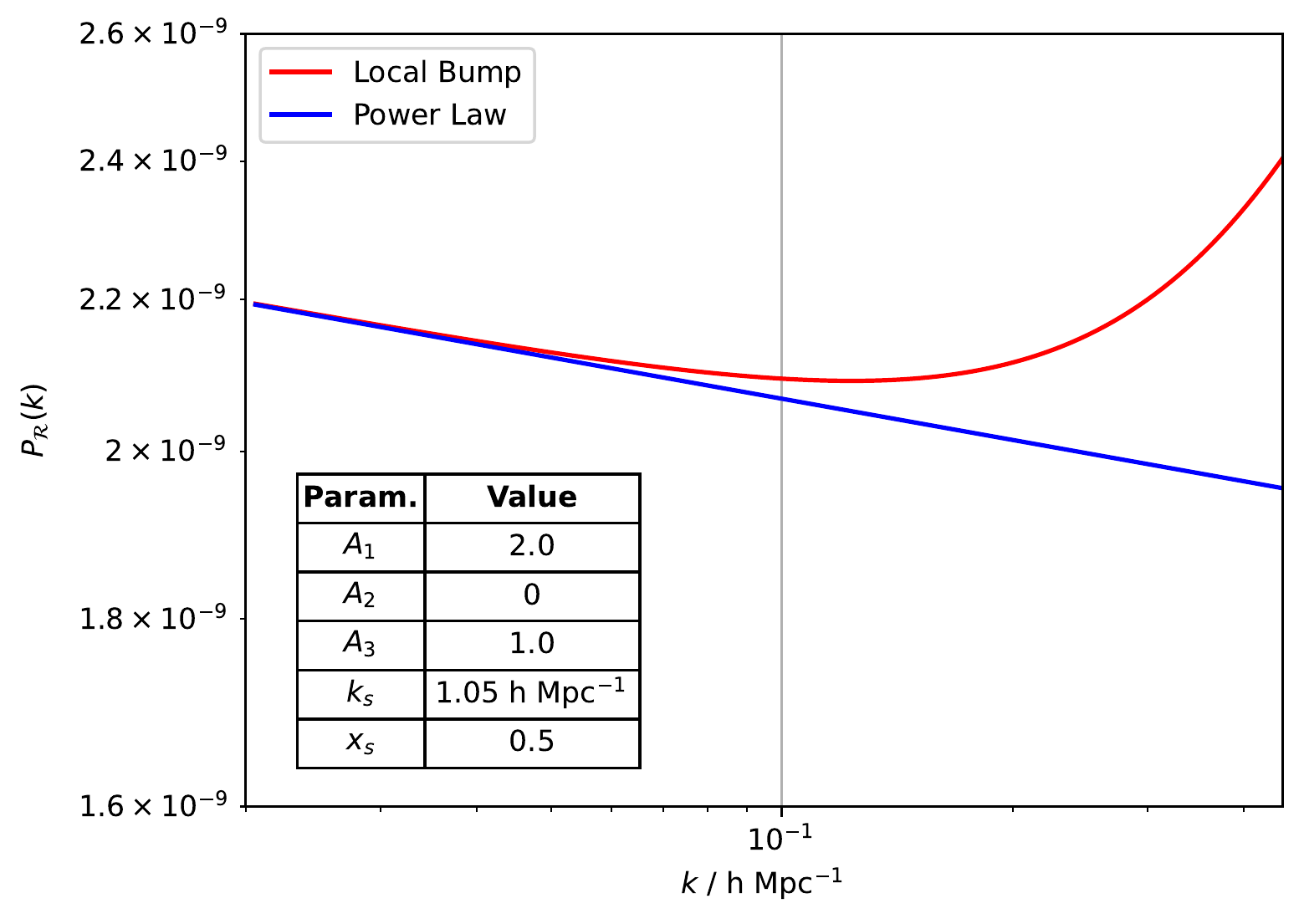}
	 \caption{Local bump feature model of the primordial power spectrum and the power law primordial power spectrum (blue). The parameters used to generate the bump feature are given in the table inside the figure. Note the increase of power at the scale $k = 0.45 \text{ h}\text{ Mpc}^{-1}$ where it reaches 20\%. At large scales, the feature model tends to the SM power law.}
	 \label{fig:LocalBumpFeature}
	\end{figure}

We study a local feature in $P_{\mathcal{R}}(k)$ following the template in \cref{ModifiedPPS}. We tune the parameters of this template to have a power excess relative to the featureless model of 5\% at $k \approx 0.2 \text{ h}\text{Mpc}^{-1}$ and of 20\% at $k \approx 0.4 \text{ h}\text{ Mpc}^{-1}$. We call this feature template `bump'. The primordial spectrum for this bump feature, and the parameters used to generate it, are shown in \cref{fig:LocalBumpFeature}.

\begin{figure}[h]
\centering 
\includegraphics[width=.79\textwidth]{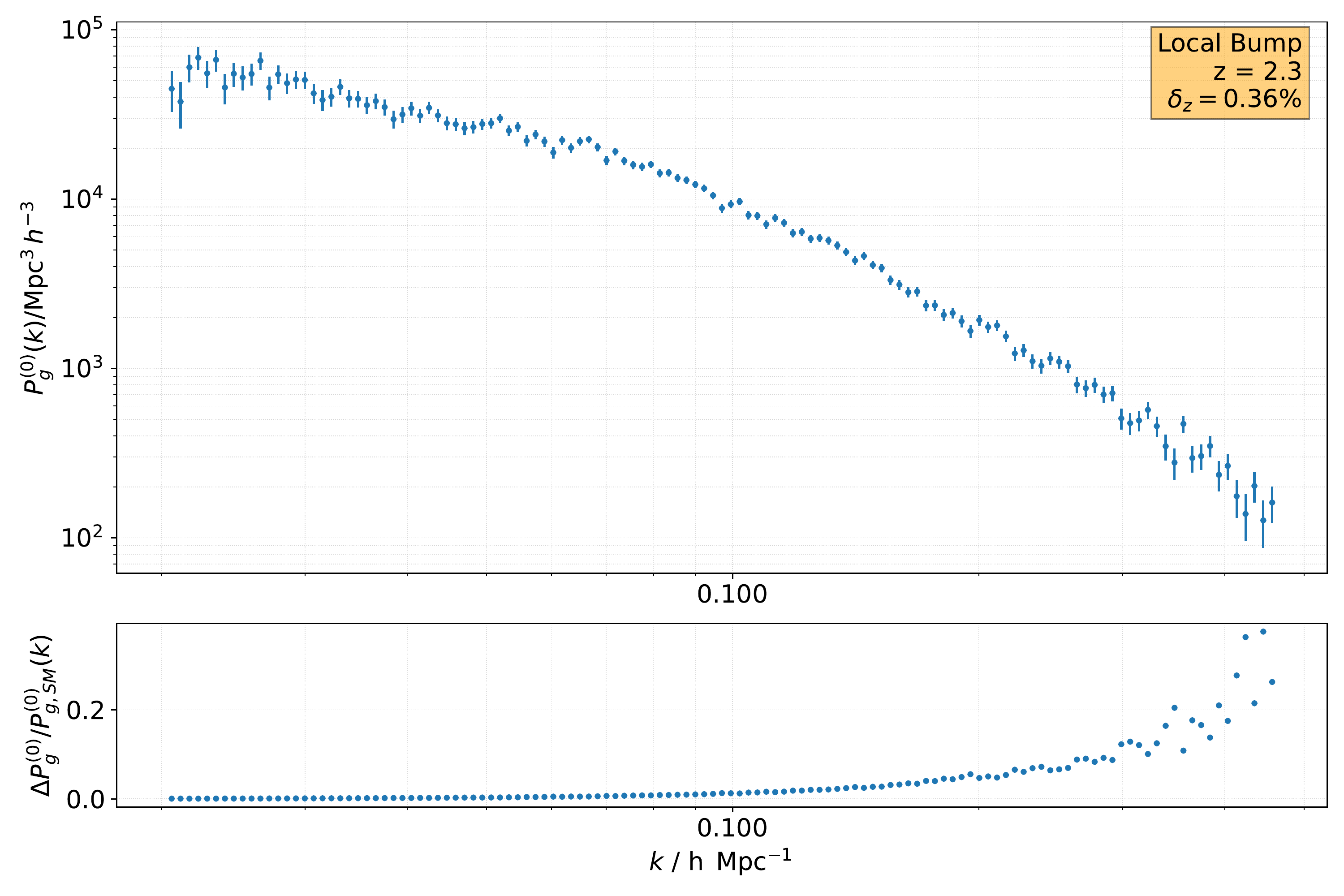}
\caption{Top panel: Simulated monopole galaxy power spectrum $P_g^{(0)}(k)$ of high-$z$ objects performed for the local bump feature template. Bottom panel: relative difference with respect to the Standard Model monopole galaxy power spectrum $P_{g,\text{SM}}^{(0)}(k)$.}
	 \label{fig:LocalBumpRealizations}
\end{figure}

We use the same redshift bin than for the SM. The corresponding simulated power spectrum shows the slight increase of power at the smallest scales of the template, as plotted in \cref{fig:LocalBumpRealizations}.

\begin{figure}[h]
\centering 
\includegraphics[width=.59\textwidth]{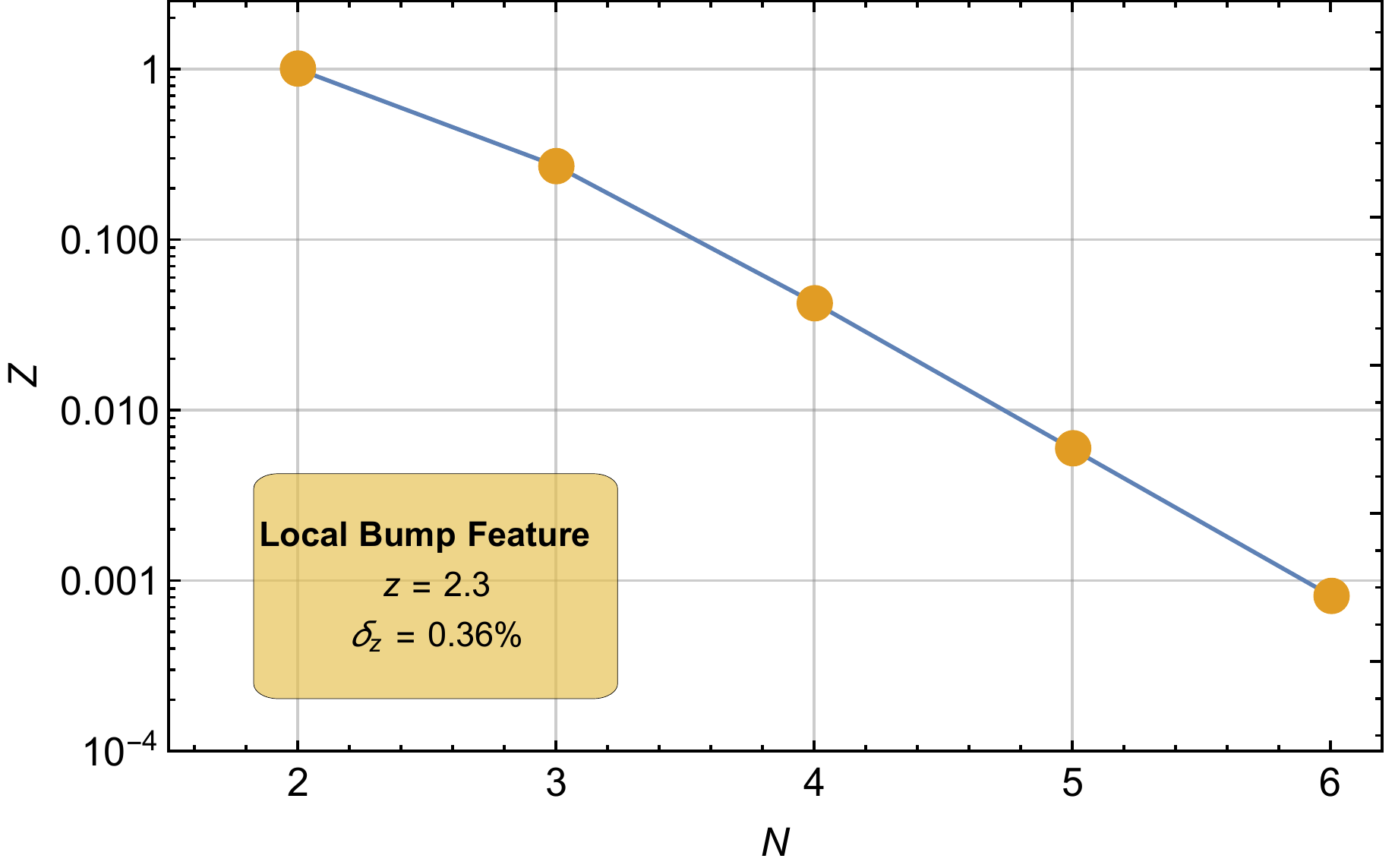}
\caption{Evidence $Z$ of the local bump feature template reconstructions for each of the $N$ knots configurations considered.}
	 \label{fig:LocalBumpEvidences}
\end{figure}

The global test for the bump feature is based on the ratio of evidences that are shown in \cref{fig:LocalBumpEvidences}. $Z_3$ is three times as high as in the case of the SM and $Z_4$ forty times higher. Nevertheless, the maximum evidence is still obtained for $N=2$ and the ratio $Z_3/Z_2$ is below $0.3$. 

\begin{figure}[h]
\centering 
\includegraphics[width=.79\textwidth]{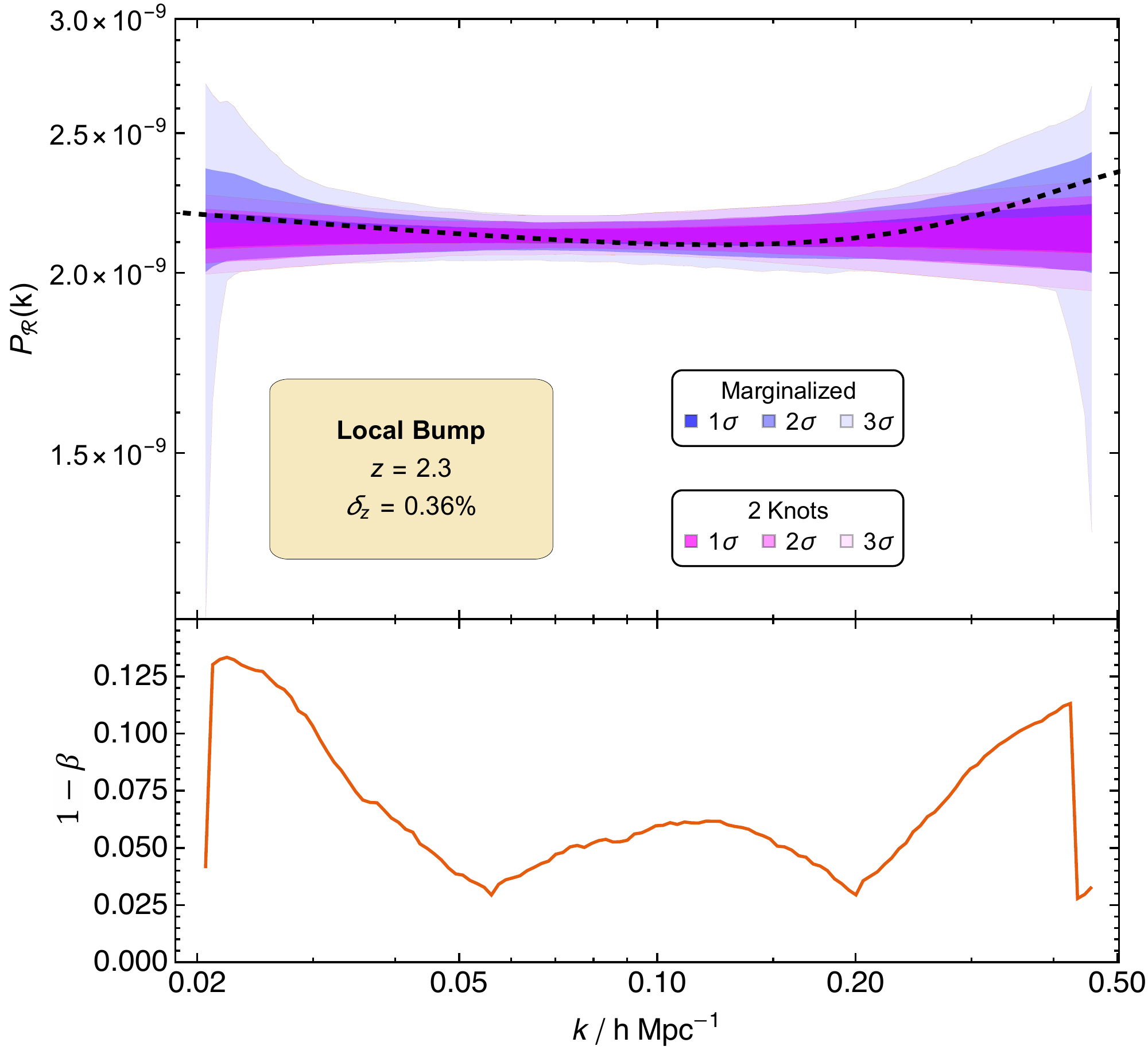}
\caption{Contours of the high-$z$ $P_{\mathcal{R}}(k)$ reconstructions for the local bump feature simulations for the $N = 2$ case (magenta) and for the case of marginalized probability over $N$ knots (blue). We include in dashed black the bump feature template that sourced the simulations. In the bottom panel the values of the power $1 - \beta$ obtained from the hypothesis test are plotted.}
	 \label{fig:LocalBumpContoursAndHypoTest}
\end{figure}

The $P_{\mathcal{R}}(k)$ reconstruction is represented in the top panel of \cref{fig:LocalBumpContoursAndHypoTest}, whereas the hypothesis test is represented in the panel below. Although there is a minor increase at the extreme scales of the explored range, the power of the test remains below the threshold of $1-\beta = 0.5$ for all scales.

To summarize, for the bump feature template, even for the redshift bin with best $S/N$, we can not detect the feature, as shown in the third column of \cref{tab:TableLocalFeature}. The application of our methodology to the bump feature provides the same qualitative results as for the SM case.

\subsection{Local oscillatory feature}

\begin{figure}[h]
	 \centering
	    \includegraphics[width=0.65\textwidth]{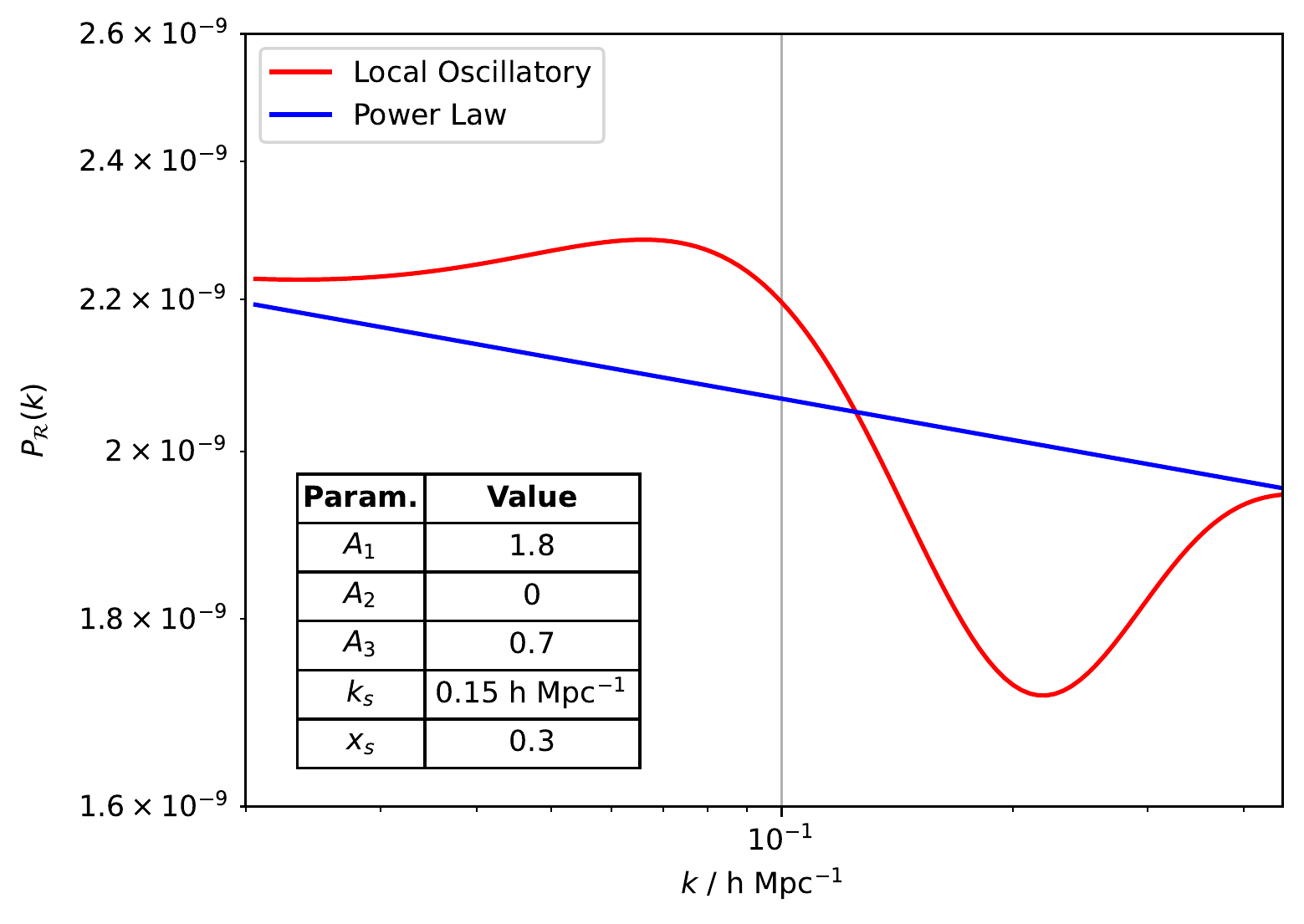}
	 \caption{Local oscillatory feature model of the primordial power spectrum (in red) and the power law primordial power spectrum (in blue). The parameters used to generate this oscillatory feature are given in the table inserted in the figure. The power excess peaks at $k = 0.07 \text{ h}\text{ Mpc}^{-1}$, reaching a $10 \%$ power excess, and is followed by a trough with a minimum power deficit of near $10 \%$ at $k = 0.20 \text{ h}\text{ Mpc}^{-1}$. The model tends to the power law at large scales.}
	 \label{fig:LocalOscillatoryFeature}
	\end{figure}

We examine how our methodology performs with another local feature in the primordial power spectrum $P_{\mathcal{R}}(k)$. The template for generating this feature is the same as for the bump one (see \cref{ModifiedPPS}). We now choose values of the different parameters to generate a power excess and a deficit of amplitude $\approx 10 \%$. We call this feature ``oscillatory", as it completes an oscillation in the range of scales considered. This feature is still local since at large scales the model tends to the power law. The primordial spectrum for the local oscillatory feature, and the parameters used to generate it, are shown in \cref{fig:LocalOscillatoryFeature}.

\begin{figure}[h]
\centering 
\includegraphics[width=.79\textwidth]{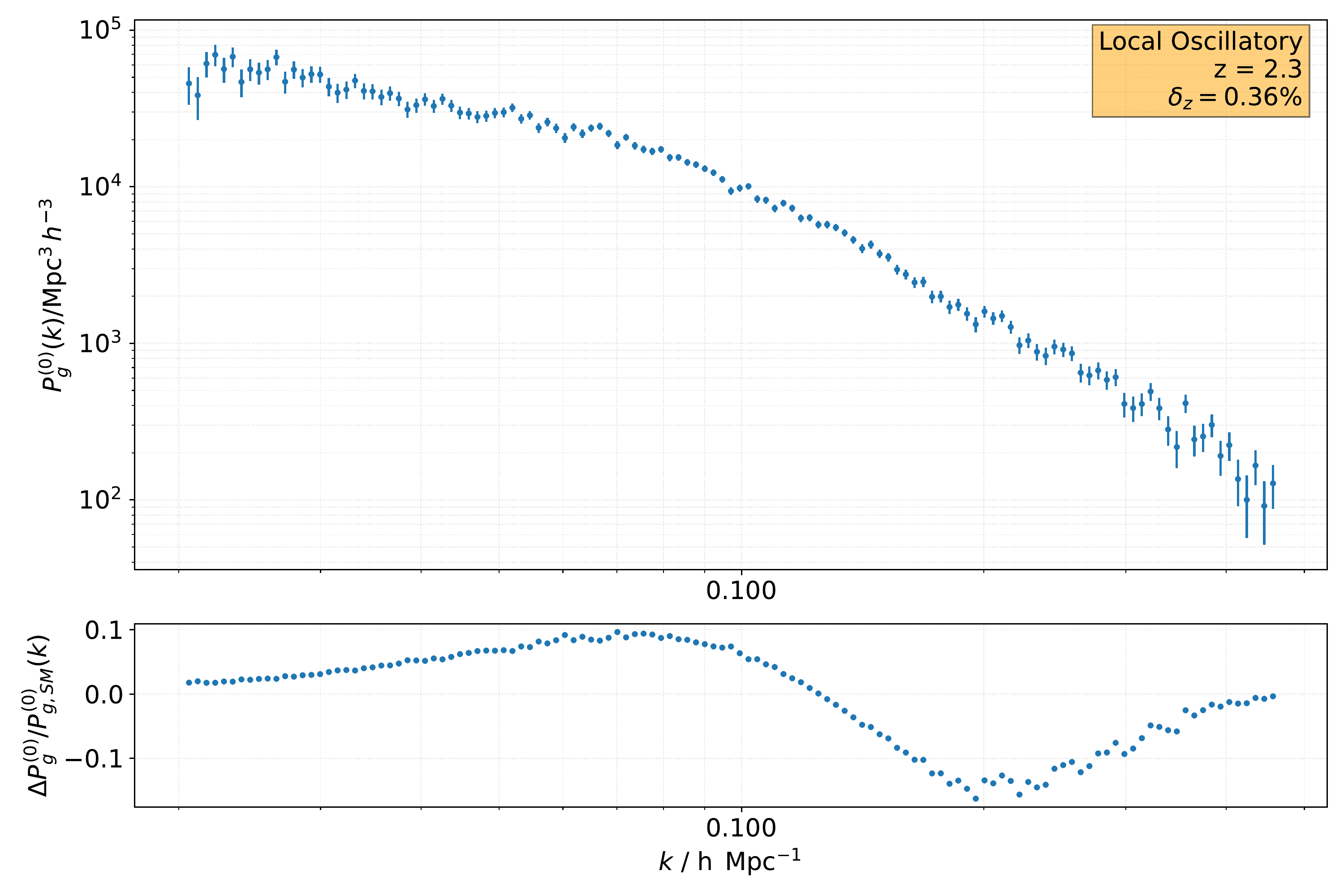}
\caption{Top panel: Simulated monopole galaxy power spectrum $P_g^{(0)}(k)$ of high-$z$ objects performed for the local oscillatory feature template. Bottom panel: relative difference with respect to the Standard Model monopole galaxy power spectrum $P_{g,\text{SM}}^{(0)}(k)$.}
	 \label{fig:LocalOscillatoryRealizations}
\end{figure}

In \cref{fig:LocalOscillatoryRealizations} the $P_g^{(0)}(k)$ simulation for the local oscillatory feature template is shown. We keep the same favorable bin $z = 2.3$ (see \cref{fig:SN}). The simulated $P_g^{(0)}(k)$ shows the power excess and deficit coming from the primordial feature.

\begin{figure}[h]
\centering 
\includegraphics[width=.59\textwidth]{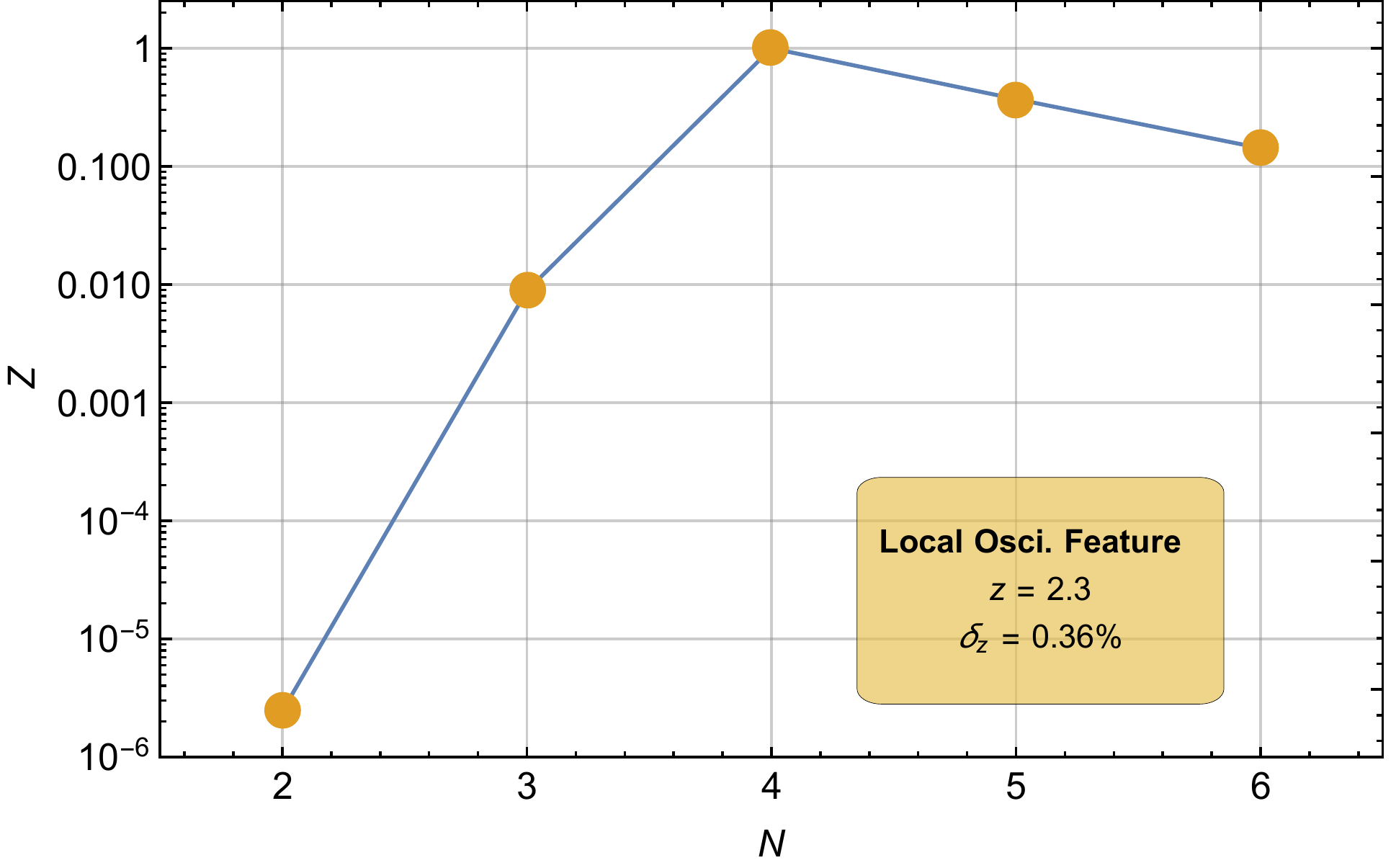}
\caption{Evidence $Z$ of the oscillatory feature reconstructions of high-$z$ objects for each of the $N$ knots configuration considered.}
	 \label{fig:LocalOscillatoryEvidences}
\end{figure}

The evidences of the global test appear in \cref{fig:LocalOscillatoryEvidences}. The minimum evidence is for two knots, $N=2$, with values below $10^{-5}$ relative to the highest evidence, which is $N = 4$. Significant values of the evidence $\gtrsim 10 \%$ are found at $N=5,6$.

\begin{figure}[h]
\centering 
\includegraphics[width=.79\textwidth]{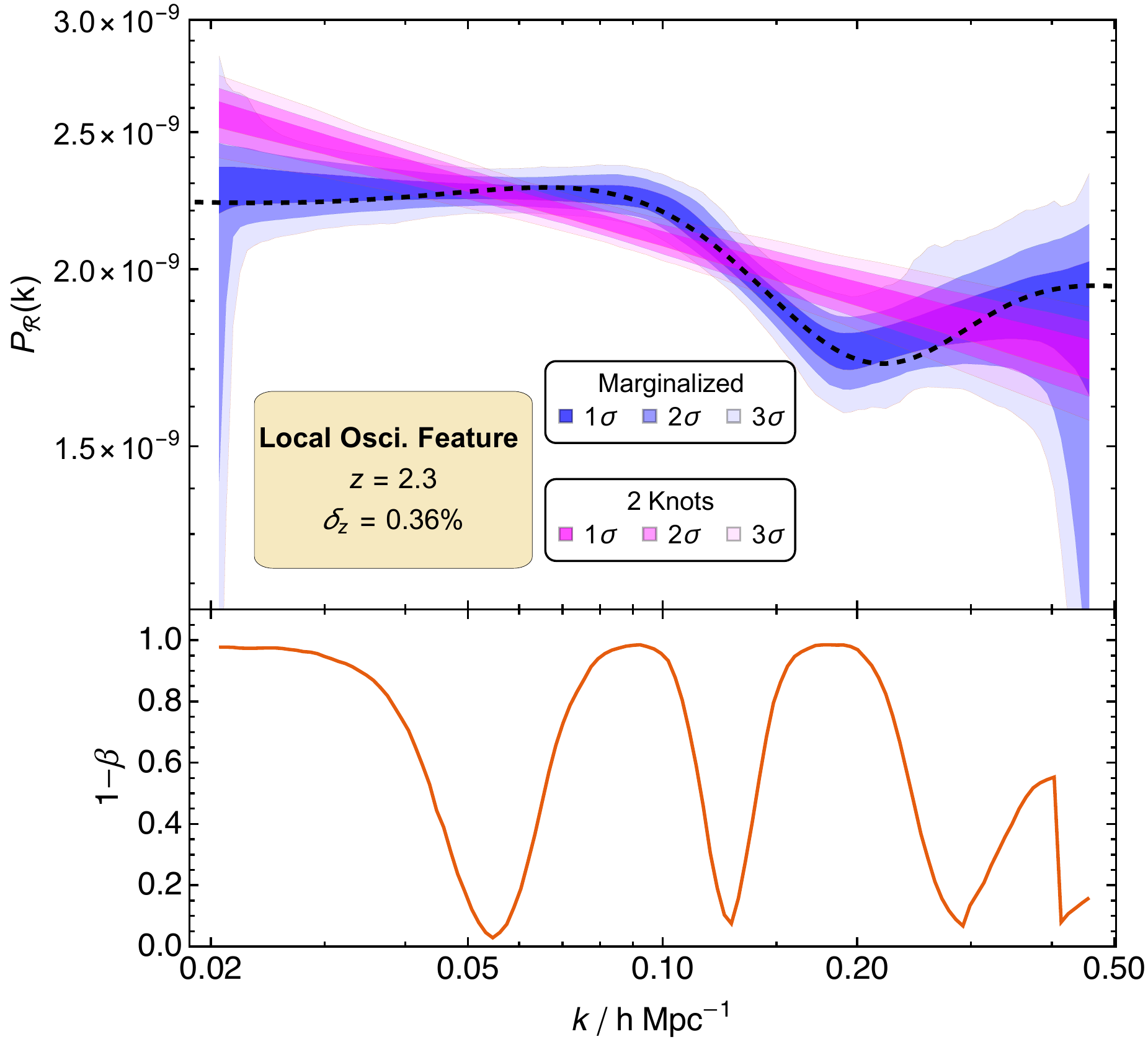}
\caption{Contours of $P_{\mathcal{R}}(k)$ reconstructions of high-$z$ objects for the oscillatory feature simulations for the $N = 2$ case (magenta) and for the case of marginalized probability over $N$ knots (blue). The oscillatory feature template of the primordial power spectrum is shown as a dashed black line. In the bottom panel of each figure the values of the power $1 - \beta$ obtained from the hypothesis test are plotted.}
	 \label{fig:LocalOscillatoryContoursAndHypoTest}
\end{figure}

The reconstruction for the oscillatory feature are shown in the top panel of \cref{fig:LocalOscillatoryContoursAndHypoTest}. The contours follow a $N = 4$ shape mainly (i.e., with two changes of slope), accounting for a complete oscillation and leading to a decisive detection. The hypothesis test displayed at the bottom panel of \cref{fig:LocalOscillatoryContoursAndHypoTest} also provides clear detection since there are three scale ranges where $1 - \beta > 0.95$, being $1 - \beta > 0.6$ at the largest scales (see the last two columns of \cref{tab:TableLocalFeature}).

\subsection{Global oscillatory feature}

We have tested our methodology with the power law and the bump and oscillatory feature templates, which are local features that approach the power law at large scales. We now consider a different type of feature in $P_{\mathcal{R}}(k)$ which is global, i.e. $P_{\mathcal{R}}(k)$ deviates from the power law at all scales in the form of oscillations. This kind of feature was motivated by a better fit to the Planck data \cite{ReconParametrico1, ReconParametrico2}, although a modulation is usually required. In those works the oscillations are located at scales from $k \approx 0.04 \text{ h}\text{ Mpc}^{-1}$ to $k \approx 0.4 \text{ h}\text{ Mpc}^{-1}$, where lensing is significant. The oscillations help to alleviate cosmological tensions, such as those in $H_0$, $\sigma_8$ or the lensing amplitude $A_l$. These global oscillations in the primordial power spectrum are predicted by inflation models with non-Bunch-Davies initial conditions \cite{ModelMartin2001,ModelMartin2003,ModelBozza2003} or with axion monodromy \cite{ModelFlauger2017}. 

\begin{figure}[h]
	 \centering
	    \includegraphics[width=0.72\textwidth]{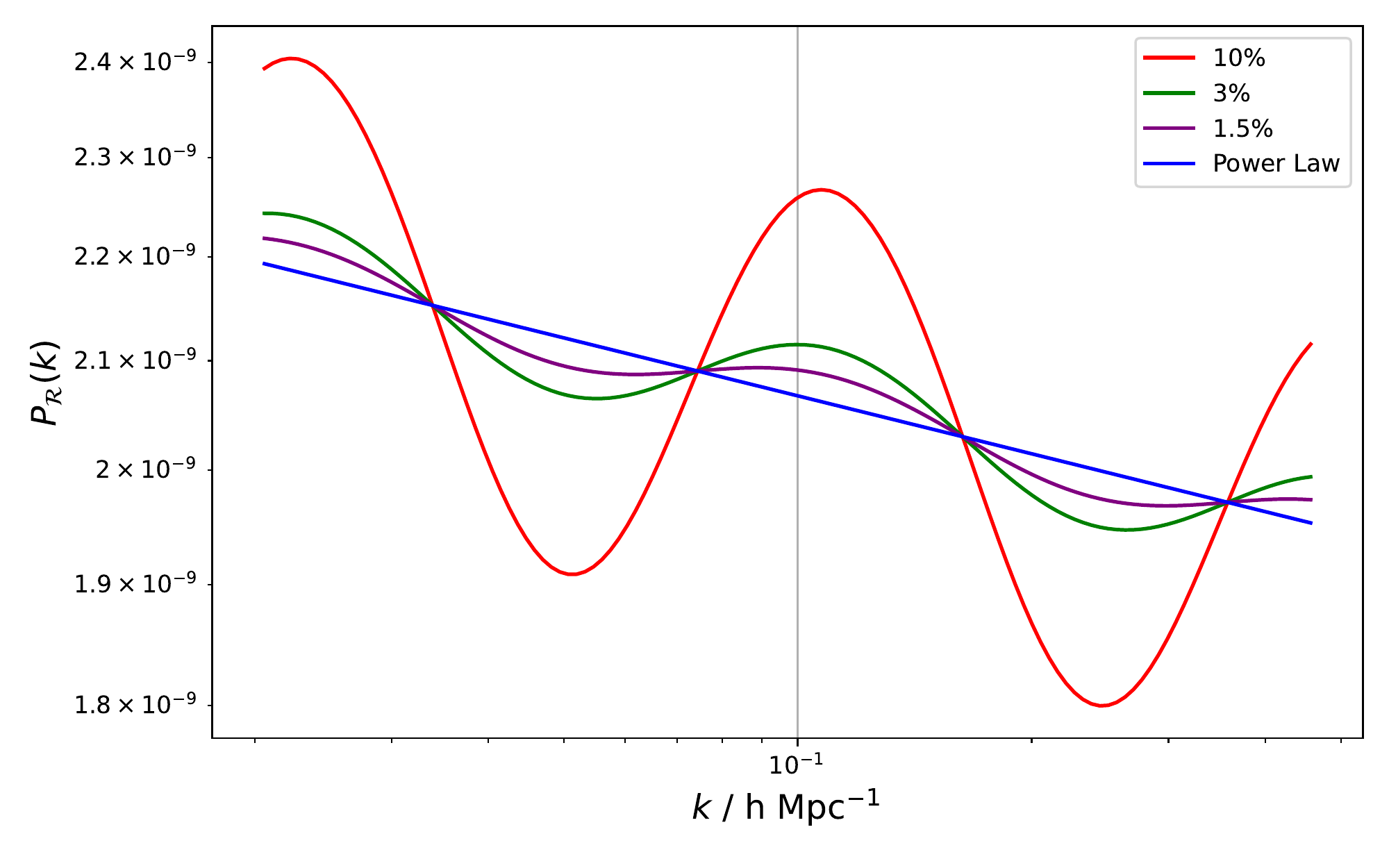}
	 \caption{Global oscillatory feature template of \cref{ModifiedPPSLogLog} used as primordial power spectrum for three different values of its amplitude parameter $A_{\text{log}}$: 0.1 (red), 0.02, (green) and 0.01 (purple), corresponding to $10\%$, $3\%$ and $1.5\%$ power deviations w.r.t. the power law (blue). Their frequency and phase are kept fixed at $\omega_{\text{log}} = 4.0$ and $\phi_{\text{log}} = 0$ respectively. Note the bump peaking near $k = 0.1 \text{ h}\text{ Mpc}^{-1}$ and the two deficits of power at scales $k \approx 0.05 \text{ h}\text{ Mpc}^{-1}$ and $ k\approx 0.25 \text{ h}\text{ Mpc}^{-1}$.}
	 \label{fig:GlobalOscillatoryFeature}
	\end{figure}

The model used is oscillatory in log-log scale 
and provides global oscillations for the range of scales that we use ($k \sim 0.01-0.1 \text{ h} \text{ Mpc}^{-1}$). It is parametrized including a modulation in the power law primordial power spectrum $P_{\mathcal{R}}^{\text{SM}}$: 
	\begin{equation}\label{ModifiedPPSLogLog}
\mathcal{P}_{\mathcal{R}}(k)= \mathcal{P}_{\mathcal{R}}^{\text{SM}}(k) \left[1+A_{\text{log}} \text{Cos}\left(\omega_{\text{log}} \text{Log} \left({\frac{k}{k_0}}\right)+\phi_{\text{log}}\right)\right],
	\end{equation}
where $A_{\text{log}}$, $\omega_{\text{log}}$ and $\phi_{\text{log}}$ are the amplitude, frequency and phase of the oscillations, respectively. \Cref{fig:GlobalOscillatoryFeature} shows this modified primordial spectrum template, with the parameters used for our specific feature listed in a table inside.
	
\begin{figure}[h]
\centering 
\includegraphics[width=.49\textwidth]{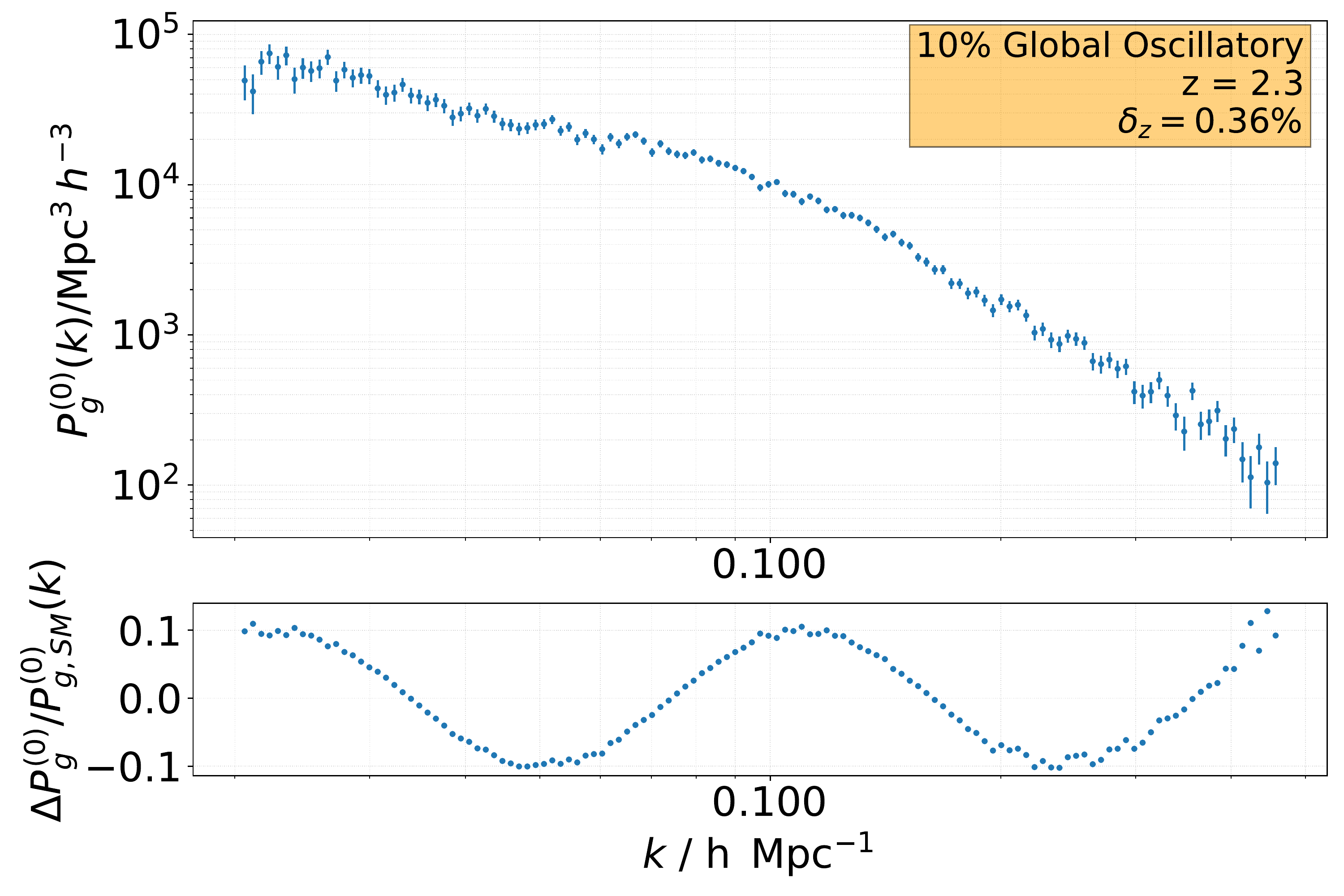}
\hfill
\includegraphics[width=.49\textwidth]{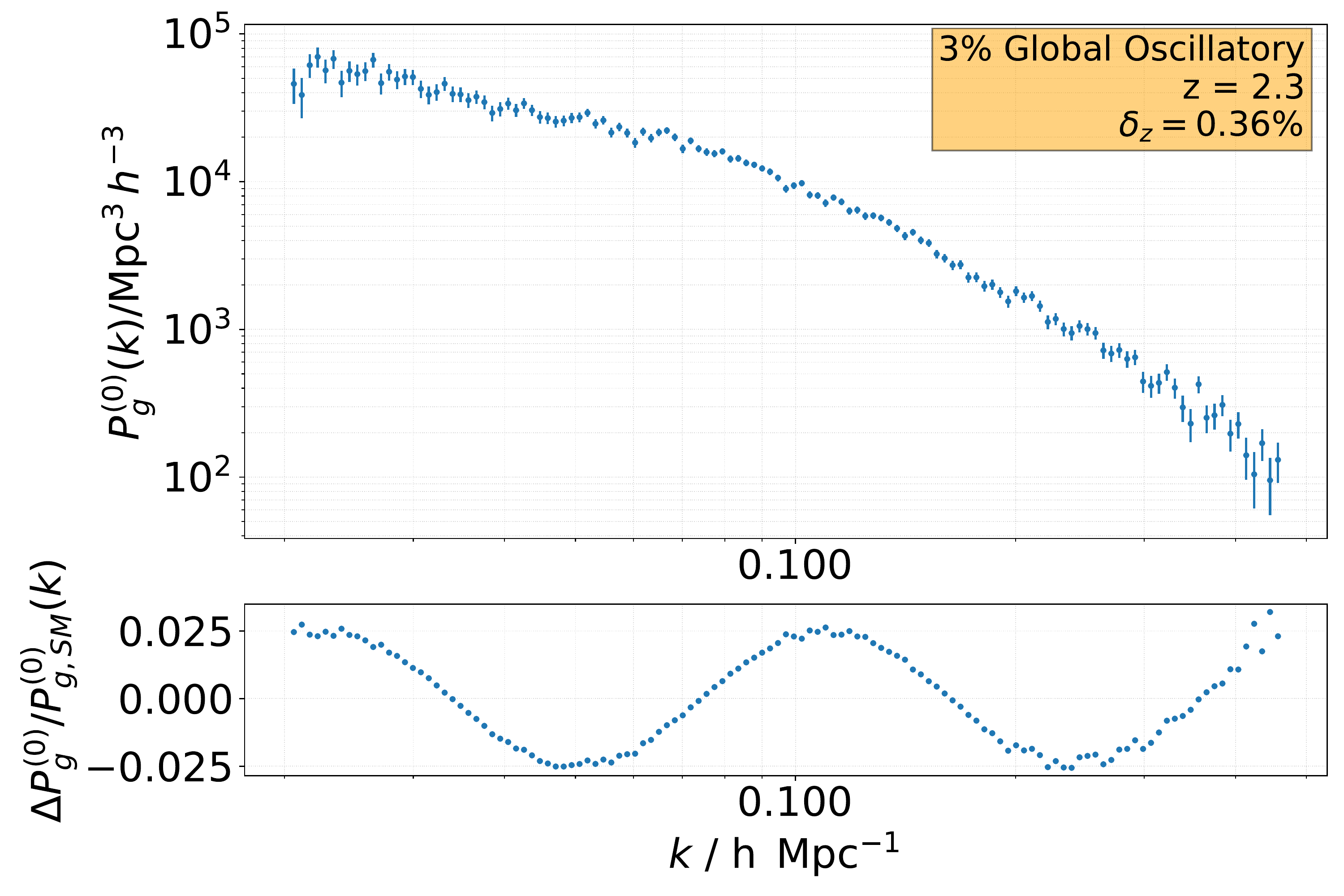}
\caption{Upper panels: Simulated monopole galaxy power spectrum $P_g^{(0)}(k)$ for the high-$z$ survey at the bin $z=2.3$ performed for the global oscillatory feature template. Left: amplitude of the feature $A_{\text{log}} = 0.1$, corresponding to 10\% deviations. Right: amplitude of the feature $A_{\text{log}} = 0.02$, corresponding to 3.\% deviations. Bottom panels: relative differences with respect to the Standard Model monopole galaxy power spectrum $P_{g,\text{SM}}^{(0)}(k)$.}
	 \label{fig:GlobalOscillatoryRealization}
\end{figure}

The simulated $P_g^{(0)}(k)$ for the global oscillatory template are shown in the top panels of \cref{fig:GlobalOscillatoryRealization} for two of the three values of the amplitude considered, $A_{\text{log}} = 0.1,0.02, 0.01$. The $P_g^{(0)}(k)$ oscillations across multiple scale ranges can be clearly seen in the bottom panels.

\begin{figure}[h]
\centering 
\includegraphics[width=.49\textwidth]{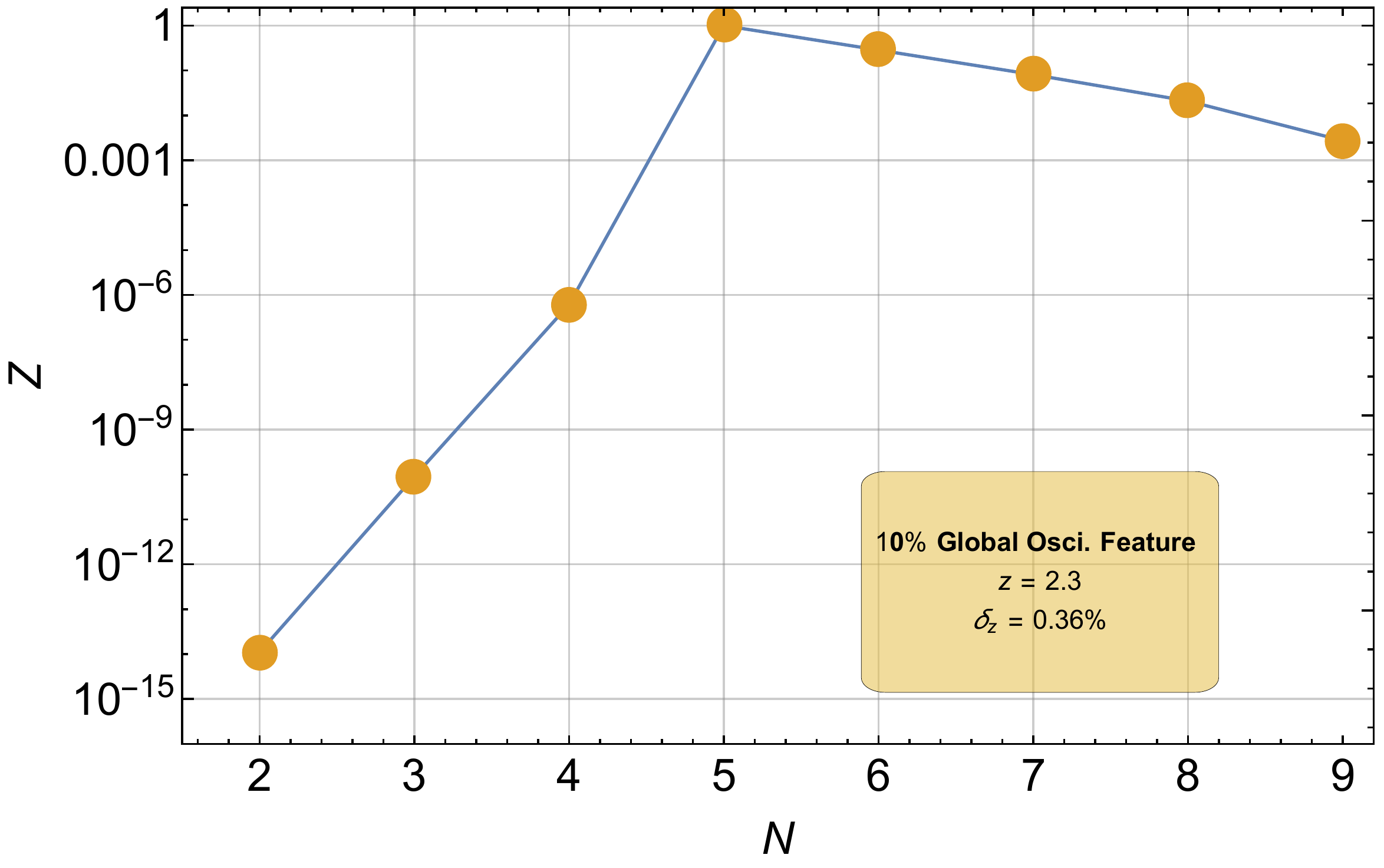}
\hfill
\includegraphics[width=.49\textwidth]{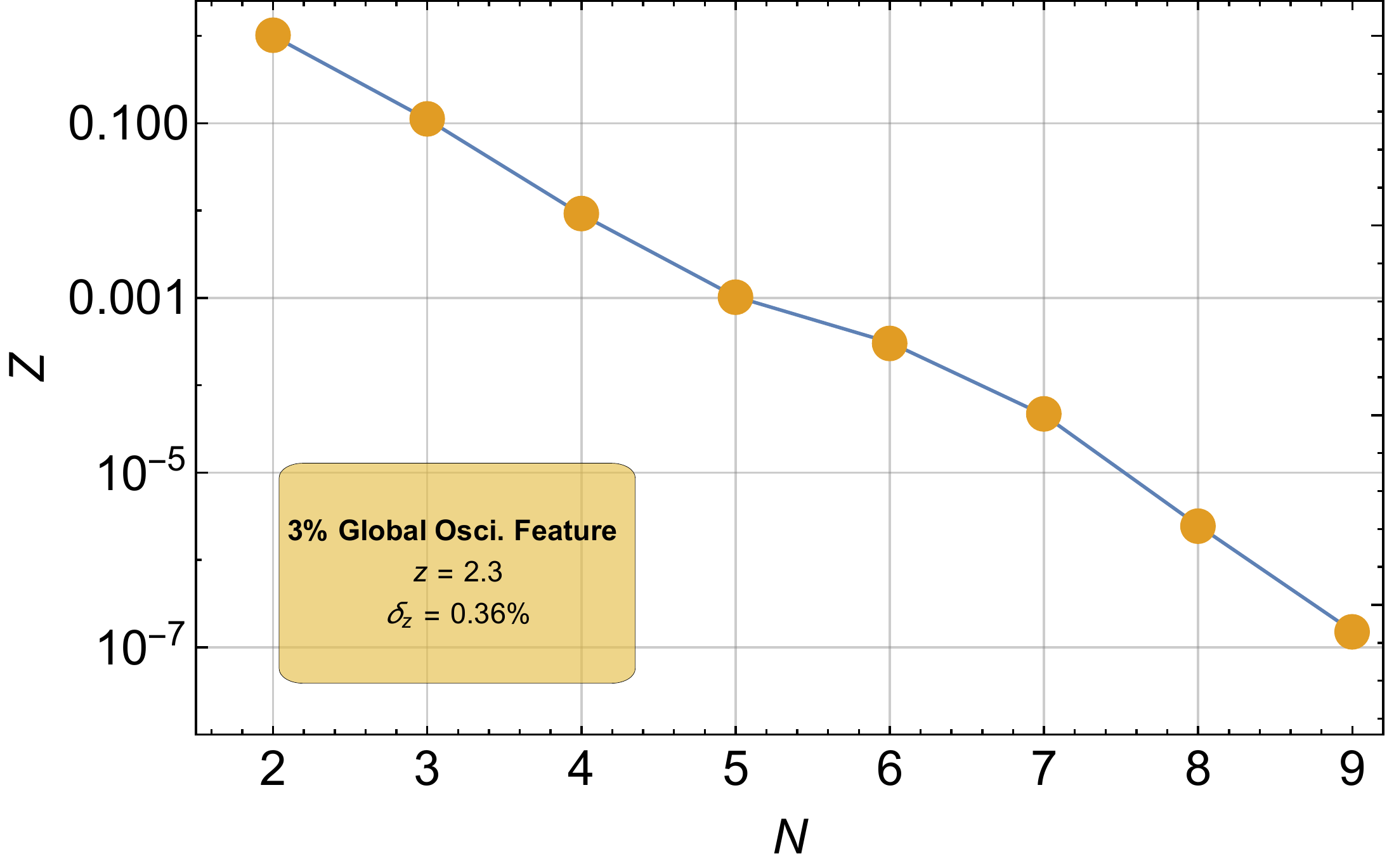}
\caption{Evidence $Z$ of the global oscillatory template reconstructions for each of the $N$ knots configurations considered in the high-$z$ survey with a single bin $z = 2.3$. Left panel: evidences for the reconstruction with 10\% power deviations. Right panel: evidences for the reconstructions with 3\% power deviations.}
	 \label{fig:GlobalOscillatoryEvidences}
\end{figure}

\Cref{fig:GlobalOscillatoryEvidences} shows their global tests. For the $A_{\text{log}} = 0.1$ case, $N=5$ has the highest evidence, with contributions above the 1\% coming from $N \in [6,8]$. The $N \in [2,4]$ configurations have negligible contributions, being $N=2$ the lowest. This indicates a decisive evidence for feature detection for this amplitude. For the $A_{\text{log}} = 0.02$ feature, $N=2$ has the largest contribution, with $Z_3 \approx 0.1 Z_2$, and the rest of evidences following a quasi-exponential fall. This yields an inconclusive result on the global test. \Cref{tab:TableGlobalFeature} summarizes the results of the global test for these two amplitudes, including also a third one of $A_{\text{log}} = 0.01$.

\begin{figure}[h]
\centering 
\includegraphics[width=.49\textwidth]{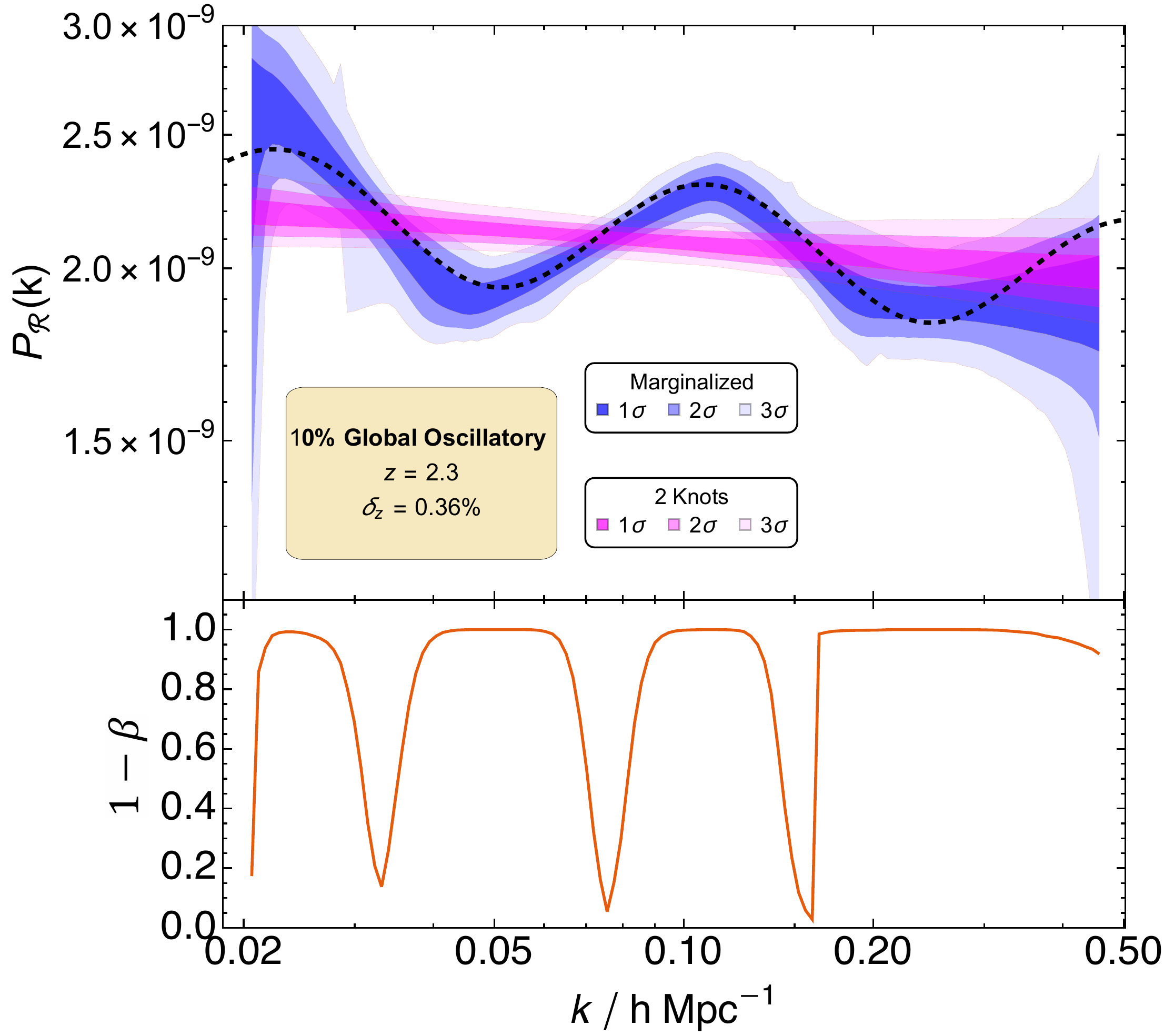}
\hfill
\includegraphics[width=.49\textwidth]{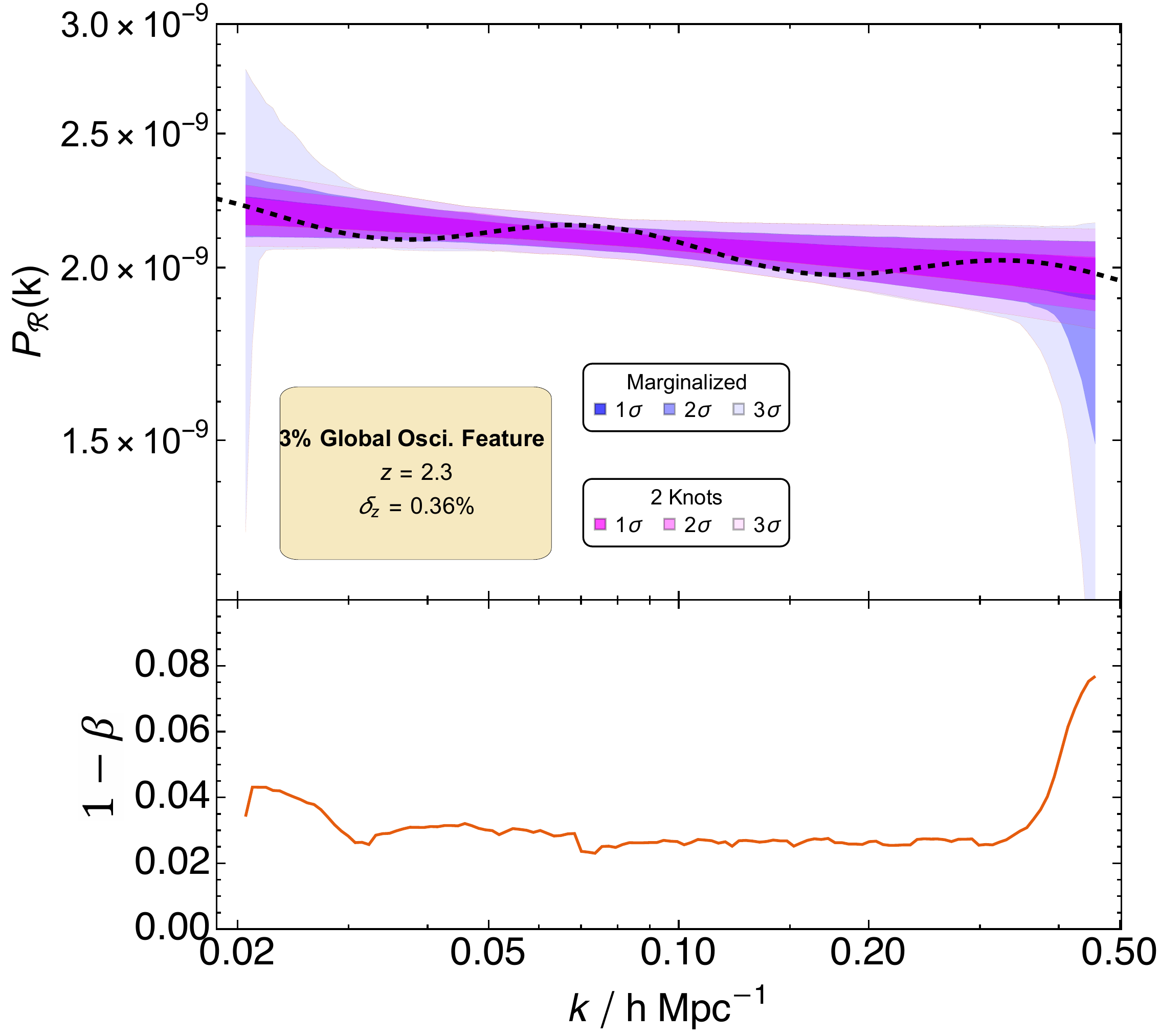}
\caption{Contours of the $P_{\mathcal{R}}(k)$ reconstructions of the global oscillatory feature template with the high-$z$ survey specifications at the bin $z=2.3$ for the $N = 2$ case (magenta) and for the case of marginalized probability over $N$ knots (blue). The oscillatory feature template is shown as a dashed black line. In the bottom panel of each figure the value of the power $1-\beta$ used for detecting features in the hypothesis test. Left figure: reconstructions with $A_{\text{log}} = 0.1$, corresponding to 10\% power deviations w.r.t. the power law. Right figure: reconstructions with $A_{\text{log}} = 0.02$, corresponding to 3\% power deviations w.r.t. the power law.}
	 \label{fig:GlobalOscillatoryContoursAndHypoTest}
\end{figure}

For the $A_{\text{log}} = 0.1$ feature the marginalized contours, represented in the left panel of \cref{fig:GlobalOscillatoryContoursAndHypoTest}, deviate strongly from the power law according to the global oscillatory template, as expected from the global test. The hypothesis test, in the left bottom panel, indicates values of $1-\beta$ very close to $1$ in those ranges of $k$ where the oscillations are located.

The right panel of \cref{fig:GlobalOscillatoryContoursAndHypoTest} shows the reconstructions for the smaller amplitude $A_{\text{log}} = 0.02$. For this weaker oscillations, the marginalized reconstruction follows the power law one, showing no appreciable deviations. Consequently, no detection of the feature is found with the hypothesis test. 
 
\Cref{tab:TableGlobalFeature} summarizes all the results of the global and local tests applied to this feature template. The global oscillatory feature studied can be clearly detected with our methodology for the high-$z$ scenario in both global and local tests when the amplitude causes $10$ \% power deviations w.r.t. the power law. If the deviations are reduced to the $3$ \%, we did not find any deviation from the power law.
\subsection{Summary of the results for low-$z$ and high-$z$ objects}

We explore the impact of sampling the cosmological parameters according to Planck DR 3 uncorrelated Gaussian posteriors in \cref{sec:AppendixComparisonCosmology}. As it is shown in that section, the impact of varying the cosmological parameters on the reconstructions is small. As a result, the outcomes of the global and local tests are very similar in both cases. Therefore, in addition to the previous results for high-$z$ objects, we also reconstruct $P_{\mathcal{R}}(k)$ for both low-$z$ (ELGs) and high-$z$ objects, but we keep the cosmological parameters fixed in order to reduce the computational load.

In \cref{tab:TableLocalFeature,tab:TableGlobalFeature} we summarize the detection status for the considered local and global features respectively, based on the evidences and hypothesis test results.

For a sampled cosmology using the high redshift $z=2.3$ bin we could not detect the local bump feature with a power excess of 5\% at $k \approx 0.2 \text{ h}\text{ Mpc}^{-1}$, but the global and local oscillatory ones are detected with decisive statistical evidence for deviations of the 10\% w.r.t. the power law. When the global oscillatory model is reduced to 3\% deviations, no feature is detected.

For a fixed cosmology we compute reconstructions even combining the information of different redshift bins. The smallest power deviations that we are able to detect with substantial evidence, are $\approx 2\%$ for all the considered features combining either low or high $z$ redshift bins, in scenarios with different redshift bins and photometric errors. As shown in these tables, the local bump feature can only be detected when combining all the redshift bins of the high-$z$ objects. In this case the detection is achieved with decisive evidence, as we also do for all local and global oscillatory realizations with deviations larger than $2\%$.
 \begin{table}[h]
    \centering
    \begin{threeparttable}
        \input{TableLocalFeature}
        \caption{Feature detection status of the SM power law, local bump feature, and local oscillatory feature templates of $P_{\mathcal{R}}(k)$ for four different scenarios of low/high redshifts $z$ and low/high photometric errors $\delta_z$. The status is classified according to the global and local tests applied, which are based on the evidence comparison and the hypothesis test power respectively. We classify the global test results according to the Jeffreys criterion. For the hypothesis test we claim 'none' detection when the power of the test $1-\beta < 0.5$ for all scales, 'hint' of detection when at least at one scale $0.5 < 1-\beta < 0.95$, and 'detection' when at any scale $1-\beta > 0.95$.}
        \label{tab:TableLocalFeature}
        \begin{tablenotes}
            \item[*] Results obtained with a fixed cosmology.
        \end{tablenotes}
    \end{threeparttable}
\end{table}
 \begin{table}[h]
    \centering
    \begin{threeparttable}
        \input{TableGlobalFeature}
        \caption{Feature detection status of the global oscillatory feature template of $P_{\mathcal{R}}(k)$ with three different amplitudes: $A_{\text{log}} = 0.1$, $A_{\text{log}} = 0.02$, and $A_{\text{log}} = 0.01$, corresponding to $10\%$, $3\%$, and $1.5\%$ power deviations. The status of the feature detection is classified with both global and local indicators as in \cref{tab:TableLocalFeature}. The smallest power deviation detected with this feature is 2\%, with a substantial status.}
        \label{tab:TableGlobalFeature}
        \begin{tablenotes}
            \item[*] Results obtained with a fixed cosmology.
        \end{tablenotes}
    \end{threeparttable}
\end{table}

\section{Conclusions}
\label{sec:Conclusions}

We have presented a flexible methodology that reconstructs the primordial power spectrum $P_{\mathcal{R}}(k)$ in a model independent way using Bayesian inference. The main advantages of our methodology are that it reconstructs $P_{\mathcal{R}}(k)$ in a non-parametric way, without assuming any specific model. This is done by sampling an arbitrary number $N$ of knots without any prior in $k$-space, which allows a quantitative comparison of different reconstructions based on the evidence $Z$. Models with lower $Z$ are penalized, contributing less to the marginalized probability over $N$ of the $P_{\mathcal{R}}(k)$ reconstructions.

This methodology is applied to detect deviations from the Standard Model primordial power spectrum. For this, we use a global indicator (evidence comparison) and a local test (hypothesis test), as they provide complementary information. The global test compares the overall fit of different $N$ knot configurations and determines the preferred number of knots. The local test examines each scale separately and identifies where deviations from the power law model ($N = 2$) occur. The combination of different redshift bins can also help to detect features as it increases the signal-to-noise ratio.

We have tested this methodology on simulated spectra by sampling the cosmological parameters $\{H_0,\Omega_b,\Omega_c\}$ following Planck DR 3 uncorrelated Gaussian posteriors. The methodology shows a good performance in all scenarios considered, with or without features, and is sensitive to the signal-to-noise ratio of the different surveys. In addition, we have quantified the impact of varying the cosmological parameters on the reconstructions obtained with a fixed cosmology, finding small differences. Due to this, we also reconstruct with fixed cosmology in order to reduce computational time.

We apply our methodology on different feature templates for $P_{\mathcal{R}}(k)$: a local template from which we generate bump (\cref{fig:LocalBumpFeature}) and oscillatory (\cref{fig:LocalOscillatoryFeature}) features, and a global template generating an oscillatory feature (\cref{fig:GlobalOscillatoryFeature}). For simulations of the local bump, we detect clear deviations from the power law only at high-$z$ and with the combination of all redshift bins (keeping the cosmology fixed in this latter case). For simulations of both local and global oscillations of 10\% power deviations, we detect and reconstruct the features convincingly. 

For a first application to real data, we used a simple semi-empirical description of non-linearities and a BAO modelling, following \cite{SDSSDR04}, to reconstruct $P_{\mathcal{R}}(k)$ from the SDSS LRG 04 catalogue (\cref{fig:SDSSLRG04bestfit}). The evidence substantially supports the power law model over higher $N>2$ knots configurations, and the local test shows no significant deviations at any scale.

The next step is to apply this methodology to more recent galaxy surveys such as BOSS \cite{SDSSIIIBOSS, BOSSDR12} and the forthcoming J-PAS \cite{JPASSpecifications}, DESI \cite{DESISpecifications} and Euclid surveys \cite{EuclidSpecifications}, previously incorporating non-linear perturbative effects, BAO modelling and more advanced redshift-space distortions models \cite{RSDScoccimarro,TNS1}. In order to maximize the sensitivity that can be obtained with this methodology, combinations of the information from different redshift bins should be applied, particularly those with a higher signal-to-noise ratio. Also the inclusion of Planck DR 3 posteriors with correlations between the sampled parameters as cosmological priors and a covariance matrix for the $k$-bins will be explored in applications to recent surveys, something not applied in the present work in order to reduce the computational effort.

Finally, we can envisage several extensions of this methodology that will be addressed in the near future: the use of higher order splines to interpolate between knots, allowing smoother reconstructions; the inclusion of higher order multipoles, $\ell > 0 $, of the galaxy power spectrum in the reconstruction, such as the quadrupole or the hexadecapole; the incorporation of weak lensing data and its modelling, which complements the clustering information and allows smaller scales to be explored; and the extension of the methodology to include CMB data in addition to the LSS ones.

\acknowledgments

We thank J. J. Blanco-Pillado and Antonio L. Maroto for interesting discussions during the development of this work and, the later, together with Carlos Hernández-Monteagudo, for providing estimates of the densities of low-$z$ objects from the miniJPAS survey; Carolina Queiroz and L. Raul Abramo for providing estimates of QSO densities; Fabio Finelli for interesting discussions on possible applications of the method; Will J. Handley for an interesting discussion on CosmoChord at the beginning of the project, and Jesús Torrado for yet another fruitful discussion on Cobaya and PolyChord. The authors also acknowledges programming support from M. Ruiz-Granda, C. Gimeno-Amo, and J. Villafañe-Calvo. The authors also acknowledge Santander Supercomputación support group at the University of Cantabria who provided access to the supercomputer Altamira Supercomputer at the Institute of Physics of Cantabria (IFCA-CSIC), member of the Spanish Supercomputing Network, for performing simulations/analyses. GMS, EMG and AMC thank the Spanish AEI and MICIU for the financial support provided under the project with reference PID2019-110610RB-C21 and acknowledge support from Universidad de Cantabria and Consejería de Universidades, Igualdad, Cultura y Deporte del Gobierno de Cantabria via the \textit{Instrumentación y ciencia de datos para sondear la naturaleza del universo} project, as well as from Unidad de Excelencia María de Maeztu
(MDM-2017-0765). GMS acknowledges financial support from the Formación de Personal Investigador (FPI) programme, ref. PRE2018-085523, associated to the Spanish Agencia Estatal de Investigación (AEI, MICIU) project ESP2017-83921-C2-1-R. GCH acknowledges support through the ESA research fellowship programme. We acknowledge the use of CosmoMC \cite{LewisBridle2002}, CosmoChord \cite{CosmoChordRepository}, PolyChord \cite{MandatoryPolyChord1,MandatoryPolyChord2}, CAMB \cite{CAMB}, GetDist \cite{GetDist}, and Cobaya \cite{CobayaMandatory1,CobayaMandatory2}. The Cobaya code has been created with the help of the python packages \texttt{NumPy} \cite{Numpy}, \texttt{Matplotlib} \cite{Matplotlib} and \texttt{SciPy} \cite{Scipy}.

\appendix
\section{Solution of the label switching problem}

\label{sec:AppendixSwitching}

The label switching problem appears in Bayesian analysis of models with multiple indistinguishable parameters whose order is arbitrary. The permutation of any of these parameters is equivalent \cite{SwitchingLabelProblem}, leading to highly multi-modal posterior distributions in models with many such parameters.

As described in \cref{subsec:reconstructions}, we sample $N-2$ coordinates for the normalized scales $x_i$, with $2 \leq i \leq N-1$ (in the case of one or two knots, no $k$ coordinates are sampled). The variables $x_i$ are indistinguishable parameters that the sampler can order arbitrarily, causing the label switching problem. A change of coordinates of these scales $x_i$ into a $(N-2)$-dimensional hypertriangle of coordinates $x_i^{\prime}$ solves the problem \cite{SwitchingLabelProblem}:
\begin{equation}\label{SwitchingLabelsolution}
x_{i}^{\prime}=x_{i-1}^{\prime}+\left(1-x_{i-1}^{\prime}\right)\left[1-\left(1-x_{i}\right)^{\frac{1}{N+1-i}}\right],
	\end{equation}
with $x'_1 = x_1 = 0$ and $x'_N = x_N = 1$. This transformation can be interpreted as an ordering of the $k$s for sampling, removing the degeneracy of $\mathcal{L}$ with respect to its different $k$ permutations and thus avoiding this problem.

\section{Sample selection in PolyChord}
\label{sec:AppendixPolyChord}

\begin{figure}[t]
	 \centering
	    \includegraphics[width=0.44\textwidth]{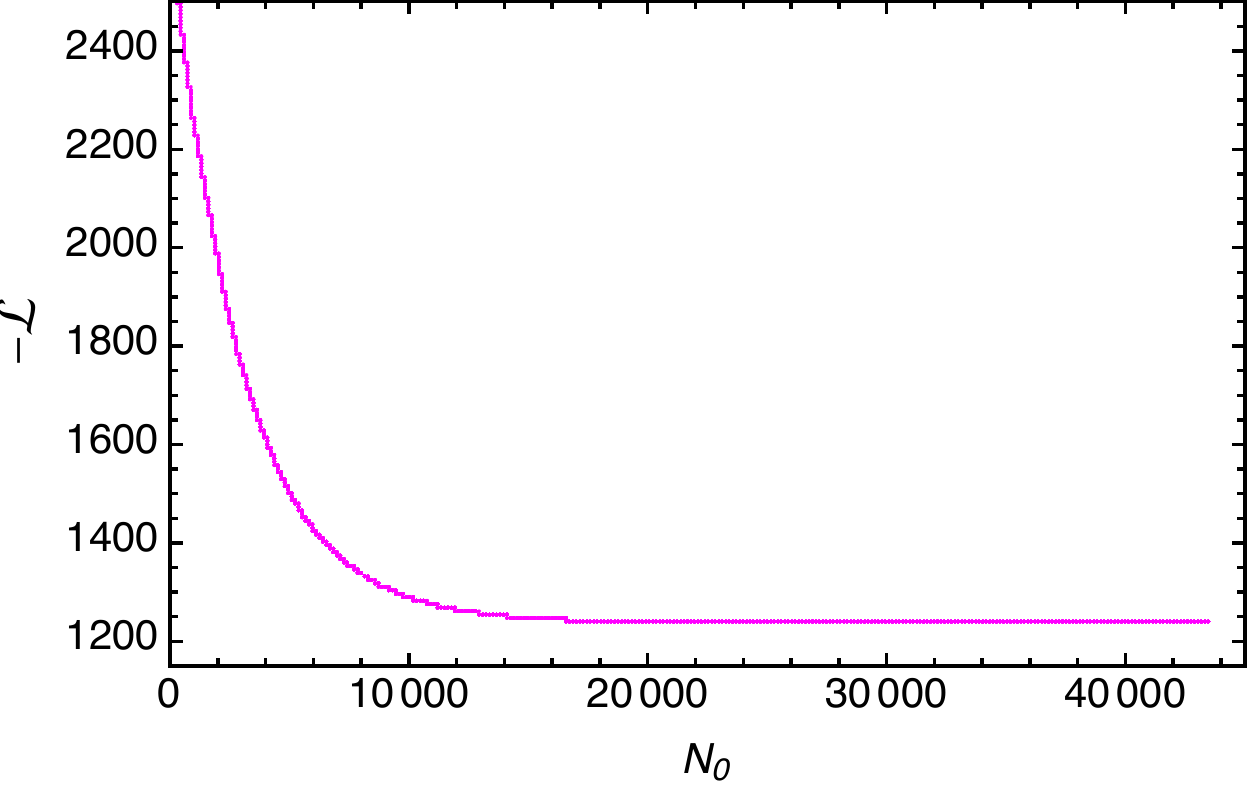}
	    \includegraphics[width=0.44\textwidth]{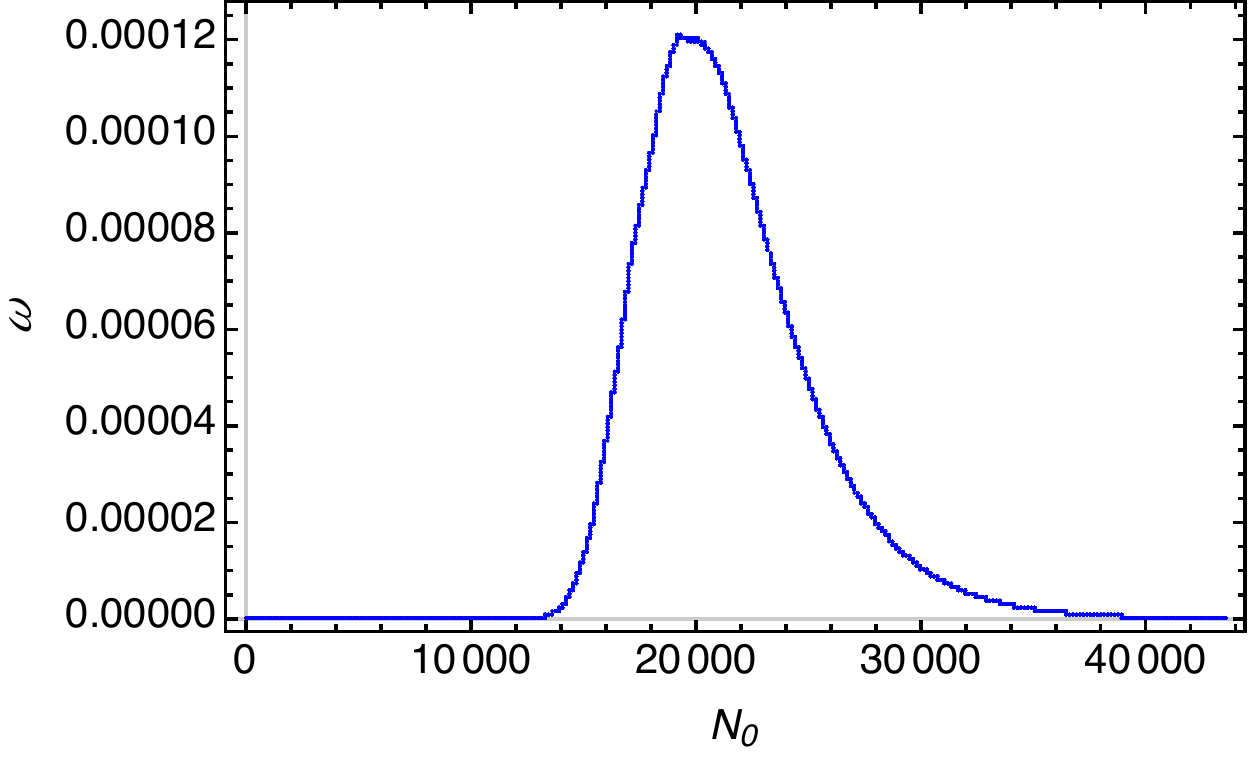}
   	 \caption{Left panel: minus likelihood $-\mathcal{L}$ against time $N_0$. Note how $\mathcal{L}$ increases monotonically with the advance of the step. Right panel: normalized importance weights $\omega$ against $N_0$. Note that the maximum in the weights is found at steps when the Markov chain is halfway completed, i.e at $N_0 \approx N_{0,\text{max}}/2$.}
   	 \label{fig:likewvsN0}
	\end{figure}

In this paper, we use the sampler PolyChord to sample the primordial power spectrum. At each step of the sampling, a set of parameters determining the power spectrum is obtained. The chains represent samples of the primordial power spectrum with a probability according to the posterior. In this method, the likelihood $\mathcal{L}$ increases monotonically as the sampling progresses (see left panel of \cref{fig:likewvsN0}). PolyChord provides the importance weights $\omega$ \cite{MandatoryPolyChord2}, that once normalized can be interpreted as the probability of that particular point of the parameter space. These weights increase with time $N_0$ until they reach a maximum value, then they drop by a few orders of magnitude (see right panel of figure \cref{fig:likewvsN0}). Thus, the reconstructions with the highest $\omega$  won't have the highest $\mathcal{L}$ values. However, they correspond to the tail where $\mathcal{L}$ is barely improved at each step.

We filter the $i$ obtained reconstructions for each number of knots $N$ to remove the burn-in and to exclude reconstructions with a low likelihood value. For that, we use a Monte Carlo method that selects reconstructions according to their values of $\omega_{i,N}$. If $\omega_{i,N}$ is larger than a random value from 0 to 1, that reconstruction is kept. Otherwise, it is discarded. Due to the shape of importance weights through the chain step $\omega(N_0)$ (the highest $\mathcal{L}$ points do not have the largest $\omega$, see the correspondence in \cref{fig:likewvsN0}), the reconstructions with the highest likelihood also fail this Monte Carlo filtering in most cases, and the ones with the lowest likelihood fail in almost all cases. The retained reconstructions $P_{\mathcal{R}}(k)$ have high $\omega$ values while still having high enough $\mathcal{L}$ values. We label by $P_{\mathcal{R},j,N}$ the reconstructions that passed through this filter.

\section{Local feature template}

\label{sec:AppendixModifiedPPS}

The parametrization of the primordial power spectrum used as the local bump and local oscillatory feature template has a localized oscillatory burst \cite{MirandaHuModeloFeature}. Steps in the warp or potential, over which the inflaton rolls in much less than an e-fold, generate oscillations in the power spectrum that can be modelled with this parametrization \cite{MirandaHuModeloFeature}. It is written as a perturbation of the standard power law $P_{\mathcal{R}}(k)$:
	\begin{equation}\label{ModifiedPPS}
\text{log} \left( \mathcal{P}_{\mathcal{R}}(k) \right) =\exp \left[\text{log} \left(\mathcal{P}_{\mathcal{R}}^{\text{SM}}(k)\right) +\mathcal{I}_{0}(k) +  \text{log} \left(1+\mathcal{I}_{1}^{2}(k)\right)\right],
	\end{equation}
where the first and second order terms $\mathcal{I}_{0}(k)$ and $\mathcal{I}_{1}(k)$ are:
	\begin{equation}\label{I0Term}
\mathcal{I}_{0}= {\left[\mathcal{A}_{1} \mathcal{W}_{1}^{(0)}\left(k / k_{\mathrm{s}}\right)+\mathcal{A}_{2} \mathcal{W}_{2}^{(0)}\left(k / k_{\mathrm{s}}\right)\right.} \\
\left.+\mathcal{A}_{3} \mathcal{W}_{3}^{(0)}\left(k / k_{\mathrm{s}}\right)\right] \mathcal{D}\left(\frac{k / k_{\mathrm{s}}}{x_{\mathrm{s}}}\right),
	\end{equation}
	\begin{equation}\label{I1Term}
\mathcal{I}_1=\frac{1}{\sqrt{2}}\left\{\frac{\pi}{2}\left(1-n_{\mathrm{s}}\right)+\left[\mathcal{A}_{1} \mathcal{W}_1^{(1)}\left(k / k_{\mathrm{s}}\right)+\mathcal{A}_2 \mathcal{W}_2^{(1)}\left(k / k_{\mathrm{s}}\right)+\mathcal{A}_3 \mathcal{W}_3^{(1)}\left(k / k_{\mathrm{s}}\right)\right] \mathcal{D}\left(\frac{k / k_{\mathrm{s}}}{x_{\mathrm{s}}}\right)\right\} ,
	\end{equation}
with $\mathcal{D}(x)$ being the dumping function, $\mathcal{D}(x)=\frac{x}{\sinh x}$.

This model has five parameters: three amplitudes $\{\mathcal{A}_1,\mathcal{A}_2,\mathcal{A}_3\}$, the scale $k_s$ where the oscillations start, and the damping parameter $x_s$. The set of window functions $W_{i}^{(j)}$ are given below.

 The burst in $\mathcal{P}_{\mathcal{R}}(k)$ corresponds to a sudden step-like feature in the inflation potential  \cite{ModelAdams2001} or the sound speed \cite{ModelAchucarro2010, ModelInflationSpeedSound}. Specifically, this parametrization describes a $\tanh$-step in the inflationary potential and in the warp term of a DBI model \cite{MirandaHuModeloFeature}.
  
The window functions used in the modified power spectrum model of \cref{ModifiedPPS} are:
	\begin{equation}\label{W:W01}
W_{1}^{(0)}(x)=\frac{1}{2 x^{3}}\left[\left(18 x-6 x^{3}\right) \cos 2 x+\left(15 x^{2}-9\right) \sin 2 x\right],
	\end{equation}
	\begin{equation}\label{W:W02}
\mathcal{W}_{2}^{(0)}(x)=\frac{3}{2 x^{3}}\left[\sin (2 x)-2 x \cos (2 x)-x^{2} \sin (2 x)\right],
	\end{equation}	
	\begin{equation}\label{W:W03}
\mathcal{W}_{3}^{(0)}(x)=\frac{1}{x^{3}}\left[6 x \cos (2 x)+\left(4 x^{2}-3\right) \sin (2 x)\right],
	\end{equation}	
	\begin{equation}\label{W:W11}
W_{1}^{(1)}(x)=-\frac{1}{x^{3}}\left\{3(x \cos x-\sin x)\left[3 x \cos x+\left(2 x^{2}-3\right) \sin x\right]\right\},
	\end{equation}
		\begin{equation}\label{W:W12}
	W_{2}^{(1)}(x)=\frac{3}{x^{3}}(\sin x-x \cos x)^{2},
	\end{equation}
	\begin{equation}\label{W:W13}
	\mathcal{W}_{3}^{(1)}(x)=-\frac{1}{x^{3}}\left[3+2 x^{2}-\left(3-4 x^{2}\right) \cos (2 x)-6 x \sin (2 x)\right].
	\end{equation}

\section{Posteriors sample}
\label{sec:AppendixPosteriors}

\begin{figure}[t]
	 \centering
	    \includegraphics[width=0.999\textwidth]{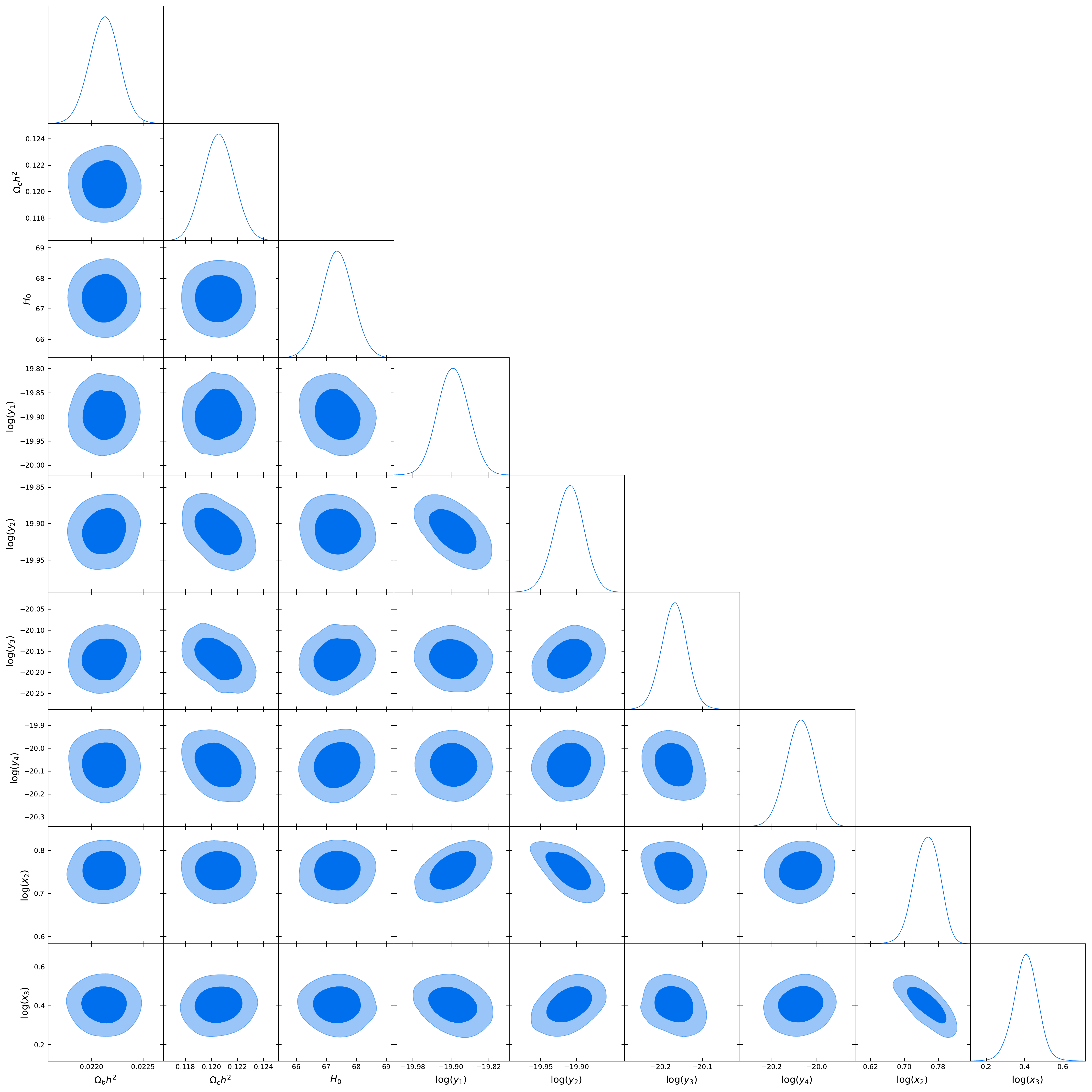}
	 \caption{Triangle plot of the posterior distribution for the 4 knots case of the local oscillatory feature reconstruction. The figure was obtained with the statistical python package \texttt{GetDist}.}
	 \label{fig:PosteriorsTriangle}
	\end{figure}

In \cref{fig:PosteriorsTriangle} we provide a triangle plot of the posterior distributions for the sampled parameters in the 4 knots case of the local oscillatory feature reconstruction studied in \cref{subsec:local_bump_feature}. No strong correlation or degeneracy among the parameters is present, specially between the cosmological and knot ones. Similar posteriors are obtained for other cases of features and number of knots.

\section{Comparison between fixed and sampled cosmology for the $P_{\mathcal{R}}(k)$ reconstructions}

\label{sec:AppendixComparisonCosmology}

In this appendix we quantify the impact of varying the cosmological parameters on the $P_\mathcal{R}(k)$ reconstructions obtained with a fixed cosmology.

\begin{figure}[h]
\centering 
\includegraphics[width=.49\textwidth]{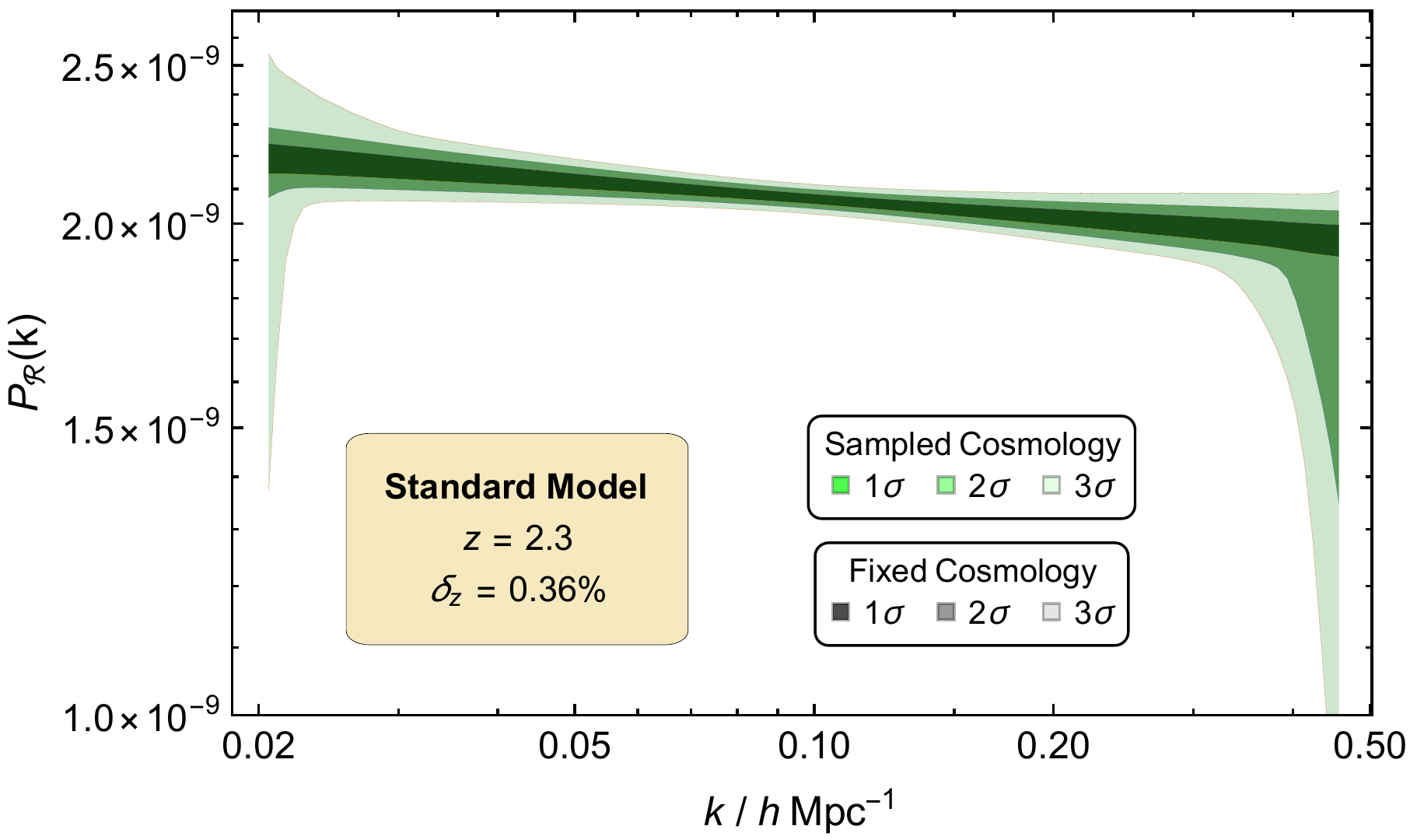}
\hfill
\includegraphics[width=.49\textwidth]{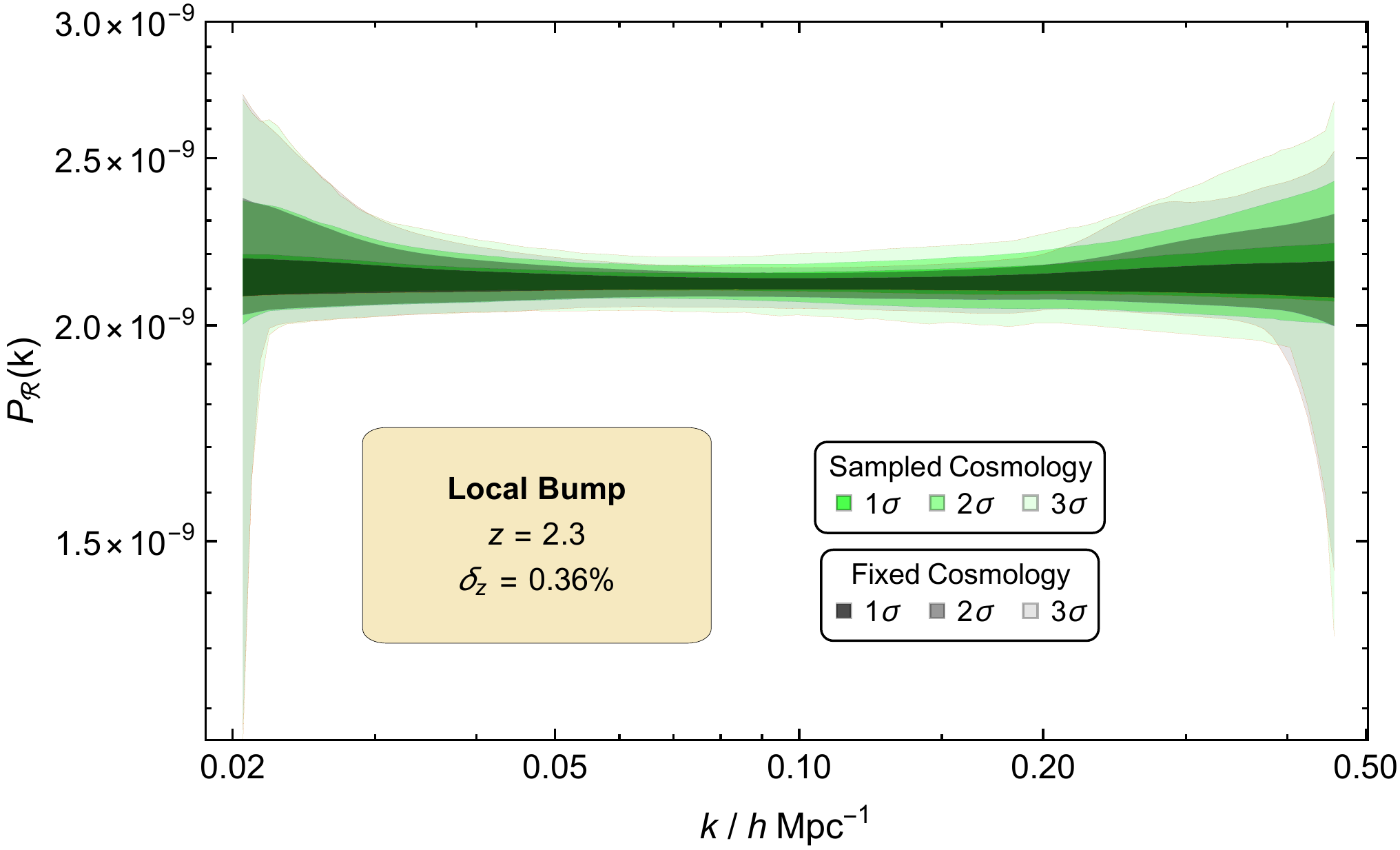}
\hfill
\includegraphics[width=.49\textwidth]{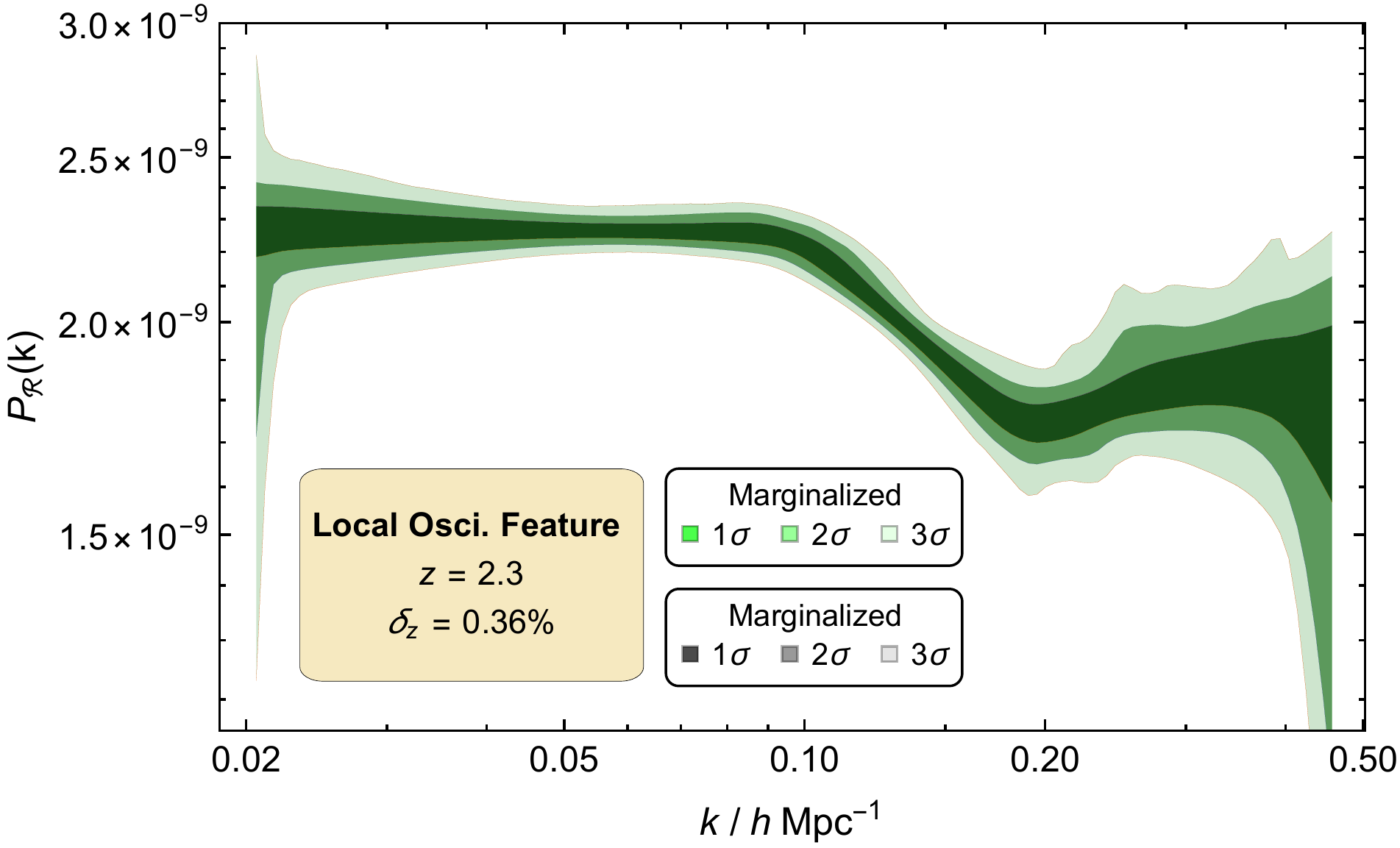}
\caption{Reconstructions of $P_{\mathcal{R}}(k)$ sampling the cosmological parameters $\{H_0,\Omega_b,\Omega_c\}$ (green) and keeping them fixed (black). Top figures: reconstructions for the SM (left) and the local bump feature (right). Bottom figure: reconstruction for the local oscillatory feature.}
	 \label{fig:CosmoSampling}
\end{figure}

The results of this comparison are presented in \cref{fig:CosmoSampling} for high-$z$ objects in the cases of the Standard Model and local features.
In all cases we obtain increments of the 1$\sigma$ contours smaller than 20\%. For the specific case of the local oscillatory feature deviations up to a factor of 2 appear at the smallest scales for the 3$\sigma$ contours (lower panel). For the rest of scales the differences are much smaller, even smaller than 1 \%.

\begin{figure}[t]
	 \centering
	   \label{fig:CosmoSamplingEvidences}
	    \includegraphics[width=0.44\textwidth]{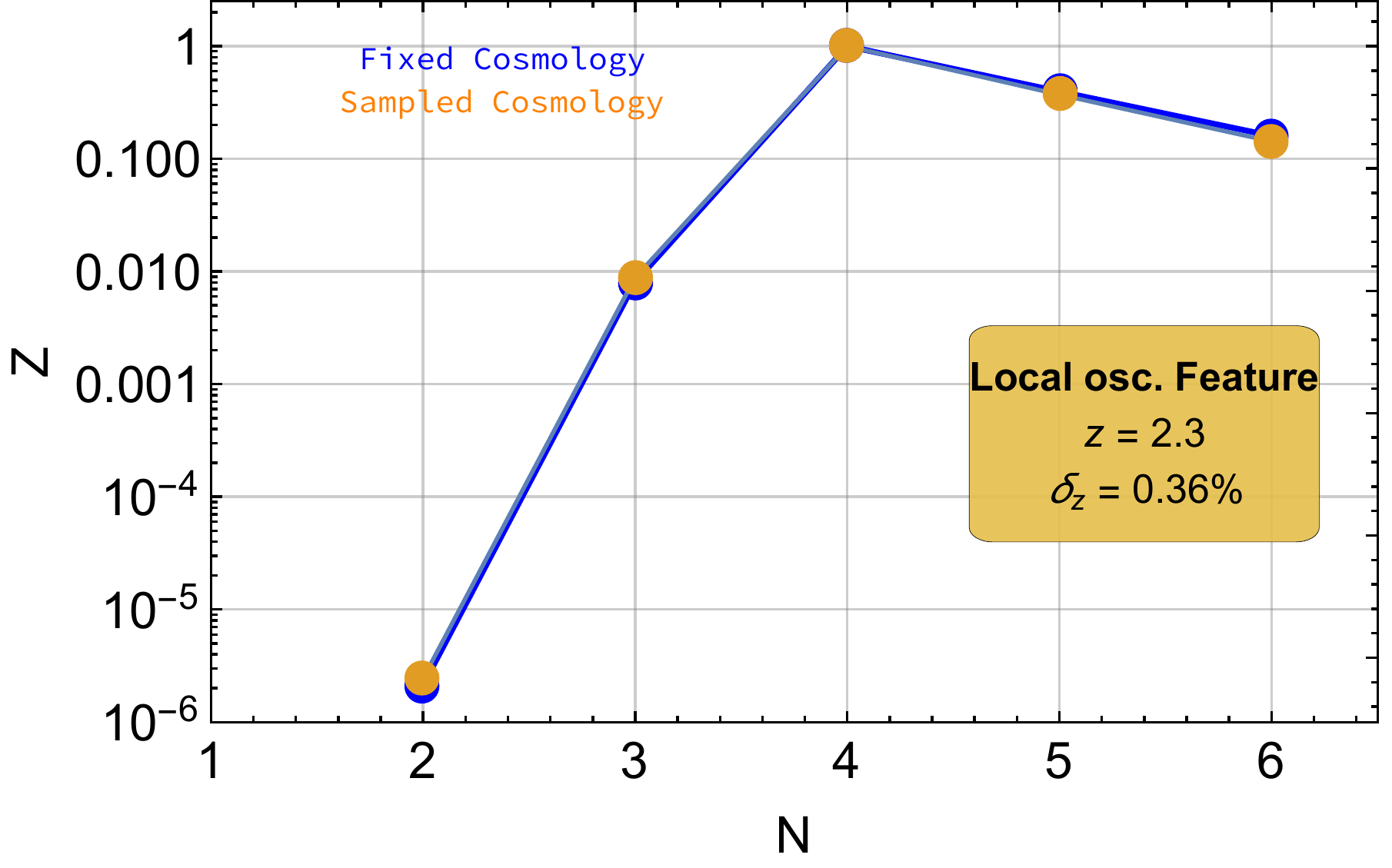}
	   \label{fig:CosmoSamplingLocal}
	    \includegraphics[width=0.44\textwidth]{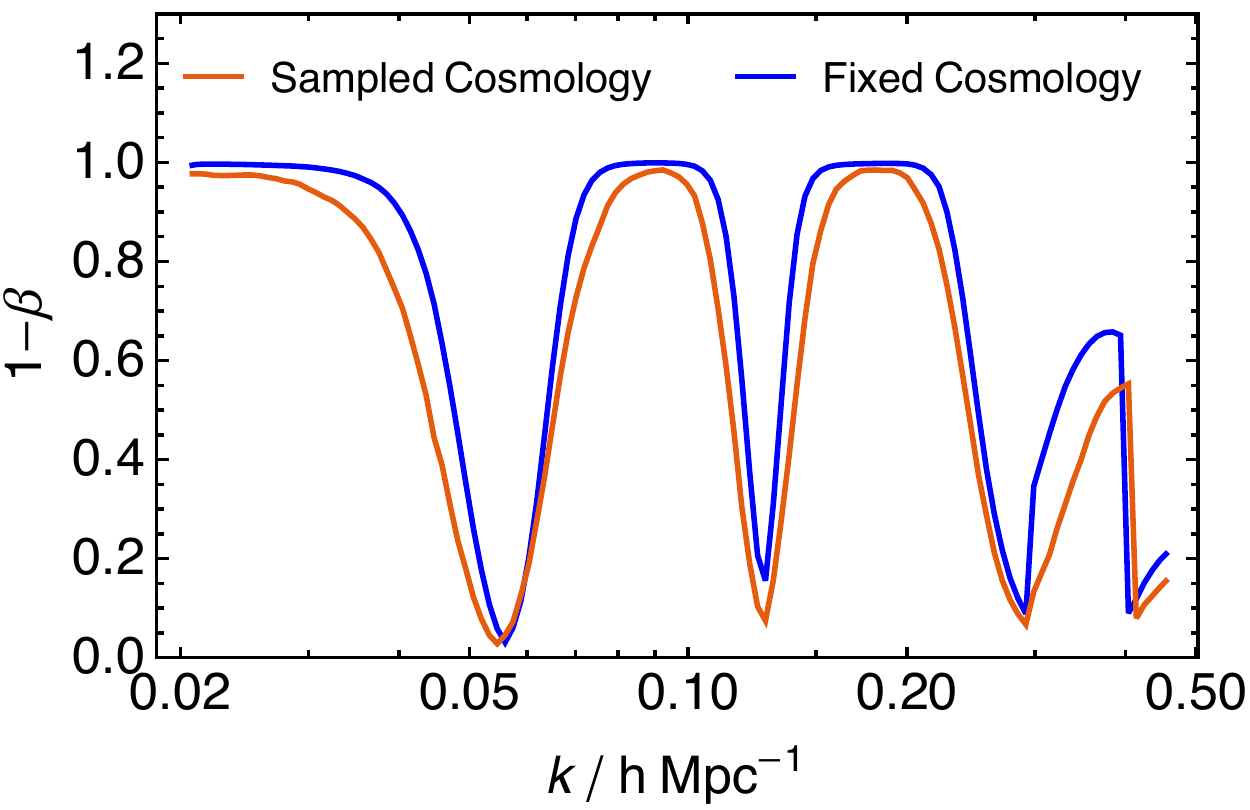}
   	 \caption{Comparison of tests for a sampled cosmology (orange) and a fixed one (blue). Left figure: global test. Right figure: local test.}
   	 \label{fig:CosmoSamplingEvidencesAndLocal}
	\end{figure}

Regarding the impact in the evidences and the hypothesis test, we obtain small changes, as expected from the high similarity in the reconstructions. This is shown in \cref{fig:CosmoSamplingEvidencesAndLocal}, where in the left panel we compare the values of the evidences and in the right panel the hypothesis tests. The values of the evidences are very similar.
In relation to the power of the test (right panel), the maximum deviations for a sampled cosmology are similar, and the feature detection intervals appear at the same scales but with reduced widths.

\section{Application to real data: SDSS LRG 04}
\label{sec:AppendixApplicationSDSSLRG04}

\subsection{The SDSS LRG 04 data}

We present a first application of our methodology to real LSS data. We choose the catalogue provided by the Sloan Digital Sky Survey Luminous Red Galaxies 04 data release (SDSS LRG 04). The catalogue \cite{SDSSLRG04} contains a sample of 58360 luminous red galaxies, selected\footnote{In \cite{SDSSLRG04} another SDSS DR 04 sample is given: the SDSS Main DR 04. This catalogue is about six times smaller than the SDSS LRG 04, and even though the occupation of a larger volume and a more strongly clustered LRGs, the number of LRGs is an order of magnitude lower, leading to a SDSS Main DR 04 $P_g(k)$ LRG catalogue having uncertainties about 6 times bigger. Thus, we use just the SDSS LRG 04 sample in order to have a better $S/N$ in the galaxy power spectrum.} from the 4th data release of the SDSS \cite{SDSSDR04}. These galaxies have a redshift range of $0.155 < z < 0.474$, and they form a particularly clean and uniform galaxy sample, consisting of luminous early-types galaxies at all redshifts.

Two improvements in the construction of the galaxy power spectrum model for the SDSS LRG 04 are introduced compared to our previous simulations model used in \cref{sec:Simulations}. Its galaxy power spectrum model incorporates a BAO modelling and a semi-empirical description of non-linearities:
	\begin{equation}\label{GPSSDSSLRG04}
P_g^{(0)}(k) =  b^2 \frac{1+Q_{nl} k^2}{1+A_g k} P_{\text{dewiggled}}(k),
	\end{equation}
where $b = 1.89$ is the galaxy bias and $Q_{nl}$ and $A_g$ are parameters accounting for non-linearities \cite{Ag14, QnlAg}. Non-linearities are introduced in a semi-empirical way through the factor $ \frac{1+Q_{nl} k^2}{1+A_g k}$. According to \cref{GPSSDSSLRG04}, $P_{\text{dewiggled}}(k)$ accounts for the BAO signal, written as:
	\begin{equation}\label{SDSSdewiggled}
P_{\text{dewiggled}} =  W_{\text{SDSS}}(k) P_m(k)+[1-W_{\text{SDSS}}(k)] P_{\text {nowiggle}}(k),
	\end{equation}
where the matter power spectrum is calculated with CAMB, and $P_{\text{nowiggle}}(k)$ is the `no-wiggle' power spectrum defined in \cite{ref89SDSS}, suppressing oscillations of $P_m(k)$. $W_{\text{SDSS}}(k)$ is a window function chosen as:
	\begin{equation}\label{SDSSwindowfunction}
W_{\text{SDSS}}(k) \equiv e^{-\left(k / k_{*}\right)^{2} / 2},
	\end{equation}
with $k_{*}$ being the wiggle suppression scale defined in \cite{ref88SDSS}. This BAO modelling mimics the smoothing of the power spectrum caused by the non-linear matter clustering.  

The analysis of the SDSS LRG 04 data is performed with fixed cosmological parameters (also compatible with Planck DR 3), and relies on a likelihood function that assumes a cosmology independent and diagonal covariance matrix. The covariance matrix is derived from simulations \cite{SDSSLRG04}.

\begin{figure}[t]
	 \centering
	   \label{fig:PkSDSSAjusteLinearLog}
	    \includegraphics[width=0.44\textwidth]{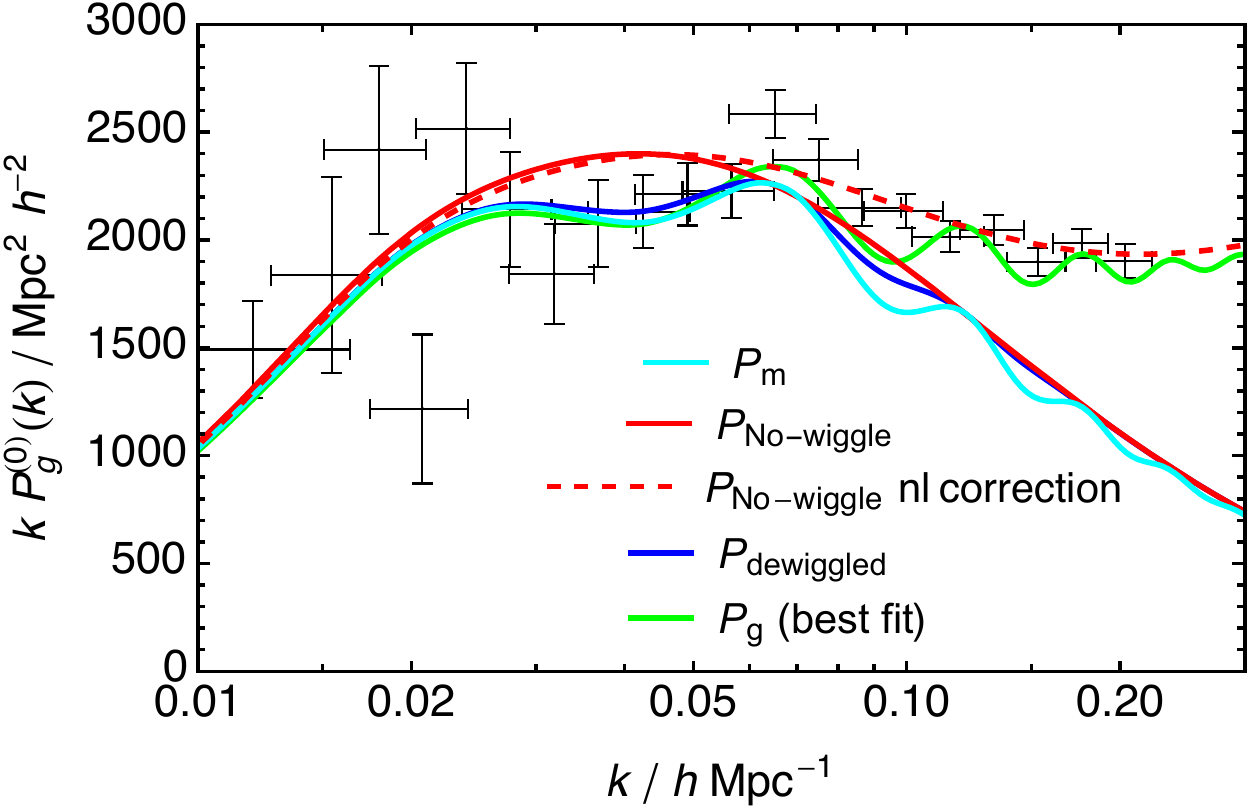}
	   \label{fig:PkSDSSAjusteLogLog}
	    \includegraphics[width=0.44\textwidth]{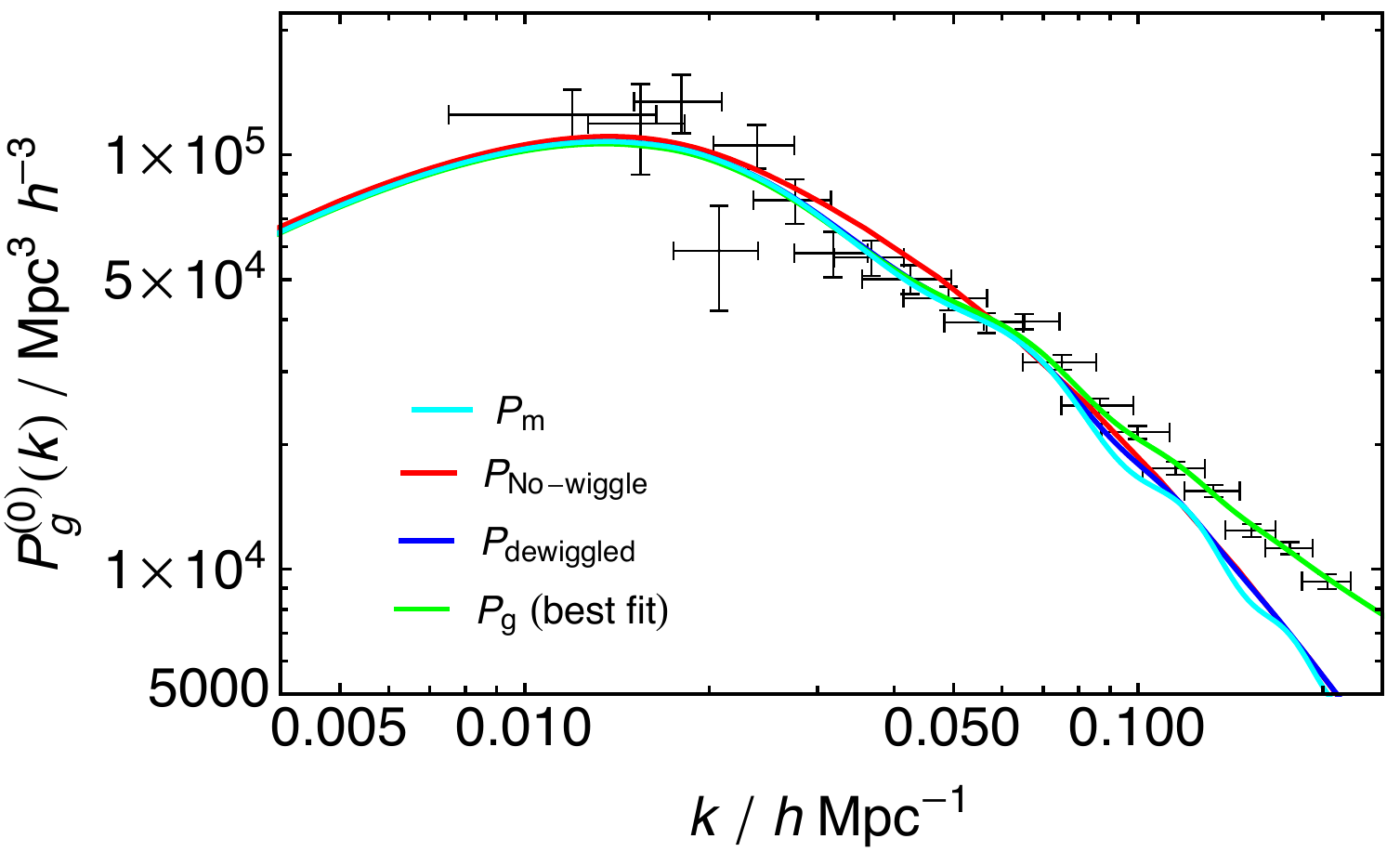}
   	 \caption{Galaxy power spectrum data for the SDSS LRG 04 (black points) in semi-logarithmic scale represented as $k P_g^{(0)}(k)$ (left panel) and logarithmic scale (right panel). In green the employed best fit of $P_g^{(0)}(k)$ is plotted, in cyan the matter power spectrum $P_m(k)$, in red $P_{\text{nowiggle}}(k)$, and in blue $P_{\text{dewiggled}}(k)$. The uncertainties on $P_{g}^{(0)}(k)$, indicated by the vertical bars, are 1$\sigma$ and they are uncorrelated, even though the horizontal error bars overlap \cite{SDSSLRG04}. Note that the models without the non-linear correction $\frac{1+Q_{nl} k^2}{1+A_g k}$ are not able to properly track the non-linearity of data at scales bigger than $k \approx 0.09$ h/Mpc. In the left panel the effect of the non-linear correction is shown explicitly with two plots of the no wiggle spectrum $P_{\text{nowiggle}}(k)$: without the non-linear correction (continuous red line) and with it (dashed red line).}
   	 \label{fig:SDSSLRG04bestfit}
	\end{figure}

\Cref{fig:SDSSLRG04bestfit} shows $P_m(k)$, $P_{\text{nowiggle}}(k)$, $P_{\text{dewiggled}}(k)$ and $P_g(k)$ along with the SDSS LRG 04 data, ensuring that the constructed $P_g(k)$ reproduces the best fit in \cite{SDSSLRG04}.
The linear matter power spectrum provides a reasonably good parametrization of the shape of the monopole galaxy power spectrum at large scales. At small scales $k \sim 0.1 $ linear theory fails to fit the data accurately, as shown in the left panel of \cref{fig:SDSSLRG04bestfit}. The non-linearities alter the broad shape of the matter power spectrum and wash out baryon wiggles on small scales, among other effects. 
In particular, most of the departure from the linear regime at the small scales is caused by considering the underlying matter power spectrum given by the multiple galaxies sharing the same dark matter halo instead of the one given by the dark matter halos themselves (see \cite{Ag14} for more details).

In the case of SDSS LRG 04, the non-linear effects start to be significant at $k \approx 0.09$ h/Mpc (see \cref{fig:SDSSLRG04bestfit}). The addition of the BAO modelling and the non-linear correction term $\frac{1+Q_{nl} k^2}{1+A_g k}$ helps significantly in fitting the data \cite{SDSSLRG04}, although the non-linear correction has a semi-empirical origin and it is not physically derived. More realistic and physically motivated models of galaxy clustering are needed for upcoming stage IV LSS surveys, including a bias model such as in \cite{RoyMcDonald2009}, a model of redshift space distortions as in \cite{TNS1}, and a more physically motivated derivation on non-linearities as in \cite{BOSSDR12}.

\subsection{Results and statistical interpretation}
	
\begin{figure}[t]
\centering 
	    \includegraphics[width=0.45\textwidth]{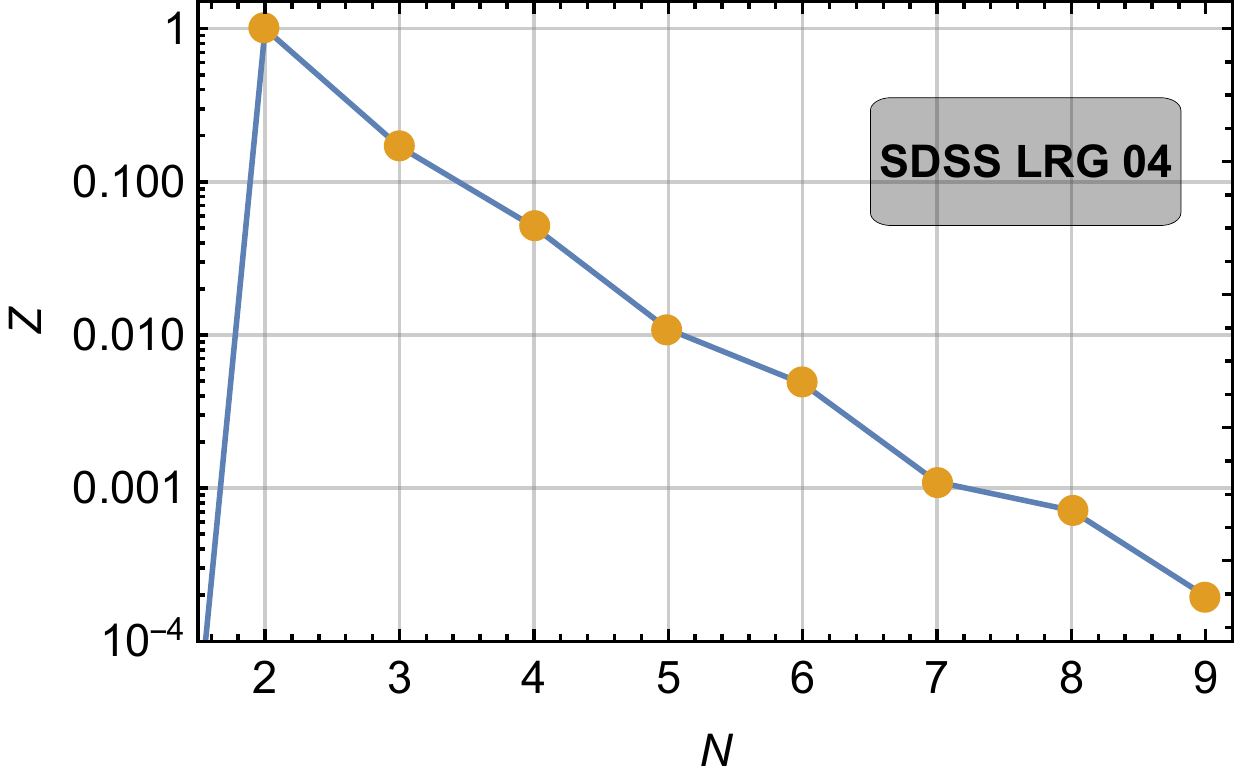}
\hfill
	    \includegraphics[width=0.45\textwidth]{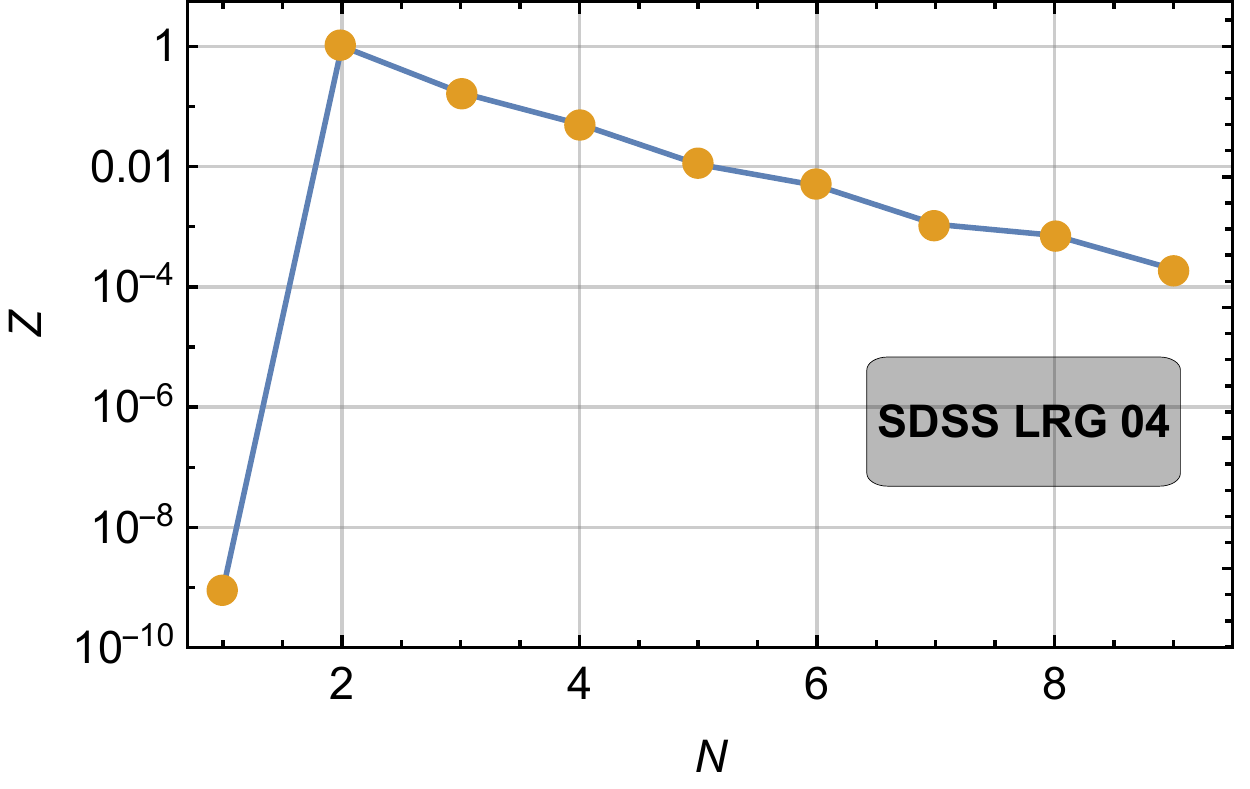}

\caption{Evidence $Z$ of the SDSS LRG 04 reconstruction for each of the $N$ knots configuration considered. Left panel: all $N$ configurations excluding $N = 1$. Right panel: all configurations with $N = 1$ case included.}
 \label{fig:SDSSEvidences}
\end{figure}

 \Cref{fig:SDSSEvidences} shows the evidences for the SDSS LRG 04. The power law model has the highest evidence $Z$. $Z_3$ contributes slightly over a 10\% relative to $Z_2$, while the evidences for configurations with higher number of knots decrease quasi-exponentially. The $N = 1$ configuration, which corresponds to a primordial power spectrum with $n_s = 1$, has an evidence many orders of magnitude smaller that $Z_2$, $Z_1 \approx 10^{-9} Z_2$, as shown in the right panel of \cref{fig:SDSSEvidences}. Thus, it is strongly disfavoured. The Jeffreys criterion favours the power law in our SDSS LRG 04 analysis.

\begin{figure}[t]
	 \centering
	    \includegraphics[width=0.68\textwidth]{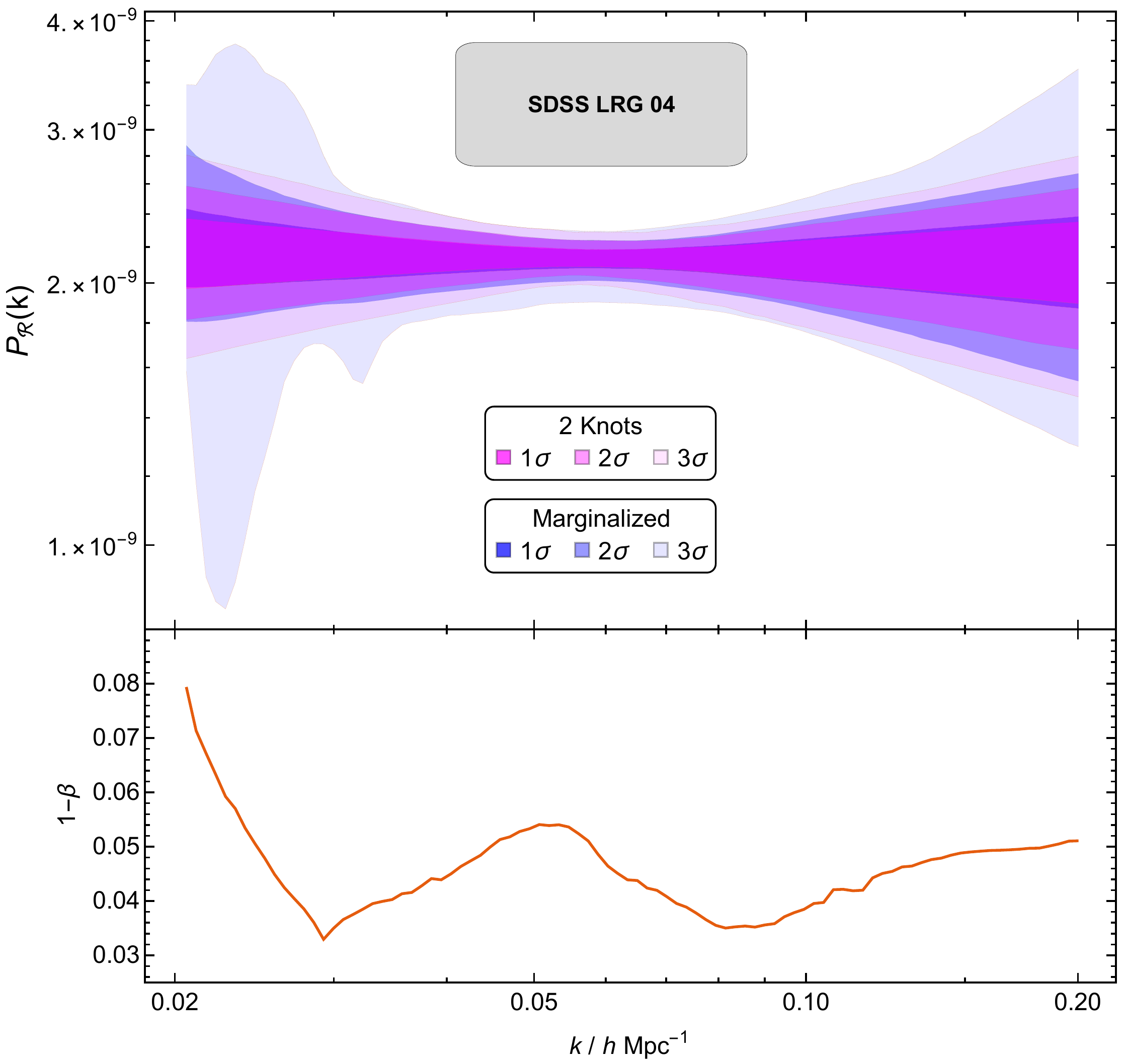}
	 \caption{Contours of $P_{\mathcal{R}}(k)$ reconstructions for the SDSS LRG 04 catalogue for the $N = 2$ case (magenta) and for the case of marginalized probability over $N$ knots (blue). In the bottom panel the values of the power $1 - \beta$ obtained from the hypothesis test are  plotted.}
	 \label{fig:SDSSContoursHypo}
	\end{figure}

The marginalized probability over $N$ and $N=2$ reconstructions are very similar (top panel of \cref{fig:SDSSContoursHypo}), except for the 3$\sigma$ confidence level contours of the marginalized one that are much wider, mainly at the largest scales. The hypothesis test does not show significant discriminating power between both distributions at any scale, being $1-\beta < 0.08$ at all scales (see bottom panel of \cref{fig:SDSSContoursHypo}), well below the 0.5 threshold. Therefore, we conclude that deviations are not significant and the power law is the preferred model for the SDSS LRG 04 $P_{\mathcal{R}}(k)$ reconstructions according to both of our tests.

\bibliographystyle{JHEP.bst}
\bibliography{bibliography}

\end{document}

%% file: PriorTable.tex
\begin{tabular}{lll} 
Sampled Parameters & Prior & Range \\
\hline
$H_0$ & Gaussian & {$67.4 \pm 0.54$ $\mathrm{km} \text{ } \mathrm{s}^{-1} \mathrm{Mpc}^{-1}$} \\
$\Omega_b h^2$ & Gaussian & {$0.02212\pm 0.0015$} \\
$\Omega_c h^2$ & Gaussian & {$0.1206 \pm 0.0012$} \\
$\text{log}(y_i)$ & uniform & {$[-23,-19]$} \\
$\text{log}(x_i)$ & uniform & {$[0,1]$} \\ \\

Fiducial Cosmology  & Value \\

\hline
\centering
$H_\text{fid}$ & {$67.4$} $\mathrm{km} \text{ } \mathrm{s}^{-1} \mathrm{Mpc}^{-1}$ \\
$\Omega_{b,\text{fid}} h_{\text{fid}}^2$ & {$0.02212$} \\
$\Omega_{c,\text{fid}} h_{\text{fid}}^2$  & {$0.1206$} \\
$A_{s,\text{fid}}$  & {$2.0905 \times 10^{-9}$} \\
$n_{s,\text{fid}}$ & {$0.9626$} \\
$k_0$  & {$0.05$} $\mathrm{Mpc}^{-1}$  \\
$\tau$ & {$0.07$} \\
\hline
\end{tabular}

%% file: DensitiesLow.tex
\begin{tabular}{| c| c| c| }
 \hline
\textbf{Low-z} & Low-$\delta_z$ & High-$\delta_z$    \\
 \hline
$z$-bin & $\hat{n}/  10^{-3} h^3 \text{ Mpc}^{-3}$ & $\hat{n}/  10^{-3} h^3 \text{ Mpc}^{-3}$    \\
 \hline\hline
$0.1$ &     12       & 48      \\
$0.3$ &     6.1      & 25      \\
$0.5$ &     2.4      & 10       \\
$0.7$ &     0.75     & 3.2      \\
$0.9$ &     0.19     & 0.87     \\
 \hline
\end{tabular}

%% file: DensitiesHigh.tex
\begin{tabular}{| c| c| }
 \hline
\textbf{High-z} & Low-$\delta_z$    \\
 \hline
$z$-bin & $ \hat{n}/ 10^{-3} h^3 \text{ Mpc}^{-3}$    \\
 \hline\hline
$1.7$ &     0.030      \\
$1.9$ &     0.035     \\
$2.1$ &     0.037     \\
$2.3$ &     0.037    \\
$2.5$ &     0.033    \\
$2.7$ &     0.028    \\
$2.9$ &     0.023    \\
 \hline
\end{tabular}

%% file: TableLocalFeature.tex
\begin{tabular}{|c| c| c| c| c|c| c| }
 \hline

\textbf{Local feature} & \multicolumn{2}{|c|}{\textbf{Standard Model}} & \multicolumn{2}{|c|}{\textbf{Bump}}& \multicolumn{2}{|c|}{\textbf{Oscillation}}  \\
\hline
\textbf{Test}  &  Global &  Local   &  Global  &  Local  &  Global  &  Local  \\
\hline
 Low $z$ low $\delta_z$*  &  Negative &  None  &  Negative &  None &  Decisive &  Detection       \\
  
 Low $z$ high $\delta_z$*  &  Negative &  None  &  Negative &  None &  Decisive &  Detection      \\
  High $z$ all bins* &  \cellcolor{black!25}--- &  \cellcolor{black!25}---  &  Decisive &  Detection &  \cellcolor{black!25}--- &  \cellcolor{black!25}---       \\

 High $z$ bin $z = 2.3$ &  Negative &  None  &  Negative &  None &  Decisive &  Detection       \\

\hline
\end{tabular}

%% file: TableGlobalFeature.tex
\begin{tabular}{|c| c| c| c| c| c| c| }
 \hline
\textbf{Global feature} & \multicolumn{2}{|c|}{$A_{\text{log}} = 0.1 \text{ }(10\%)$} & \multicolumn{2}{|c|}{$A_{\text{log}} = 0.02 \text{ }(3\%)$ } & \multicolumn{2}{|c|}{$A_{\text{log}} = 0.01 \text{ }(1.5\%)$}  \\
\hline
\textbf{Test}  &  Global &  Local &  Global &  Local  &  Global &  Local       \\
\hline
 Low $z$ bin $z = 0.9$*  &  Decisive &  Detection  &  \cellcolor{black!25}--- &  \cellcolor{black!25}--- &  \cellcolor{black!25}--- &  \cellcolor{black!25}---     \\
 
 Low $z$ all bins*  &  Decisive &  Detection &  Decisive &  Detection &  Inconclusive &  Hint       \\

High $z$ bin $z = 2.3$  &  Decisive &  Detection &  Negative &  None &  Negative &  None       \\

 \hline
\end{tabular}